\input harvmac
\input epsf

\let\includefigures=\iftrue
%
%
%
\newfam\black

\input epsf
\noblackbox
%
%
\includefigures
\message{If you do not have epsf.tex (to include figures),}
\message{change the option at the top of the tex file.}
\def\figin{\epsfcheck\figin}\def\figins{\epsfcheck\figins}
\def\epsfcheck{\ifx\epsfbox\UnDeFiNeD
\message{(NO epsf.tex, FIGURES WILL BE IGNORED)}
\gdef\figin##1{\vskip2in}\gdef\figins##1{\hskip.5in}
\else\message{(FIGURES WILL BE INCLUDED)}%
\gdef\figin##1{##1}\gdef\figins##1{##1}\fi}
\def\DefWarn#1{}

\def\figinsert{\goodbreak\midinsert}
\def\ifig#1#2#3{\DefWarn#1\xdef#1{fig.~\the\figno}
\writedef{#1\leftbracket fig.\noexpand~\the\figno}%
\figinsert\figin{\centerline{#3}}\medskip\centerline{\vbox{\baselineskip12pt
\advance\hsize by -1truein\noindent\footnotefont{\bf
Fig.~\the\figno:} #2}}
\bigskip\endinsert\global\advance\figno by1}
\else
\def\ifig#1#2#3{\xdef#1{fig.~\the\figno}
\writedef{#1\leftbracket fig.\noexpand~\the\figno}%
\global\advance\figno by1} \fi

\def\alpalp{\hbox{$\alpha$\kern -0.55em $\alpha$}}
\def\betbet{\hbox{$\beta$\kern -0.50em $\beta$}}

\def\lroverarrow#1{\raise4.2truept\hbox{$\displaystyle
\leftrightarrow\atop\displaystyle#1$}}
\def\underarrow#1{\vbox{\ialign{##\crcr$\hfil\displaystyle
 {#1}\hfil$\crcr\noalign{\kern1pt\nointerlineskip}$\longrightarrow$\crcr}}}

\def\bigvev#1{\left\langle{#1}\right\rangle}

%

\font\teneurm=eurm10 \font\seveneurm=eurm7 \font\fiveeurm=eurm5
\newfam\eurmfam
\textfont\eurmfam=\teneurm \scriptfont\eurmfam=\seveneurm
\scriptscriptfont\eurmfam=\fiveeurm

 \font\teneusm=eusm10 \font\seveneusm=eusm7 \font\fiveeusm=eusm5
\newfam\eusmfam
\textfont\eusmfam=\teneusm \scriptfont\eusmfam=\seveneusm
\scriptscriptfont\eusmfam=\fiveeusm
\def\eusm#1{{\fam\eusmfam\relax#1}}
\font\tencmmib=cmmib10 \skewchar\tencmmib='177
\font\sevencmmib=cmmib7 \skewchar\sevencmmib='177
\font\fivecmmib=cmmib5 \skewchar\fivecmmib='177
\newfam\cmmibfam
\textfont\cmmibfam=\tencmmib \scriptfont\cmmibfam=\sevencmmib
\scriptscriptfont\cmmibfam=\fivecmmib


%
\def\EUN{\eusm N}

\def\EUP{\eusm P}
\def\BR{\eusm R}
\def\p{\partial}
\def\tilde{\widetilde}
\def\bar{\overline}
\def\hat{\widehat}

\def\half{{1 \over 2}}

\def\l{\left}
\def\r{\right}

\def\dd{\dot}
\def\ld{\lambda}

\def\T{{\rm T}}

\def\R{{\rm R}}
\def\H{{\rm H}}

\def\P{{\rm P}}
\def\K{{\rm K}}
\def\Spin{{\rm Spin}}

\def\A{{\rm A}}

\def\Ch{{\rm Ch}}
\def\v{{\rm v}}

\def\Hom{{\rm Hom}}

\def\td{{\hbox{Td}}}

\def\Tr{{{\rm Tr~ }}}

\def\ad{{\rm ad~ }}

\def\Re{{\rm Re\hskip0.1em}}
\def\Im{{\rm Im\hskip0.1em}}

\def\W{{\rm W}}
\def\v{{\rm v}}
\def\x{{\rm x}}
\def\b{{\rm b}}
\def\d{{\rm d}}
\def\a{{\rm a}}
\def\c{{\rm c}}
\def\B{{\rm B}}

\def\CO{{\cal O}}
\def\O{{\cal O}}

\def\CS{{\cal S}}

\def\ct{\Bbb{C}}

\def\BM{\Bbb{M}}
\def\BD{\Bbb{D}}

\def\BZ{\Bbb{Z}}
\def\BT{\Bbb{T}}
\def\BE{\Bbb{E}}
\def\BQ{\Bbb Q}
\def\BPT{\Bbb {PT}}
\def\BP{\Bbb P}

\def\BFW{{\bf W}}

\def\BFR{{\bf R}}
\def\BFS{{\bf S}}
\def\BFT{{\bf T}}
\def\BFW{{\bf W}}
\def\BFH{{\bf H}}
\def\BFQ{{\bf Q}}
\def\BFP{{\bf P}}
\def\BFV{{\bf V}}

\def\RR{{\BFR}}

\def\t{\times}
\def\M{{\rm M}}

\def\cta{{\bf Acknowledgements}:\ }
\overfullrule=0pt

\Title{} {\vbox{\centerline{Superconformal Field Theory In Six
Dimensions}
\bigskip \centerline{And Supertwistor}}}
\smallskip
\centerline{Tong Chern$^\dagger$}
\smallskip
\centerline{\it Institute of High Energy Physics, Chinese Academy of
Sciences} \centerline{\it P. O. Box 918(4), 100049 Beijing,
China}\medskip \centerline{\it $^\dagger$ tongchen@mail.ihep.ac.cn}

\bigskip

\smallskip
\smallskip
\input amssym.tex

\noindent
We studied the quantum dynamics of six dimensional $\EUN=(2, 0)$
superconformal field theory (the QNG theory). We developed the
spinor technique for six-dimensional quantum field theories. By
combining this technique with the canonical quantization procedure,
we can overcome the subtlety of the chiral nature of $\EUN=(2, 0)$
free tensor multiplet and work out its quantum mechanical theory. We
then studied the $\BFT^2$ compactification of the QNG theory and
argued that the resulting four dimensional quantum field theory is
indeed the $\EUN=4$ super-Yang-Mills theory with the $SL(2, \BZ)$
duality coming from the mapping class symmetries of $\BFT^2$. We
also investigated the BPS self-dual string excitations by proposing
a CFT description to their world sheet theory. At last, we
constructed the super-twistor space $\hat{\BP\BT}$ that corresponded
to the superconformal group $U^*Sp(4|2, \H)\subset OSp(8|4, \ct)$,
encoded the full free tensor multiplet into it and made some
speculations on the possible super-twistor formulation of the QNG
theory. \Date{May, 2009}

\listtoc \writetoc

\newsec{Introduction}
\seclab\intro

\nref\wittenoncft{E. Witten, ¡°Conformal field theory in four and
six dimensions,¡± Prepared for Symposiumon Topology, Geometry and
Quantum Field Theory (Segalfest), Oxford, England, United Kingdom,
24-29 Jun 2002 [arXiv:math.RT/0712.0157];
E. Witten, ¡°Geometric Langlands From Six Dimensions,¡± [arXiv:hep-th/0905.2720]. }%
\nref\witten{E. Witten, ¡°Some comments on string dynamics,¡±
 in Future Perspectives In String Theory,
ed. I. Barset. al., 501, World Scientific 1996 [arXiv:hep-th/9507121].}%
\nref\adscftreport{O. Aharony, S. S. Gubser, J. M. Maldacena, H.
Ooguri and Y. Oz, ¡°Large N field theories, string theory and
gravity,¡± Phys. Rept. 323 (2000) 183 [arXiv:hep-th/9905111]; J. M.
Maldacena, ¡°TASI 2003 lectures on AdS/CFT,¡± Sep 2003. 45pp.
Presented at Theoretical Advanced Study Institute in Elementary
Particle Physics (TASI 2003): Recent Trends in String Theory,
Boulder, Colorado, 1-27 Jun 2003. Published
in *Boulder 2003, Progress in string theory* 155-203 [arXiv:hep-th/0309246].}%
\nref\DKutasov{A. Giveon and D. Kutasov, ¡°Brane dynamics and gauge
theory,¡± Rev. Mod. Phys. 71 (1999) 983.}
\nref\olivemontonen{C.
Montonen and D. Olive, ¡°Magnetic monopoles as gauge particles?¡±
Phys. Lett. B 72 (1977) 117; P. Goddard, J. Nuyts, and D. Olive,
¡°Gauge theories and magnetic charge,¡± Nucl. Phys. B 125 (1977) 1;
H. Osborn, ¡°Topological charges for N = 4 supersymmetric gauge
theories and monopoles of spin 1,¡± Phys. Lett.
B 83 (1979) 321.}%
\nref\Stromingeropenbrane{A. Strominger, ¡°Open p-branes,¡±
 Phys. Lett. B 383 (1996) 44 [arXiv:hep-th/9512059].}%
\nref\wittenfive{E. Witten, ¡°Five-branes and M-theory on an
orbifold,¡± Nucl. Phys. B 463 (1996) 383 [arXiv:hep-th/9512219].}%
\nref\bfss{T. Banks, W. Fischler, S. H. Shenker, L. Susskind, M
theory as a matrix model: a conjecture, Phys. Rev. D 55 (1997) 5112
[arXiv:hep-th/9610043].}%
\nref\cftinmatrix{M. Berkooz, M. Rozali and N. Seiberg, ¡°Matrix
description of M-theory on $T^4$ and $T^5$,¡± Phys. Lett. B 408
(1997) 105 [arXiv:hep-th/9704089]; M. Rozali, ¡°Matrix theory and U
duality in seven dimensions,¡± Phys. Lett. B
400 (1997) 260 [arXiv:hep-th/9702136].}%
\nref\adscft{J. M. Maldacena, ¡°The large N limit of superconformal
field theories and supergravity,¡± Adv. Theor. Math. Phys. 2 (1998)
231 [Int. J. Theor. Phys. 38 (1999) 1113] [arXiv:hep-th/9711200]; S.
S. Gubser, I. R. Klebanov and A. M. Polyakov,¡°Gauge theory
correlators from non-critical string theory,¡± Phys. Lett. B 428
(1998) 105 [arXiv:hep-th/9802109]; E. Witten, ¡°Anti De Sitter space
and holography,¡± Adv. Theor. Math.
Phys. 2 (1998) 253 [arXiv:hep-th/9802150].}%
\nref\nahm{W. Nahm, ¡°Supersymmetries and their representations,¡±
Nucl. Phys. B 135 (1978) 149.}%
\nref\seiberg{N. Seiberg, ¡°Note on theories with 16 supercharges,¡±
Nucl. Phys. Proc. Suppl. 67 (1998) 158 [arXiv:hep-th/9705117].}%
\nref\wittenmaction{E. Witten, ¡°Five-brane effective action in
M-theory,¡± J. Geom. Phys. 22 (1997) 103 [arXiv:hep-th/9610234]; E.
Witten, ¡°Duality relations among topological effects in string
theory,¡± J. High Energy Phys. 05 (2000) 031 [arXiv:hep-th/9912086].}%
\nref\dufflu{M. Duff and J. Lu. , Nucl. Phys. B 416 (1994) 301.
[arXiv:hep-th/9306052].} \nref\twistorwitten{E. Witten,
¡°Perturbative gauge theory as a string theory in twistor space,¡±
Commun. Math. Phys. 252 (2004) 189 [arXiv:hep-th/0312171].}
\nref\twistorstring{F. Cachazo and P. Svr\v{c}ek, ¡°Lectures on
Twistor Strings and Perturbative Yang-Mills Theory,¡± PoS RTN2005
(2005) 004 [hep-th/0504194].} \nref\twistorberk{N. Berkovits, ¡°An
alternative string theory in twistor space for N=4
super-Yang-Mills,¡± Phys. Rev. Lett. 93 (2004) 011601
[arXiv:hep-th/0402045].} \nref\matrixcft{O. Aharony, M. Berkooz, S.
Kachru, N. Seiberg, E. Silverstein, ¡°Matrix description of
interacting theories in six dimensions,¡± Adv. Theor. Math. Phys. 1
(1998) 148 [arXiv:hep-th/9707079]; O. Aharony, M. Berkooz, N.
Seiberg, ¡°Light-cone description of (2, 0) superconformal theories
in six
dimensions,¡± Adv. Theor. Math. Phys. 2 (1998) 119 [arXiv:hep-th/9712117].}%
\nref\henningson{P. Arvidsson, E. Flink and M. Henningson, ¡°Thomson
scattering of chiral tensors and scalars against a self-dual
string,¡± J. High Energy Phys. 12 (2002) 010 [arXiv:hep-th/0210223];
P. Arvidsson, E. Flink and M. Henningson, ¡°The (2,0) supersymmetric
theory of tensor multiplets and self-dual strings in six
dimensions,¡± J. High Energy Phys. 05 (2004) 048
[arXiv:hep-th/0402187]; M. Henningson, ¡°Self-dual strings in six
dimensions: anomalies, the ADE-classification, and the world-sheet
WZW-model,¡± Commun. Math. Phys. 257 (2005) 291
[arXiv:hep-th/0405056]; E. Flink, ¡°Scattering amplitudes for
particles and strings in six-dimensional (2,0) theory, ¡± Phys. Rev. D 72 (2005) 066010 [arXiv:hep-th/0505015].}%
\nref\Proeyen{P. Claus, R. Kallosh and A. Van Proeyen, M5-brane and
superconformal (0,2) tensor multiplet in 6 dimensions, Nucl. Phys.
B, 518 (1998) 117, [arXiv: hep-th/9711161]. }\nref\minwalla{S.
Minwalla, "Restrictions imposed by superconformal invariance on
quantum field theories," Adv. Theor. Math. Phys. 2(1998) 781-846,
[arXiv:hep-th/9712074].} \nref\Connell{C. Cheung, D. O'Connell,
Amplitudes and Spinor-Helicity in Six Dimensions, JHEP, [arXiv:
hep-th/0902.0981v2] .} \nref\henningsonstates{M. Henningson, ¡°The
Quantum Hilbert Space Of A Chiral Two Form In D = (5+1)
Dimensions,¡± J. High Energy Phys. 0203 (2002) 021 [arXiv:hep-th/0111150].}%
\nref\flink{E. Flink, ¡°Scattering amplitudes for
particles and strings in six-dimensional (2,0) theory, ¡± Phys. Rev. D 72 (2005) 066010 [arXiv:hep-th/0505015].}%
\nref\haw{P.S. Howe, G. Sierra and P.K. Townsend, Supersymmetry in
six dimensions, Nucl. Phys. B221 (1983) 331.} \nref\hitchin{N.
Hitchin, ¡°What is a gerbe,¡± Notices Amer. Math.
Soc. 50 (2003) 218.}%

The six dimensional $\EUN=(2,0)$ superconformal quantum field theory
-- named as the quantum non-Abelian gerbe (QNG) theory recently
\wittenoncft -- is an important but still mysterious theory. This
theory was originally found \witten\ by considering the Type-IIB
superstring theory at an $\rm{A-D-E}$ singularity of the $\rm{K}3$
compactification. The QNG theory is believed to be a well defined
quantum field theory in high spacetime dimensions and can serve as a
starting point to understand the dynamics of various
four-dimensional gauge theories (see the references in
\adscftreport\DKutasov). In particular, the $\BFT^2$
compactification of it turns out to be the four-dimensional $\EUN=4$
super-Yang-Mills theory with complex coupling constant $\tau$
determined by the complex structure of $\BFT^2$, and the
Montonen-Olive duality of the four-dimensional $\EUN=4$
super-Yang-Mills theory \olivemontonen\ was interpreted as the
$SL(2,\BZ)$ transformations of $\BFT^2$(for the simply-laced gauge
groups) \witten. Besides, the QNG theory is also crucial for $M$
theory. This theory is believed to be the exact theory that controls
the world volume dynamics of almost coincident parallel $M_5$-branes
\Stromingeropenbrane\wittenfive. And, this theory is conjectured to
be the dual Matrix theory \bfss\ describing $M$ theory compactified
on $\BFT^4$ \cftinmatrix. Moreover, the QNG theory is also the dual
conformal field theory(CFT) of $M$ theory on $AdS_7\times S^4$
\adscft\ (see \adscftreport\ for an introduction).

The QNG theory has many exotic but fascinating properties. {\it
Firstly}, this theory is in some certain senses unique. It is the
unique superconformal quantum field theory that lives in maximum
spacetime dimension ($D=6$) with maximum supersymmetries --
according to Nahm's classification to the superconformal algebra
\nahm. And it is expected to be an isolated interacting fixed point
theory of the six-dimensional renormalization group flow (see, for
example, \seiberg). The superconformal symmetry group of this theory
is $U^*Sp(4|2, \H)\subset OSp(8|4, \ct)$, where $OSp(8|4, \ct)$ is
the complexification of $U^*Sp(4|2, \H)$. {\it Secondly}, the QNG
theory is a chiral theory with the chiral supersymmetry $\EUN=(2,
0)$, this chiral nature makes its Lagrangian description to be a
quite subtle problem. In fact, even in Abelian case an ordinary
Lagrangian description for the tensor field $H$ does not exist
\foot{Nevertheless, the quantum mechanical theory of the Abelian
gerbe theory does exist and a beautiful approach to construct its
partition function -- by identifying the right theta function -- has
been proposed \wittenmaction .}.  {\it Thirdly}, the theory's
non-Abelian nature introduces more subtleties to its mysterious
quantum dynamics. For example, after compactifying on $\BFT^2$, the
theory's non-Abelian gerbe group $G$ will become the non-Abelian
gauge group $G$ of the resulting four-dimensional $\EUN=4$
super-Yang-Mills theory. Here, $G$ must be the tensor product of
some simple laced Lie groups (in types as $\rm{A}-\rm{D}-\rm{E}$)
and some copies of $U(1)$ group.

The purpose of the present paper is to try to develop a systematical
method to approach the QNG theory. Our method is based on the union
of four different elements. {\it The first two elements} are based
on a careful analysis to the superconformal symmetry $U^*Sp(4|2,
\H)\subset OSp(8|4, \ct)$ (and its chiral primary representations
and operators), and on a combination of the spinor method and the
canonical quantization procedure. By using this combination
extensively, we can overcome the chiral nature of the free tensor
multiplet and get its quantum mechanical theory successfully.

{\it The third element} is to get a glance at the non-Abelian nature
of the QNG theory, by perturbing it down to four spacetime
dimensions \wittenoncft. Our arguments go as follows: Firstly, we
deform the QNG theory to a generic point at the moduli space. We
then compactify the theory on a torus $\BFT^2$ and demonstrate
explicitly that the Mantonen-Olive dualities of the four-dimensional
Abelian gauge theory come from the $SL(2, \BZ)$ mapping class
symmetries of $\BFT^2$. We also show that the four dimensional
superconformal symmetry $PSU(2, 2|4)\subset PSL(4|4, \ct)$ and its
central extensions come from the $\BFT^2$ compactification of
$U^*Sp(4|2, \H)\subset OSp(8|4, \ct)$ and its central extensions.
Especially, the BPS states which become massless at the
singularities of the moduli spaces of the two theories match. Thus,
we argued that the enhanced non-Abelian gauge symmetry of the
four-dimensional theory should be a manifestation of the
corresponding non-Ablian nature of the QNG theory, which is the
conjectured connection between the QNG theory and the
four-dimensional $\EUN=4$ super-Yang-Mills theory.

{\it The last element} is to study the selfdual string excitations
\dufflu\ and their tensionless limits. After perturbing the QNG
theory to a generic point of the moduli space. There are self-dual
string excitations which are coupled to the tensor multiplets. The
tensions of these self-dual strings are proportional to the
expectation values of the scalars of tensor multiplet.  At the
moduli space singularities, these self-dual strings become
tensionless, the interactions between these tensionless strings and
the tensor multiplets may be the origins of the non-Abelian nature
of the QNG theory.

In the present paper, motivated by the elegancy of the spinor
technique and the success of the twistor-string theory
\twistorwitten\twistorstring\twistorberk\ for the four-dimensional
$\EUN=4$ perturbative super-Yang-Mills theory, we try to unify the
above four elements into a unique formulation of the QNG theory in
terms of the variables of $\hat{\BT}$. Here, $\hat{\BT}$ is the
supertwistor space that corresponds to the supergroup $OSp(8|4,
\ct)$. We constructed $\hat{\BT}$ and studied the action of the
superconformal symmetry group $OSp(2, 6|2)\subset OSp(8|4, \ct)$ on
it. And we also made some efforts to translate the information
concerning the tensor fields and the self-dual strings in spacetime
to the related data in $\hat{\BT}$. Finally, we argue that all the
information of the QNG theory can be appropriately encoded into the
supertwistor space, although the specific formulation is presently
unknown to the author.

Our tentative approach may provide a supplement to the beautiful
previous ones, which include, for examples, the proposed Matrix
theory description \matrixcft\ for the discrete light cone
quantization (DLCQ) of the QNG theory, and the recent proposals
\henningson\ to investigate the interactions between a free tensor
multiplet and the selfdual strings.

The paper is organized as follows. In section 2, we study the
superconformal algebra of $OSp(2, 6|2)\subset OSp(8|4, \ct)$ (see
\Proeyen\ for an investigation with ordinary spacetime indices) and
its chiral primary representations. Especially, we include a
discussion to the spectrum of scalar chiral primary operators and
their dimensions/R-charges relationship \minwalla\ . We present the
quantum mechanical theory of the Abelian gerbe theory in section 3.
At the same section, we also calculate the quantum anomalies of the
quantum Abelian gerbe theory, with the results agreeing with known
results. Section 4 is devoted to giving a first look at the
non-Abelian nature of the QNG theory by investigating the
six-dimensional origin of the four-dimensional $\EUN=4$
super-Yang-Mills theory in terms of the Hamiltonian formalism. In
section 5, we study the self-dual string excitations and their
tensionless limits. Section 6 towards a super-twistor formulation
for the QNG theory.

\newsec{Superconformal Algebra And Chiral Primary Operators}
\seclab\superalgebra
\bigskip\noindent{\it{$\EUN=(2,0)$ Superconformal Algebra}}

The $\EUN=(2,0)$ six-dimensional superconformal field theory has a
symmetry of supergroup $U^*Sp(4|2, \H)\simeq{OSp}(2,6|2)$ whose
bosonic part is $\Spin(2,6)\t Sp(2,\H)_{\BR}$. $\Spin(2,6)$ is the
two-fold covering of the conformal group $SO(2, 6)$ of
(1+5)-dimensional Minkowski space $\W$. $Sp(2,\H)_{\BR}\simeq
\Spin(5)_{\BR}$ is the $\BR$-symmetry group. And the fermionic
generators of ${OSp}(2,6|2)$ are in the $(\bf{8},\bf{4})$
representation of $\Spin(2,6)\times Sp(2,\H)_{\BR}$.

One can view $Sp(2,\H)_{\BR}$ as the intersection of $SU(4)_{\BR}$
and $Sp(4,\ct)$, $Sp(2,\H)_{\BR}\simeq SU(4)_{\BR}\cap Sp(4,\ct)$,
where $Sp(4,\ct)$ is the subgroup -- which commutes with a
symplectic form $\omega^{AB}$ -- of the volume preserving
automorphism $SL(4,\ct)$ of a four-dimensional complex space
$\ct^{4}_{\BR}$. Thus, in the $(\bf 4)$ representation, the
generators of $Sp(2,\H)_{\BR}$ can be written as
$\BR^{(AB)}=\half\l(\BR^{A}_{~C}\omega^{CB}+\BR^{B}_{~C}\omega^{CA}\r)$,
where the traceless Hermitian matrixes $\BR^A_{~B}$ are the
generators of $SU(4)_{\BR}$ in fundamental representation. The
isomorphism between the $({\bf{4}})$ representation and its dual can
be achieved by raising and lowing the indices with $\omega^{AB}$ and
its inverse $\omega_{AB}$. The volume form $\epsilon_{ABCD}$ of
$\ct^4_{\BR}$ turns out to be an invariant of $Sp(2,
\H)_{\BR}\subset Sp(4, \ct)$ and can be given as
$\epsilon_{ABCD}=\half\left(\omega_{AC}\omega_{BD}-\omega_{BC}\omega_{AD}\right)$.

One can decompose the conformal group $SO(2,6)$ as the direct
product of (1+5)-dimensional Lorentz group $SO(1,5)$ and the
$SO(1,1)$ group generated by the dilation $D$. Accordingly, the
fermionic generators ${(\bf{8}, \bf{4})}$ of $OSp(2,6|2)$ can be
decomposed into the $Q$-charges $Q^{\alpha}_{~A}$ and $S$-charges
$S_{\alpha A}$, with \eqn\dqds{\eqalign{i[D,
Q^{\alpha}_{~A}]&={+\half}Q^{\alpha}_{~A},\cr i[D, S_{\alpha
A}]&=-{\half}S_{\alpha A}.}} Clearly, $Q^{\alpha}_{~A}$ transforms
as a chiral spinor of $\Spin(1, 5)$ and $S_{\alpha A}$ transforms as
an anti-chiral spinor.

Now, we'd like to recall some familiar properties of the spinors of
(1+5)-dimensional spacetime. Firstly, the complexification
$\Spin(6,\ct)$ of the Lorentz group $\Spin(1,5)$ has the four
dimensional chiral spinor and anti-chiral spinor representations,
denoted as $S^+$ and $S^-$. Secondly, $S^+$ and $S^-$ are dual to
each other. Thus, given a chiral spinor $\lambda^{\alpha}\in S^+$
and an anti-chiral spinor $\mu_{\alpha}\in S^-$, one can form an
invariant \eqn\ssinner{(\mu, \lambda)=\mu_{\alpha}\lambda^{\alpha}.}
Here, we have used the superscription $\alpha$ to label the indices
of the spinor in ${S}^+$, and the subscription $\alpha$ to label the
indices of the spinor in $S^-$. Finally, by noticing that $\Spin(6,
\ct)\simeq SL(S^+)=SL(4, \ct)$, one can impose a real structure
$\tau$ on $S^+$ to get the real form $\Spin(1, 5)\simeq SL(2, \H)$.
The condition can be explicitly written as, $\tau O=\bar{O}\tau$,
where the real structure $\tau$ is represented as a skew matrix
$\tau^{\alpha}_{~\beta}$ that satisfies $\tau^2=-1$, and
$O=\exp(\half \theta),\
\theta=\half\theta_{\mu\nu}\gamma^{\mu}\wedge\gamma^{\nu}$, is an
arbitrary element of $\Spin(1, 5)$. Here, $\gamma^{\mu}$ are the
bases of the (1+5)-dimensional Clifford algebra, $\gamma^{\mu}\wedge
\gamma^{\nu}=\half
(\gamma^{\mu}\gamma^{\nu}-\gamma^{\nu}\gamma^{\mu})$, and
$\theta_{\mu\nu}$ are the parameters of the Lorentz group $\Spin(1,
5)\simeq SL(2, \H)$.

One can then pick up a real structure $\hat{\tau}$ on the definition
super vector space ${\ct}^{8|4}=(S^+, S^-, \ct^4_{\BR})$ of the
complexified superconformal group $OSp(8|4, \ct)$. $\hat{\tau}$ acts
on the fermionic generators $Q^{\alpha}_{~A}$ and $S_{\alpha A}$ as
follows (the symplectic-Majorana-Weyl conditions)
\eqn\superchargesmw{\eqalign
{Q^{\alpha}_{~A}&=\tau^{\alpha}_{~\beta}\omega_{AB}\bar{Q}^{\beta
B}\cr S_{\alpha
A}&=\tau_{\alpha}^{~\beta}\omega_{AB}\bar{S}_{\beta}^{~B},}} where
$\tau_{\alpha}^{~\beta}$ is the inverse of $\tau^{\alpha}_{~\beta}$.

Some further relationships between the representations of $\Spin(6,
\ct)$ will be helpful for us. Firstly, one recalls that the vector
representation $V_{\ct}$ of $\Spin(6, \ct)$ can be constructed as
the wedge product of two chiral (or anti-chiral) spinors
$V_{\ct}\simeq S^+\wedge S^+$ ($V_{\ct} \simeq S^-\wedge S^-$).
Thus, an arbitrary $\Spin(6, \ct)$ vector $A^{\mu}$ (or $A_{\mu}$)
can be rewritten as $A^{\alpha\beta}$ (or $A_{\alpha\beta}$), with
$A^{\alpha\beta}=\gamma^{\alpha\beta}_{~~\mu}A^{\mu},
A_{\alpha\beta}=A_{\mu}\gamma^{\mu}_{~\alpha\beta}$ (the
$\alpha\beta$ superscriptions of $\gamma^{\alpha\beta}_{~~\mu}$ and
the $\alpha\beta$ subscriptions of $\gamma^{\mu}_{~\alpha\beta}$
both are antisymmetric), where $\gamma^{\alpha\beta}_{~~\mu}$ are
the gamma matrices that map the anti-chiral spinors to the chiral
spinors, $\gamma^{\mu}_{~\alpha\beta}$ are gamma matrices that map
chiral spinors to anti-chiral spinors, and the six-dimensional
Clifford algebra is realized as
$\{\gamma^{\alpha\xi}_{~~\mu}\gamma_{\nu
\xi\beta}+\gamma^{\alpha\xi}_{~~\nu}\gamma_{\mu
\xi\beta}\}=2\delta^{\alpha}_{~\beta}\eta_{\mu\nu}$, where
$\eta_{\mu\nu}$ is the metric of (1+5)-dimensional Minkowski space
$\rm W$ with signature $(+ - - - - -)$. Further more, from the
volume form of $SL(4, \ct)\simeq\Spin(6, \ct)$, one has two natural
$\Spin(6, \ct)$ invariant tensors
$\epsilon_{\alpha\beta\gamma\delta}$ and
$\epsilon^{\alpha\beta\gamma\delta}$, which are just the spinor
notational correspondences of the metric tensor $\eta_{\mu\nu}$ and
$\eta^{\mu\nu}$, respectively. Hence, one can lower and raise the
spinor indices of an arbitrary $\Spin(6, \ct)$ vector by using
$\epsilon_{\alpha\beta\gamma\delta}$ and
$\epsilon^{\alpha\beta\gamma\delta}$. For examples,
$A_{\alpha\beta}=\half
\epsilon_{\alpha\beta\gamma\delta}A^{\gamma\delta}$,
$A^{\alpha\beta}=\half
\epsilon^{\alpha\beta\gamma\delta}A_{\gamma\delta}$, and
$\eta_{\mu\nu}A^{\mu}A^{\nu}=(1/4) A_{\alpha\beta}A^{\alpha\beta}$.

The $Q-Q$ anti-commutators and the $S-S$ anti-commutators of the
superconformal algebra can now be written as
\eqn\qqss{\eqalign{\{Q^{\alpha}_{~A},
Q^{\beta}_{~B}\}&=\omega_{AB}\P^{\alpha\beta}\cr \{S_{\alpha A},
S_{\beta B}\}&=\omega_{AB}\K_{\alpha\beta},}} where
$\P^{\alpha\beta}=\P^{\mu}\gamma^{\alpha\beta}_{~\mu}$ are the
translations, and
$\K_{\alpha\beta}=\K_{\mu}\gamma^{\mu}_{~\alpha\beta}$ are the
special conformal transformations. One can also write out the $\P-S$
and $\K-Q$ commutators \eqn\pkqs{\eqalign{i[\P^{\alpha\beta},
S_{\gamma
A}]&=Q^{\alpha}_{~A}\delta^{\beta}_{~\gamma}-Q^{\beta}_{~A}\delta^{\alpha}_{~\gamma}\cr
i[\K_{\alpha\beta}, Q^{\gamma}_{~A}]&=S_{\alpha
A}\delta_{\beta}^{~\gamma}-S_{\beta A}\delta_{\alpha}^{~\gamma}.}}
Clearly, $\P^{\alpha\beta}$ are the operators with conformal weight
$1$ and $\K_{\alpha\beta}$ are the operators with weight $-1$, that
is $i[D, \P^{\alpha\beta}]=\P^{\alpha\beta}$ and $i[D,
\K_{\alpha\beta}]=-\K_{\alpha\beta}$. $\hat{\tau}$ naturally acts on
$\P^{\alpha\beta}$ and $\K_{\alpha\beta}$ as
$\P^{\alpha\beta}=\tau^{\alpha}_{~\gamma}\tau^{\beta}_{~\delta}\bar{\P}^{\gamma\delta}$
and
$\K_{\alpha\beta}=\tau_{\alpha}^{~\gamma}\tau_{\beta}^{~\delta}\bar{\K}_{\gamma\delta}$.

The other nontrivial commutators or anti-commutators of the
superconformal algebra involve the generators $J_{\mu\nu}$ of the
Lorentz group $\Spin(1,5)$. In terms of the spinor notations, one
get a traceless matrix $J^{\alpha}_{~\beta}$ which acts on the
chiral spinor space $S^+$ as $J^{\alpha}_{~\beta}=\half
J_{\mu\nu}(\gamma^{\mu}\wedge \gamma^{\nu})^{\alpha}_{~\beta}$. The
real condition that should be imposed on $J^{\alpha}_{~\beta}$ is
obvious. The conformal weight of $J^{\alpha}_{~\beta}$ is zero,
which means that $[D, J^{\alpha}_{~\beta}]=0$. The $\P-\K$
commutators can be given as \eqn\pkc{i[\P^{\alpha\beta},
\K_{\gamma\xi}]=\delta^{\alpha}_{~\gamma}J^{\beta}_{~\xi}-
\delta^{\beta}_{~\gamma}J^{\alpha}_{~\xi}+
\delta^{\beta}_{~\xi}J^{\alpha}_{~\gamma}-\delta^{\alpha}_{~\xi}J^{\beta}_{~\gamma}
-\delta^{\alpha\beta}_{~~\gamma\xi}D,} where
$\delta^{\alpha\beta}_{~~\gamma\xi}=\half
(\delta^{\alpha}_{~\gamma}\delta^{\beta}_{~\xi}-\delta^{\beta}_{~\gamma}\delta^{\alpha}_{~\xi})$.
And the $J-\K$ commutators are $i[J^{\alpha}_{~\beta},
\K_{\gamma\xi}]=\delta^{\alpha}_{~\gamma}\K_{\beta\xi}-\delta^{\alpha}_{~\xi}\K_{\beta\gamma}$.
Other commutators involving $J^{\alpha}_{~\beta}$ form the familiar
algebra of (1+5)-dimensional Lorentz group.

Since $Q^{\alpha}_{~A}$ and $S_{\alpha A}$ transform as the chiral
and anti-chiral spinors of $\Spin(6, \ct)\simeq SL(4, \ct)$, one has
$i[J^{\alpha}_{~\beta},
Q^{\gamma}_{~A}]=\delta^{\gamma}_{~\beta}Q^{\alpha}_{~A}$ and
$i[J^{\alpha}_{~\beta}, S_{\gamma
A}]=-\delta^{\alpha}_{~\gamma}S_{\beta A}$. The $Q-S$
anti-commutators can now be written as \eqn\qs{\{Q^{\alpha}_{~A},
S_{\beta B}\}=\half
J^{\alpha}_{~\beta}-i\delta^{\alpha}_{~\beta}\left(\half\omega_{AB}D
-\BR_{(AB)}\right),} where the relative factors of the various terms
of the right hand side may be fixed by using the Jacobi identities.

\bigskip\noindent{\it{Chiral Primary Operators}}

One can organize the local operators of the QNG theory as various
irreducible representations of the superconformal group $OSp(2,
6|2)$.  This may be achieved by performing the radial quantization,
classifying the quantum states according to their different $OSp(2,
6|2)$ symmetries, and then using the state-operator correspondence
of conformal field theory to map these states to their corresponding
operators. In some details, one continues the (1+5)-dimensional
theory from the Minkowski space $\W$ to a six-dimensional Euclidean
space $\RR^6$, the superconformal group is then rotated to $OSp(1,
7|2)$. One then pick out a point of $\RR^6$ as the origin and cut
out a tiny hole of infinite small radius around it. This procedure
breaks the conformal symmetry $\Spin(1, 7)$ into $\Spin(1, 1)\times
\Spin(6)$, where $\Spin(1, 1)$ is the group of the dilations along
the radial direction and $\Spin(6)$ is the rotation group around the
origin. Finally, one takes the generator of $\Spin(1, 1)$ as the
Hamiltonian, quantizes the theory along the radial direction, and
constructs the Hilbert space on $\BFS^5$ surrounding the origin.
State-operator correspondence tells us that an operator inserted at
the origin will create a quantum state in the Hilbert space, and
inversely, the shrinking of a given state on $\BFS^5$ (into the
origin) will define an operator at the origin. And in the present
paper, with the hope of not confusing the reader, we'll use the same
notations to denote the generators of $OSp(2, 6|2)$ and the
generators of $OSp(1, 7|2)$. Thus, $D$ is used to denote the
Hamiltonian in the radial quantization. And, since what we are
considering is a unitary quantum theory, $\K_{\alpha\beta}$ and
$S^{A}_{~\alpha}$ are the Hermitian conjugations of
$\P^{\alpha\beta}$ and $Q^{\alpha}_{~A}$.

In radial quantization, by repeatedly acting the raising operators
$Q^{\alpha}_{~A}$, $\P^{\alpha\beta}$ on the state of highest weight
(superconformal primary), which is annihilated by the lowering
operators $S^{A}_{~\alpha}$, $\K_{\alpha\beta}$, one can construct
the full superconformal module associating with the primary state.
Some of the states of the superconformal module are $Q$ descendants
only, but not $\P$ descendants. These states are conformal primaries
of the conformal group $\Spin(2, 6)$. By using the state-operator
correspondence, the superconformal primary states will be mapped to
superconformal primary operators, and the conformal primaries will
be mapped to conformal primary operators.

For the ordinary superconformal small representations, at least 8 of
the 16 $Q$-charges will annihilate the superconformal primary state.
Hence, these representations include $2^8=256$ conformal primaries
at most. The primary operators of these small representations are
called chiral primary operators. A special property of chiral
primary operators is that their conformal weights are uniquely
determined by their $\BR$-symmetries.

Both the R-symmetries of the scalar chiral primary operators for the
QNG theory and their dimensions/$\BR$-charges relationship are well
known, and can be determined as follows. One considers a scalar
chiral primary operator ${\cal O}^{(n_1, n_2)}$, which transforms as
the representation of $Sp(2, \H)_{\BR}$ with highest weight $(n_1,
n_2)$, where $n_1, n_2\geq 0$ are integers. We denote the correspond
state of ${\cal O}$ as $|\psi_{\cal O}\rangle$. By using the algebra
\qs, one then has \eqn\qschiral{\langle \psi_{\cal
O}|S_{A\beta}Q^{\alpha}_{~B}|\psi_{\cal O}\rangle=-i\left(\half
\omega_{AB}\Delta_{\cal O}-\langle \psi_{\cal
O}|\BR_{(AB)}|\psi_{\cal O}\rangle\right)\delta^{\alpha}_{~\beta},}
where $\Delta_{\cal O}$ is the eigenvalue of operator $D$ on
$|\psi_{\cal O}\rangle$ (the dimension of the chiral primary
operator ${\cal O}$). The right hand side of \qschiral\ tells us
that the $Sp(2, \H)_{\BR}$ weight of ${\cal O}^{(n_1, n_2)}$ is
$(n_1-\half \Delta_{\cal O}, n_2-\half \Delta_{\cal O})$. For chiral
operator, this weight must vanish since some combinations of
$Q^{\alpha}_{~A}$ will vanish the left hand side of \qschiral. Thus,
we arrive at the conclusion that the nontrivial scalar chiral
primary operators must be in the $(n, n), n\geq 1$ representations
of $Sp(2, \H)_{\BR}$, and the dimensions of these operators are
given as \minwalla \eqn\cweight{\Delta_{\cal O}=2n.}

On the other hand, for $G=U(N)$ case, one can combine the
predictions, to the spectrum of chiral primary operators, of the
$\rm{AdS_7}\times \rm{S^4}/\rm{CFT_6}$ correspondence and of the
DLCQ description of $SU(N)$ QNG theory \matrixcft, which will tell
us that the scalar chiral primary operator ${\cal O}^{(n)}, 1\leq
n\leq N$ falls into the $n-$th order symmetric traceless irreducible
representation of $SO(5)$ with conformal weight $\Delta_{\O}=2n$.
These results agree with the results of the above paragraph, since
$SO(5)\simeq Sp(2, \H)_{\BR}/\BZ_2$ and the fundamental
representation of $SO(5)$ is identical to the $(1, 1)$
representation of $Sp(2, \H)_{\BR}$.

We now consider two notable examples of ${\cal O}^{(n)}$ with $n=1$
and $n=2$, in some more details. In these cases the small
representations are even more shorter. For the $n=1$ case. ${\cal
O}^{(1)}$ and its descendants form the free tensor multiplet of
$\EUN=(2, 0)$ superalgebra, which takes value in the decoupled
diagonal $U(1)$ part of $U(N)$. This free tensor multiplet consists
of a scalar fields $\Phi^{AB}$, four chiral fermions
$\Psi^{\alpha}_{~A}$, and a self-dual tensor $H^{(\alpha\beta)}$,
where the primary operator $\Phi^{AB}$ takes the $(1, 1)$
representation of $Sp(2, \H)_{\BR}\simeq \Spin(5)_{\BR}$,
$\Psi^{\alpha}_{~A}$ and $H^{(\alpha\beta)}$ are the $Q$-descendant
and $Q^2$ descendant of $\Phi^{AB}$, respectively. More properties
of these fields and their descendants will be discussed in next
section, explicitly.

The free tensor multiplet is itself a somewhat subtle theory for its
chiral nature. On the other hand, the quantum Abelian gerbe theory
-- consists of several copies of free tensor multiplet -- is the IR
effective theory of the QNG theory at a generic point of the moduli
space. Thus, in a certain sense the free tensor multiplets can be
viewed as one of the two elements of QNG theory -- the other
elements may be the self-dual strings. And before going into QNG
theory, we will firstly work out the quantum mechanics theory of the
free tensor multiplet in section 3.

For the $n=2$ case. The chiral primary operators $T^{(AB, CD)}$ have
conformal weight $4$, where the $AB$ (and $CD$) are antisymmetric,
while between the pairs of $AB$ and $CD$ are symmetric. And the
traceless condition imposes constraint $T^{(AB,
CD)}\epsilon_{ABCD}=0$. These operators are the only relevant
deformations of the interacting QNG theory and they preserve some of
the supersymmetries. The associated representation is ultra short
since its conformal primaries are descendants of $T^{(AB, CD)}$ with
no more than five-raising operators $Q$. The bosonic conformal
primaries include the $\BR$-symmetry currents
${J_{\BR}}^{\alpha\beta}_{~(AB)}={J_{\BR}}^{\mu}_{~(AB)}\gamma^{\alpha\beta}_{~\mu}$
of the QNG theory, the currents $J^{(\alpha\beta)}_{~AB}={1\over
3!}J^{\mu\nu\rho}_{~AB}(\gamma_{\mu}\wedge \gamma_{\nu}\wedge
\gamma_{\rho})^{(\alpha\beta)}$ of self-dual strings which satisfy
$J^{(\alpha\beta)}_{~AB}\omega^{AB}=0$, and the energy-momentum
tensor $T^{(\alpha\beta, \gamma\delta)}=\half
T^{\mu\nu}(\gamma^{\alpha\gamma}_{~\mu}\gamma^{\beta\delta}_{~\nu}
+\gamma^{\beta\gamma}_{\nu}\gamma^{\alpha\delta}_{\nu})$ of the QNG
theory. $J^{\alpha\beta}_{\BR, (AB)}$ and $J^{(\alpha\beta)}_{~AB}$
are dimension five operators. They are $Q^2$ descendants of $T^{(AB,
CD)}$. And the dimension six operators $T^{(\alpha\beta,
\gamma\delta)}$ are $Q^4$ descendants of $T^{(AB, CD)}$. All these
operators should commute with appropriate action of the real
structure $\hat{\tau}$.
\bigskip
\noindent{\it{The Moduli Space}}

A general QNG theory is an intrinsically strong interacting
conformal field theory--with superconformal group $OSp(2,6|2)$ that
we have studied. The theory has a rank $r$ gerbe group $G$, which is
in type of $\rm{A-D-E}$ series of the Lie groups. QNG theory lies on
the singularity points of the moduli space ${\cal{M}}_r$.

${\cal{M}}_r$ is isomorphic to the $W_G$ orbifold $\RR^{5r}/W_G$ of
$\RR^{5r}$, where $W_G$ is the Weyl group of $G$. Locally,
${\cal{M}}_r$ can be parameterized by the vacuum expectations of the
scalars $\Phi^{AB}$ that take values in the Cartan subalgebra of the
Lie algebra of $G$. And globally, ${\cal{M}}_r$ should be
parameterized by the vacuum expectations of scalar chiral primary
operators $\O^{(n)}, 1\leq n\leq r$ that we have studied.

The singularity of ${\cal{M}}_r$, that defines the QNG theory,
deserves some extra treatments. There, arguably, the vacuum
expectations of $\O^{(n)}$ should vanish. This vanishing is due to
their positive dimensions, which is guranteed by \cweight, since in
a CFT only the dimension zero operator can get the vacuum
expectation.

One can perturb the QNG theory away from the singularity to a
generic point of ${\cal{M}}_r$. This will break the non-Abelian
group $G$ down to their maximal torus $\BFT_G \subset G$.
Constrained by the sixteen-supercharges of the theory, the resulting
low energy effective theory, ignoring the high derivative
corrections, will be described by $r$ free tensor multiplets of the
$\EUN=(2, 0)$ supersymmetries, which form a quantum Abelian gerbe
theory with Abelian gerbe group $\BFT_G$. This theory is a free
field theory with superconformal group $OSp(2, 6|2)$.

\newsec{Quantum Abelian Gerbe Theory}
\seclab\qagt

We now turn to investigate the quantum mechanics of the Abelian
gerbes, which is one of the crucial elements of the QNG as we have
argued in section \superalgebra . There are two subtleties, one is
the unconventional and somewhat complicated kinematics due to the
unusual high spacetime dimensions, the other is the lack of an
ordinary Lagrangian description for the theory due to its chiral
nature.

We'll overcome the first subtlety by intensively utilizing the
spinor techniques (A parallel spin-helicity technique has been
developed and apply to the calculation of scattering amplitude of
the six-dimensional Yang-Mills field \Connell\ .). This technique
will be developed in the classical field theory context firstly, and
will then be applied to the quantization of the tensor multiplet.
The overcoming of the second subtlety is by utilizing the
Hamiltonian formalism which has been used to the quantization of the
free tensor multiplet by \henningsonstates\flink.

In this section, we'll show that, by combining the spinor technique
with the Hamiltonian formalism, the quantum mechanical theory of the
tensor multiplet will be elegantly worked out. As an illustration,
we'll calculate the OPE of the fields in the tensor multiplet. We'll
also calculate the anomalies of the tensor multiplet, compare the
results with the well known results from using other methods and
find agreements.

We'll mainly focus on the case of one free tensor multiplet. The
generalization to the full $r$ tensor multiplets associated with the
maximal torus $\BFT_G \subset G$ will be addressed in the last
subsection of the present section, where we also try to get an
understanding -- from the field theory viewpoint-- of the somewhat
mysterious $\rm{A-D-E}$ classification of the QNG theory.

\subsec{The Classical Theory Of The Tensor Multiplet}
\subseclab\tm

\bigskip\noindent{\it{The Tensor Multiplet}}

As we have mentioned in the last section, each tensor multiplet
\haw\ of the $\EUN=(2, 0)$ supersymmetries contains, one
antisymmetric two-form field $B$ \foot{In terms of mathematical
language, such a field $B$ is named as a connection of a gerbe
\hitchin.} with self-dual three-form strength $H=dB$, with $\star
H=H$, four symplectic-Majorana-Weyl fermions $\Psi^{\alpha}_{A}$
that transform as ${{S}^+}\otimes{{S}}_{\BR} =(\bf{4}, \bf{4})$
representation under $\Spin(1,5)\times Sp(2,\H)_{\BR}\subset SL(4,
\ct)\times Sp(4, \ct)$, and five scalars $\Phi_{AB}$ with
antisymmetric superscripts $A,B$. The fermions $\Psi^{\alpha}_{~A}$
should satisfy the real condition,
\eqn\fermionsmw{{\Psi}^{\alpha}_{A}=\tau^{\alpha}_{~\beta}\omega_{AB}{\bar{\Psi}}^{\beta
B}.} One can see that \fermionsmw\ is the natural extension of the
real structure $\hat{\tau}$ on the configuration space of the
fields. To see this point explicitly, one defines the appropriate
action of $\hat{\tau}$ on $\Psi^{\alpha}_{~A}$ as
$\hat{\tau}(\Psi)^{\alpha}_{~A}
=\tau^{\alpha}_{~\beta}\omega_{AB}{\bar{\Psi}}^{\beta B}$. Clearly,
\fermionsmw\ just means that $\Psi^{\alpha}_{~A}$ commute with
$\hat{\tau}$, $\Psi^{\alpha}_{~A}=\hat{\tau}(\Psi)^{\alpha}_{~A}$.
The real structure $\hat{\tau}$ can also be naturally acted on the
self-dual tensor $H^{(\alpha\beta)}$ and the scalars $\Phi_{AB}$ by
imposing the real conditions \eqn\scalarsymp{\eqalign{&
H^{(\alpha\beta)}=
\tau^{\alpha}_{~\gamma}\tau^{\beta}_{~\delta}\bar{H}^{(\gamma\delta)}
\cr &{\Phi}_{AB}=\omega_{AC}\omega_{BD}\bar{\Phi}^{CD}.}}

Since $\Phi_{AB}$ transform as a direct sum ${\bf (1)\oplus\bf(5)}$
of the ${(\bf 1)}$ and ${(\bf 5)}$ dimensional representations of
$Sp(2,\H)_{\BR}$, thus to describe the five scalars of the tensor
multiplet, one should impose a further condition to project the
${(\bf 1)}$ representation out. This condition is
$\Phi_{AB}\omega^{AB}=0$.

In terms of the spinor notation, self-duality means that $H$ can be
described as a chiral field $H^{(\alpha\beta)}$ with symmetric
superscripts $\alpha, \beta$, $H^{(\alpha\beta)}=H^{(\beta\alpha)}$.
The connection between the spinor notation $H^{(\alpha\beta)}$ and
the ordinary expression of the self-dual three-form is through
$H^{(\alpha\beta)}={1\over
3!}H_{\mu\nu\rho}(\gamma^{\mu}\wedge\gamma^{\nu}\wedge\gamma^{\rho})^{\alpha\beta}$,
where $\gamma^{\mu}\wedge\gamma^{\nu}\wedge\gamma^{\rho}$ denotes
the totally antisymmetrical product of $\gamma^{\mu}$,
$\gamma^{\nu}$ and $\gamma^{\rho}$.

The connection $B$ of the self-dual gerbe field $H$ can now be
written as $B^{\alpha}_{~\beta}$, with $B^{\alpha}_{~\beta}=\half
B_{\mu\nu}(\gamma^{\mu}\wedge\gamma^{\nu})^{\alpha}_{~\beta}$.
$H^{(\alpha\beta)}$ can then be given as
$H^{(\alpha\beta)}=\p^{\alpha\gamma}B^{\beta}_{~\gamma}+\p^{\beta\gamma}B^{\alpha}_{~\gamma}$,
where the partial differential operator $\p^{\alpha\beta}={\p/ \p
\x_{\alpha\beta} }$. Here, $\x_{\alpha\beta}$ are the spinor
notations of the spacetime coordinates $\x^{\mu}$, with
$\x_{\alpha\beta}=\x_{\mu}\gamma^{\mu}_{~\alpha\beta}$ satisfying
$\x_{\alpha\beta}=\tau_{\alpha}^{~\gamma}\tau_{\beta}^{~\delta}\bar{\x}_{\gamma\delta}$.
The self-duality of $H^{(\alpha\beta)}$ means that
$B^{\alpha}_{~\beta}$ should satisfy
$\p_{\alpha\gamma}B^{\gamma}_{~\beta}+\p_{\beta\gamma}B^{\gamma}_{~\alpha}=0$.
And the gauge transformations $B\rightarrow B+dA$ is now
$B^{\alpha}_{~\beta}\rightarrow
B^{\alpha}_{~\beta}+\left(\p^{\alpha\gamma}A_{\beta\gamma}-{1\over
4}\delta^{\alpha}_{~\beta}\p^{\gamma\delta}A_{\gamma\delta}\right)$,
where $A_{\alpha\beta}=A_{\mu}\gamma^{\mu}_{~\alpha\beta}$.

Under the action of the supersymmetry transformation
$\epsilon^{A}_{~\alpha}Q^{\alpha}_{~A}$, the component fields of the
$\EUN=(2, 0)$ tensor multiplet transform as
\eqn\sst{\eqalign{\delta\Phi_{AB}&=\left(\epsilon_{
A\alpha}\Psi^{\alpha}_{~B}+\epsilon_{
B\alpha}\Psi^{\alpha}_{~A}-\half\omega_{AB}\epsilon^{C}_{~\gamma}\Psi^{\gamma}_{~C}\right)\cr
\delta\Psi^{\alpha}_{~A}&=\epsilon^{B}_{~\beta}\left(\half
H^{(\alpha\beta)}\omega_{AB}+\p^{\alpha\beta}\Phi_{AB}\right)\cr
\delta
H^{(\alpha\beta)}&=\epsilon^{A}_{~\gamma}\left(\p^{\alpha\gamma}\Psi^{\beta}_{~A}+\p^{\beta\gamma}\Psi^{\alpha}_{~A}\right),}}
where the infinite small fermionic parameters
$\epsilon^{A}_{~\alpha}$ satisfy the real conditions
$\epsilon^{A}_{~\alpha}=\omega^{AB}\tau_{\alpha}^{~\beta}\bar{\epsilon}_{\beta
B}$. From \sst\ one can explicitly see that $\Psi^{\alpha}_{~A}$ is
the $Q$-descendant of $\Phi_{AB}$, $H^{(\alpha\beta)}$ is the
$Q^2$-descendant of $\Phi_{AB}$, and the descendants with more than
two $Q$ charges acting are not conformal primaries, just as we have
mentioned in last section.

A kind of super-coordinates expressions of the supersymmetrical
transformations will be suggestive. Thus, we'd like to present it in
this paragraph, although there are no superspace formulation for the
QNG theory known to the author. To do this, one can take the
super-coordinates of the (1+5)-dimensional $\EUN=(2,0)$ superspace
as $(\x_{\alpha\beta},\theta^{A}_{~\alpha})$, where
$\theta^{A}_{~\alpha}$ are the fermionic coordinates that satisfy
the real conditions
$\theta^{A}_{~\alpha}=\omega^{AB}\tau_{\alpha}^{~\beta}\bar{\theta}_{\beta
B}$. On the operators defined on the superspace, the translations
$\P^{\alpha\beta}$ now act as the differential operators
$i\p^{\alpha\beta}$, and the $Q$-transformations act as fermionic
differential operators
\eqn\qderivative{Q^{\alpha}_{~A}={\p~\over\p\theta^{A}_{~\alpha}}+
\theta_{A\xi}{\p~\over\p \x_{\alpha\xi}}.}

\bigskip\noindent{\it{Dirac Equation And Kinematics In (1+5)-Dimensional Spacetime}}
\subseclab\td

To simplify the kinematics associated with the classical or quantum
mechanical theory of the $\EUN=(2,0)$ tensor multiplet, we'll start
to develop the spinor techniques in this subsubsection by studying
an auxiliary Dirac equation
\eqn\diracx{\p_{\alpha\beta}\Psi^{\beta}(\x)=0.} Besides the
application to $\EUN=(2,0)$ tensor multiplet, the same technique may
also be useful to many other field theories defined in
(1+5)-dimensional spacetime.

To analyze the solutions, we Fourier transform \diracx\ to an
algebraic equation defined on momentum space
\eqn\dirac{p_{\alpha\beta}\Psi^{\beta}(p)=0,} where
$p_{\alpha\beta}=p_{\mu}\gamma^{\mu}_{~\alpha\beta}$. By using the
Clifford algebra, one has
$p^{\alpha\gamma}p_{\beta\gamma}=\delta^{\alpha}_{~\beta}p^2$.
Hence, \dirac\ implies $p^2\Psi^{\alpha}(p)=0$, which means that the
wave function $\Psi^{\alpha}(p)$ must support on the light cone
$L_C: p_{\alpha\beta}p^{\alpha\beta}=0$. We denote the maximal
compact subgroup of $\Spin(1,5)$ that preserves the light-like
vector $p_{\alpha\beta}$ as $W_p$ (the little group). Clearly,
$W_p\simeq \Spin(4)\simeq SU(2)\times SU(2)$. To classify the
solutions of the Dirac equation, we shall use the maximal torus
$H_p=U(1)\times U(1)$ of $W_p$ to label helicities. The two
solutions $\lambda^{\alpha}(p), \tilde{\lambda}^{\beta}(p)$ of
\dirac\ have helicities $(\half, 0)$ and $(-\half, 0)$,
respectively, which means that under $e^{i\theta}\times
e^{i\varphi}\in H_p$
\eqn\helicity{\eqalign{\lambda^{\alpha}(p)\rightarrow e^{i\half
\theta}\lambda^{\alpha}(p),\ \tilde{\lambda}^{\alpha}(p)\rightarrow
e^{-i\half\theta}\tilde{\lambda}^{\alpha}(p).}} Here, the convention
is for positive energy waves, the helicities of negative energy
waves are reverse to it.

It is now possible to represent the light-like momentum
$p^{\alpha\beta}$ in terms of the wave functions
$\lambda^{\alpha}(p)$ and $\tilde{\lambda}^{\alpha}(p)$ (by noticing
that $V_{\ct}=S^+\wedge S^+$), with
\eqn\pl{\eqalign{p^{\alpha\beta}&=2(\lambda^{\alpha}(p)\wedge\tilde{\lambda}^{\beta}(p))\cr
&=\left(\lambda^{\alpha}(p)\tilde{\lambda}^{\beta}(p)-\lambda^{\beta}(p)
\tilde{\lambda}^{\alpha}(p)\right).}} The null condition
$p^2=({1/4}){1\over
2}\epsilon_{\alpha\beta\gamma\delta}p^{\alpha\beta}p^{\gamma\delta}=0$
of the light like vector is satisfied, obviously. To see that \pl\
fulfills the Dirac equation \dirac\ with plane-wave solutions
$\lambda^{\alpha}(p)$ and $\tilde{\lambda}^{\alpha}(p)$, one
rewrites \dirac\ as
\eqn\redirac{p^{\alpha\beta}\wedge\Psi^{\gamma}(p)=p^{[\alpha\beta}\Psi^{\gamma]}(p)=0.}
Since $p^{[\alpha\beta}\Psi^{\gamma]}(p)=0\Leftrightarrow
\epsilon_{\alpha\beta\gamma\delta}p^{[\alpha\beta}\Psi^{\gamma]}(p)=0\Leftrightarrow
p_{\gamma\delta}\Psi^{\gamma}(p)=0$, the equivalence between
\redirac\ and \dirac\ follows. Obviously, \pl\ satisfies \redirac.
Hence,
$p_{\alpha\beta}\lambda^{\beta}=p_{\alpha\beta}\tilde{\lambda}^{\beta}=0$.

Now, given some $\lambda^{\alpha}$ and $\tilde{\lambda}^{\beta}$,
the corresponding light-like momentum $p^{\alpha\beta}$ is
automatically fixed through the following equation
\eqn\pll{p^{\alpha\beta}=2(\lambda^{\alpha}\wedge\tilde{\lambda}^{\beta}),}
where -- noticing the real condition
$p^{\alpha\beta}=\tau^{\alpha}_{~\gamma}\tau^{\beta}_{~\delta}\bar{p}^{\gamma\delta}$
satisfied by real momentum -- $\tilde{\lambda}^{\alpha}$ is the
complex conjugate of $\lambda^{\alpha}$, with
$\tilde{\lambda}^{\alpha}=\tau^{\alpha}_{~\beta}\bar{\lambda}^{\beta}$.
Conversely, given the light-like momentum $p$, $\lambda$ and
$\tilde{\lambda}$ cannot be specified through \pll\ only. The
additional information needed is equivalent to a choice of the plane
wave solutions of the chiral Dirac equation \redirac\ (or its
equivalent form \dirac ), as we have seen.

In fact, one needs to view $\lambda^{\alpha}$ (and
$\tilde{\lambda}^{\alpha}$) as the elements of vector bundle over
the space of light-like momentum, and view
$p^{\alpha\beta}\rightarrow (\lambda^{\alpha}(p),
\tilde{\lambda}^{\beta}(p))$ as the promotion of $p^{\alpha\beta}$,
with \pll\ being satisfied. This promotion can be related to the
second Hopf fibration $S^7\rightarrow S^4$. To see this connection,
one picks a time direction $\iota^{\alpha\beta}$ with
$\iota^2=(1/4)\iota_{\alpha\beta}\iota^{\alpha\beta}=1$ and keeps
the time component $p^0$ of $p$ fixed, where $p^0=p\cdot \iota=(1/4)
p_{\alpha\beta}\iota^{\alpha\beta}$. Then, all the null momentums
along different spatial directions and with fixed $|\vec{p}|=p^0$
will form a $S^4$. Correspondingly, the space of possible chiral
spinors are determined by $\half
\iota_{\alpha\beta}\lambda^{\alpha}\wedge\tilde{\lambda}^{\beta}=p^0$
modulo a scaling $\lambda\rightarrow t\lambda$,
$\tilde{\lambda}\rightarrow t^{-1}\tilde{\lambda}$, where $t\in
\ct^*$. This determines a quadratic surface $S^7$ in the complex
four-dimensional space $S^+$ of chiral spinors. Equation \pll\ then
gives us the Hopf map from $S^7$ to $S^4$.

These discussions can be summarized as follows. Firstly, one Fourier
transforms \redirac\ to the coordinates space, and gets an
equivalent form of the conventional chiral Dirac equation \diracx,
\eqn\ddirac{\p^{[\alpha\beta}\Psi^{\gamma]}(\x)=0.} One can then see
that the two plane-wave solutions of \ddirac\ with the same momentum
$p^{\alpha\beta}$ and opposite helicities can be chosen as
\eqn\planewave{\eqalign{\lambda^{\alpha}\exp{(i\half
\x_{\alpha\beta}\lambda^{\alpha}\wedge \tilde{\lambda}^{\beta})},\cr
\tilde{\lambda}^{\alpha}\exp{(i\half
\x_{\alpha\beta}\lambda^{\alpha}\wedge \tilde{\lambda}^{\beta})}.}}
Obviously, the momentum $p^{\alpha\beta}$ and $\lambda^{\alpha}$,
$\tilde{\lambda}^{\alpha}$ are connected through
$p^{\alpha\beta}=2(\lambda^{\alpha}\wedge \tilde{\lambda}^{\beta})$.

\bigskip\noindent{\it{Classical Field Theory and Plane Waves}}
\subseclab\pw

Now, we start to study the classical field theory and plane wave
expansions of the free $\EUN=(2,0)$ tensor multiplet by utilizing
the kinematics that we just studied. To study the classical theory
of tensor multiplet, all we need to do is to solve the equations of
motion of various component fields $\Phi_{AB}$,
$\Psi^{\alpha}_{~A}$, and $H^{(\alpha\beta)}$. These equations are
$\p_{\alpha\beta}\p^{\alpha\beta}\Phi_{AB}=0$,
$\p_{\alpha\beta}\Psi^{\beta}_{~A}=0$, and
$\p_{\alpha\gamma}H^{(\gamma\beta)}=0$, respectively.

We begin from $\Phi_{AB}(\x)$ which satisfies the Klein-Gordon
equation \eqn\kg{\p_{\alpha\beta}\p^{\alpha\beta}\Phi_{AB}(\x)=0.}
By Fourier transforming this equation to momentum space and noticing
that $p^2\delta(p^2)=0$ ($\delta(x)$ is the Dirac delta function),
one can solve \kg\ as
\eqn\phiab{\Phi_{AB}(\x)=(2\pi)\int_p\delta(p^2)\left({\phi_{AB}(p)\over
2z_p}+{\tilde{\phi}_{AB}(p)\over 2\bar{z}_p}\right),} where
$z_p=\exp(ip\cdot\x)$ is the plane wave with momentum $p$
propagating along spacetime $\W$ (the form of \phiab\ suggests that
field $\Phi_{AB}$ may be appropriate continued to the
complexification of $\W$), the integration $\int_p$ over momentum
space is given by $\int{d^6p/(2\pi)^6}$, and the Lorentz invariant
factor $\delta(p^2)$ is inserted to restrict the integration on the
light-cone $L_C=L_C^+\cup L_C^-$, which is the union of positive
patch $L_C^+$ $(p^0>0)$ and negative patch $L_C^-$ $(p^0<0)$. The
real conditions \scalarsymp\ can then be satisfied by identifying
the amplitude $\phi_{AB}(p)$ with the amplitude
$\tilde{\phi}_{AB}(p)$ through
\eqn\phireal{\tilde{\phi}_{AB}(p)=\omega_{AC}\omega_{BD}\bar{\phi}^{CD}(p).}
And the condition $\Phi_{AB}(\x)\omega^{AB}=0$ can be fulfilled by
requiring $\phi_{AB}(p)\omega^{AB}=0$.

Moreover -- noticing the specific form of the plane wave
$z_p=\exp(ip\cdot\x)$ on spacetime $\W$ -- one should identified the
modes $\phi_{AB}(-p)$ on $L_C^-$ with the modes
$\tilde{\phi}_{AB}(p)$ on $L_C^+$ and viceversa,
${\phi_{AB}(-p)=\tilde{\phi}_{AB}(p),
\tilde{\phi}_{AB}(-p)=\phi_{AB}(p)}$, where $p\in L_C^+$. Hence, on
real spacetime $\W$, \phiab\ can be rewritten as more familiar form
\eqn\phiaba{\Phi_{AB}(\x)=(2\pi)\int_{p\in
L_C^+}\left(\phi_{AB}(p)e^{-ip\cdot \x}+\phi_{AB}(-p)e^{ip\cdot
\x}\right),} with splitting positive frequency part and negative
frequency part.

\smallskip

\smallskip

We now turn to the equation of motion of $\Psi^{\alpha}_{~A}$, which
can be written as
\eqn\diraca{p^{[\alpha\beta}\Psi^{\gamma]}_{~A}(p)=0.} This equation
is equivalent to $p_{\alpha\beta}\Psi^{\beta}_{~A}(p)=0$ with the
same reason that has been explained in last subsubsection. Equation
\diraca\ have four solutions $u^{\alpha}_{~A}(p)$ with helicity
$(+\half, 0)$ and four solutions $\tilde{u}^{\alpha}_{~A}(p)$ with
helicity $(-\half, 0)$, both $u^{\alpha}_{~A}(p)$ and
$\tilde{u}^{\alpha}_{~A}(p)$ transform as $({\bf 4})$ representation
of $Sp(2,\H)_{\BR}$.

Clearly, after promoting the light-like momentum $p^{\alpha\beta}$
to $\lambda^{\alpha}$ and $\tilde{\lambda}^{\alpha}$ with
$p^{\alpha\beta}=2(\lambda^{\alpha}\wedge \tilde{\lambda}^{\beta})$,
wave functions $u^{\alpha}_{~A}(p)$ can be factorized as,
$u^{\alpha}_{~A}(p)=\lambda^{\alpha}\Psi^+_A$. Wave function
$\tilde{u}^{\alpha}_{A}(p)$ can be likewise factorized as
$\tilde{u}^{\alpha}_{~A}(p)=\tilde{\lambda}^{\alpha}\tilde{\Psi}^-_A$.
One can then expand $\Psi^{\alpha}_{~A}(\x)$ as
\eqn\freepsi{\eqalign{\Psi^{\alpha}_{~A}(\x)&=(2\pi)\int_p
\delta(p^2)\left[(\lambda^{\alpha}\Psi^+_A+
i\tilde{\lambda}^{\alpha}{\Psi}^-_{A})/(2z_p)+
(\tilde{\lambda}^{\alpha}\tilde{\Psi}^+_{A}+i\lambda^{\alpha}\tilde{\Psi}^-_A)
/(2\bar{z}_p)\right]},} where $\tilde{\Psi}^+_A(p)$ can be set as
$\tilde{\Psi}^+_A(p)=\omega_{AB}\bar{\Psi}_+^B(p)$ and
$\tilde{\Psi}^-_A(p)$ can be set as
$\tilde{\Psi}^-_{A}(p)=\omega_{AB}\bar{\Psi}^B_-(p)$, to fulfill the
real conditions \fermionsmw.

Obviously, one can separate the free field $\Psi^{\alpha}_{~A}$ as a
helicity $+\half$ part $\Psi^{\alpha+}_{~A}$ plus a helicity
$-\half$ part $\Psi^{\alpha-}_{~A}$, with
$\Psi^{\alpha}_{~A}(\x)=\Psi^{\alpha+}_{~A}(\x)+\Psi^{\alpha-}_{~A}(\x)$.
Here, \eqn\psipsi{\eqalign{\Psi^{\alpha+}_{~A}(\x)&=(2\pi)\int_p
\delta(p^2)\lambda^{\alpha}(p)\left({\Psi^+_A(p)\over
2z_p}+i{\tilde{\Psi}^-_{A}(p)\over 2\bar{z}_p}\right)\cr
\Psi^{\alpha-}_{~A}(\x)&=(2\pi)\int_p
\delta(p^2)\tilde{\lambda}^{\alpha}(p)\left(i{{\Psi}^-_A(p)\over
2z_p}+ {\tilde{\Psi}^+_{A}(p)\over 2\bar{z}_p}\right).}} The real
condition \fermionsmw\ now means that $\Psi^{\alpha+}_{~A}(\x)$ and
$\Psi^{\alpha-}_{~A}(\x)$ are complex conjugate to each other, with
$\Psi^{\alpha+}_{~A}(\x)=\tau^{\alpha}_{~\beta}\omega_{AB}\bar{\Psi}^{\beta
B}_{~-}(\x)$.

In separating the positive and negative energy modes of the fields
$\Psi^{\alpha+}_{~A}$ and $\Psi^{\alpha-}_{~A}$, one should make use
of the relationships of $\Psi^+_A(-p)=\tilde{\Psi}^-_{A}(p)$,
$\tilde{\Psi}^-_A(-p)=-\Psi_A^+(p)$ and
$\Psi^-_A(-p)=-\tilde{\Psi}^+_A(p)$,
$\tilde{\Psi}^+_A(-p)=\Psi^-_A(p)$, where $p\in L_C^+$. Here, we
have used $\lambda^{\alpha}(-p)=i\lambda^{\alpha}(p)$,
$\tilde{\lambda}^{\alpha}(-p)=i\tilde{\lambda}^{\alpha}(p)$ to
preserve the relation
$p^{\alpha\beta}=2(\lambda^{\alpha}\wedge\tilde{\lambda}^{\beta})$
under $p\rightarrow -p$.  One can then have
$\Psi^{\alpha+}_{~A}(\x)=(2\pi)\int_{p\in
L_C^+}\lambda^{\alpha}(p)[\Psi^+_A(p)e^{-ip\cdot\x}+i\Psi^+_A(-p)e^{ip\cdot\x}]$,
and $\Psi^{\alpha-}_{~A}(\x)=(2\pi)\int_{p\in
L_C^+}i\tilde{\lambda}^{\alpha}(p)[\Psi^-_A(p)e^{-ip\cdot\x}+i\Psi^-_A(-p)e^{ip\cdot\x}]$.

\smallskip\smallskip

Let's turn to the equations of motion of chiral field
$H^{(\alpha\beta)}$. We Fourier transform the equation
$\p_{\alpha\beta}H^{(\beta\gamma)}(\x)=0$ to momentum space,
\eqn\heom{p_{\alpha\beta}H^{(\beta\gamma)}(p)=0.} By multiplying
$p^{\delta\alpha}$ to left hand side of \heom\ and using Clifford
algebra, one can easily see that the wave function
$H^{(\alpha\beta)}(p)$ is supported on light-cone $L_C$. Equation
\heom\ has three solutions, $H_+^{(\alpha\beta)}(p)$,
$H_0^{(\alpha\beta)}(p)$ and $H_-^{(\alpha\beta)}(p)$, labeled
according to their different helicities. Under $e^{i\theta}\times
e^{i\varphi}\in H_p$, these solutions transform as
\eqn\hsolutions{\eqalign{&H^{(\alpha\beta)}_+(p)\rightarrow
e^{+i\theta}H^{(\alpha\beta)}_+(p)\cr
&H^{(\alpha\beta)}_0(p)\rightarrow H^{(\alpha\beta)}_0(p)\cr
&H^{(\alpha\beta)}_-(p)\rightarrow
e^{-i\theta}H^{(\alpha\beta)}_-(p).}}

To employ the kinematic relation
$p^{\alpha\beta}=2(\lambda^{\alpha}\wedge\tilde{\lambda}^{\beta})$,
one first rewrites \heom\ as $p^{[\alpha\beta}H^{(\gamma]\delta)}=0$
(One can also transform this equation to coordinate space and get
$\p^{[\alpha\beta}H^{(\gamma]\delta)}=0$). One can then
appropriately factorize the wave functions as
 $H^{(\alpha\beta)}_+=\lambda^{\alpha}\lambda^{\beta}H_+$,
 $H^{(\alpha\beta)}_0=(\lambda^{\alpha}\tilde{\lambda}^{\beta}
 +\lambda^{\beta}\tilde{\lambda}^{\alpha})H_0$
 and $H^{(\alpha\beta)}_-=\tilde{\lambda}^{\alpha}\tilde{\lambda}^{\beta}H_-$.

We now can expand $H^{(\alpha\beta)}(\x)$ as the summation of a
helicity $+1$ field $H^{(\alpha\beta)}_+(\x)$, a helicity $-1$ field
$H^{(\alpha\beta)}_-(\x)$ and a helicity $0$ field
$H^{(\alpha\beta)}_0(\x)$. Here, $H^{(\alpha\beta)}_+(\x)$
 and $H^{(\alpha\beta)}_-(\x)$, which should be complex conjugate to each other
$H^{(\alpha\beta)}_+(\x)=\tau^{\alpha}_{~\gamma}\tau^{\beta}_{~\delta}
\bar{H}_-^{(\gamma\delta)}(\x)$ to fulfill the real condition
\scalarsymp, are given by the expansions
 \eqn\hhh{\eqalign{H^{(\alpha\beta)}_+(\x)&=
 (2\pi)\int_{p\in L_C^+}\lambda^{\alpha}\lambda^{\beta}
 \left(H_+(p)e^{-ip\cdot \x}+\bar{H}_-(p)e^{ip\cdot \x}\right)\cr
 H^{(\alpha\beta)}_-(\x)&=
 (2\pi)\int_{p\in L_C^+}
 \tilde{\lambda}^{\alpha}\tilde{\lambda}^{\beta}
 \left(H_-(p)e^{-ip\cdot \x}+\bar{H}_+(p)e^{ip\cdot \x}\right),}}
 and the self conjugate part $H_0^{(\alpha\beta)}(\x)$, with $H^{(\alpha\beta)}_0(\x)
 =\tau^{\alpha}_{~\gamma}\tau^{\beta}_{~\delta}\bar{H}_0^{(\gamma\delta)}(\x)$,  is given as
 \eqn\selfh{H^{(\alpha\beta)}_0(\x)=(2\pi)
 \int_{p\in L_C^+}i{(\lambda^{\alpha}\tilde{\lambda}^{\beta}+
 {\lambda}^{\beta}\tilde{\lambda}^{\alpha})\over 2}\left(
 H_0(p)e^{-ip\cdot \x}+\bar{H}_0(p)e^{ip\cdot
 \x}\right).}

\subsec{The Quantum Theory Of Tensor Multiplet And OPE}

The quantization of the free tensor multiplet is subtle due to its
chiral nature. With $\Psi^{\alpha}_{~A}$ and $\Phi_{AB}$, there is
no problem, but -- as is well known -- an ordinary Lagrangian
description of $H$ field is unavailable due to its
self-duality\foot{With a single auxiliary scalar field, one can
construct a Lorentz-covariant Lagrangian formulation of the M5 brane
effective action, \nref\Sorokin{I. A. Bandos, K. Lechner, A.
Nurmagambetov, P. Pasti, D. P. Sorokin and M. Tonin, Covariant
action for the super-five-brane of M-theory, Phys. Rev. Lett. 78
(1997) 4332, [arXiv: hep-th/9701149]. }\nref\Jhschwarz{M. Aganagic,
J. Park, C. Popescu and J. H. Schwarz, World-volume action of the
M-theory five-brane, Nucl. Phys. B, 496 (1997) 191, [arXiv:
hep-th/9701166].} \nref\sorokinhamiltonian{E. Bergshoeff, D. P.
Sorokin and P. K. Townsend, The M5-brane Hamiltonian, Nucl. Phys. B,
533 (1998) 303 [arXiv: hep-th/9805065]. }see \Sorokin\ . For a
non-covariant M5-brane action in flat D=11 superspace, see
\Jhschwarz\ . The Hamiltonian formulation was considered in
\sorokinhamiltonian\ .} One can understand this point by noticing
that there is no quadratic Lorentz invariant concerning chiral field
$H^{(\alpha\beta)}$ only. For this reason, we'll adopt Hamiltonian
formalism and perform canonical quantization to the free tensor
multiplet.

To present the Hamiltonian formalism, we pick out an arbitrary
time-like Killing vector $\iota^{\alpha\beta}\p_{\alpha\beta}$, with
$\iota^2=(1/4)\iota_{\alpha\beta}\iota^{\alpha\beta}=1$, take
spacetime $\W$ as the form of $\W=\RR^1\times X$, where $X$ is a
five-dimensional spatial slice and $\RR^1$ is the time direction
generated by $\iota^{\alpha\beta}\p_{\alpha\beta}$. The associated
Hamiltonian $\BFH$ of the system is then given by \eqn\htmunv{\BFH
=\int_X T^{(\alpha\beta,
\gamma\delta)}\iota_{\alpha\beta}\iota_{\gamma\delta},} where
$T^{(\alpha\beta, \gamma\delta)}$ is the energy-momentum tensor,
which belongs to the $n=2$ chiral primary multiplet of $OSp(2, 6|2)$
as we have discussed in section \superalgebra.

Canonical quantization is not an obvious Lorentz covariant
procedure. Especially, it depends on some time direction
$\iota^{\alpha\beta}$ chosen arbitrarily. But, we argue that the
quantum mechanical theory, got by canonically quantizing the free
tensor multiplet, is in fact Lorentz covariant, since the
correlation functions and operator product expansions turn out to be
independent of $\iota^{\alpha\beta}$ as one will see.

Now, we try to write out the energy-momentum tensor
$T^{(\alpha\beta, \gamma\delta)}$ of the free tensor multiplet,
explicitly. Up to quadratic terms of $H^{(\alpha\beta)}$,
$T^{(\alpha\beta, \gamma\delta)}$ must be given as $T^{(\alpha\beta,
\gamma\delta)}\sim {c\over
2}\left(H^{(\alpha\gamma)}H^{(\beta\delta)}-H^{(\beta\gamma)}H^{(\alpha\delta)}\right)+......$,
where the normalization constant $c$ will be fixed as $1/4\pi$, and
the ellipsis stands for the terms that concerning $\Phi_{AB}(\x)$
and $\Psi^{\alpha}_{~A}(\x)$ (these terms are omitted since they are
a little complicated and only contribute usual quadratic terms -- of
$\Phi_{AB}(\x)$ and $\Psi^{\alpha}_{~A}(\x)$ -- to $\BFH$). One can
check this expression by noticing the symmetries of the indices of
$T^{(\alpha\beta, \gamma\delta)}$ and the traceless conditions
\eqn\traceless{T=T^{(\alpha\beta,
\gamma\delta)}\epsilon_{\alpha\beta\gamma\delta}=0,} which is
enforced by the conformal invariance (But, as we'll discuss, the
Weyl rescaling of the $\EUN=(2, 0)$ tensor multiplet is an anomalous
symmetry, thus, for a general background $T$ is in general quantum
mechanically non-vanishing. In that situations, the conformal
invariance will be destroyed.).

The full Hamiltonian of the free tensor multiplet can now be written
as \eqn\shamiltionian{\eqalign{{\bf H}&={1\over 4\pi}\int_X
\left(\iota_{\alpha\beta}\iota_{\gamma\delta}H^{(\alpha\gamma)}H^{(\beta\delta)}
+\Psi^{\alpha}_{~A} \vec{\p}_{\alpha\beta}\Psi^{\beta A}\right)\cr &
+{1\over 16\pi}\int_X \left(\Pi^{AB}\Pi_{AB}-{1\over
4}\vec{\p}^{\alpha\beta}\Phi^{AB}
\vec{\p}_{\alpha\beta}\Phi_{AB}\right),}} where the terms concerning
$\Psi^{\alpha}_{~A}$ and scalars $\Phi^{AB}$ are usual. $\Pi^{AB}$
is given by time derivative of $\Phi^{AB}$, $\Pi^{AB}=\iota\cdot
\p\Phi^{AB}=(1/4)\iota^{\alpha\beta}\p_{\alpha\beta}\Phi^{AB}$, and
$\vec{\p}_{\alpha\beta}=\p_{\alpha\beta}-\iota_{\alpha\beta}\iota\cdot\p$
are partial derivatives along the spatial section $X$. Obviously,
$\vec{\p}_{\alpha\beta}$ satisfies
$\iota^{\alpha\beta}\vec{\p}_{\alpha\beta}=0$. Here, the
normalization has been appropriately chosen to make the theory
consistent, quantum mechanically.

We now substitute the modes expansions of the free tensor multiplet,
studied in subsection \pw, into the Hamiltonian $\BFH$
\shamiltionian. The terms concerning five scalars will be, by
substituting \phiab, \eqn\modeHs{\int_{\vec{p}}
|(p\cdot\iota)|(1/4)\bar{\phi}^{AB}(p)\phi_{AB}(p),} where
$\int_{\vec{p}}=\int d^5\vec{p}/[(2\pi)^6(2p^0)]$ is given by
carrying out the $p^0=p\cdot \iota$ integration in $\int
_p\delta({p^2})\theta(p^0)$, which will enforce $p^0=|\vec{p}|>0$.
And we have dropped the zero point energy of the scalar fields since
the total zero point energy of the full tensor multiplet will be
exactly canceled out, due to the boson-fermion pairing of the
supersymmetry. Clearly, in the canonical quantization,
$\phi_{AB}(p)$ and $\bar{\phi}^{AB}(p)$ should be identified as the
annihilation and creation operators, respectively, with the
quantization conditions \eqn\scc{[\phi_{AB}(p),
\bar{\phi}^{CD}(q)]=(2\pi)^6(2p^0)\delta({\vec{p}-\vec{q}})\delta^{CD}_{~AB},}
where
$\delta^{AB}_{~CD}=\half({\delta^{A}_{~C}\delta^{B}_{~D}-\delta^{B}_{~C}\delta^{A}_{~D}})$.
One can check \scc\ by using it to calculate the commutators
$[\phi_{AB}(p), \BFH]$. The result is $[\phi_{AB}(p), \BFH]=(p\cdot
\iota)\phi_{AB}(p)$, which is just the Fourier transformations of
the Heisenberg equation $i(d\Phi_{AB}/ dt)=[\Phi_{AB}, \BFH]$.

Similarly, after substituting the mode expansion of the fermions
into $\BFH$, and with the help of the relation
$p^{\alpha\beta}=\lambda^{\alpha}\tilde{\lambda}^{\beta}
-\lambda^{\beta}\tilde{\lambda}^{\alpha}$, one will have the terms
of the fermion modes
\eqn\modeHf{\int_{\vec{p}}|(p\cdot\iota)|\left(\bar{\Psi}^A_+
\Psi^+_A+\bar{\Psi}^A_-\Psi^-_A\right).} Clearly, this term means
that $\Psi^+_{~A}(\vec{p})$ and $\Psi^-_{~A}(\vec{p})$ should be
quantized as the annihilation operators and their complex conjugate
are the creation operators, with \eqn\sfcc{\{\Psi^+_A(\vec{p}),
\bar{\Psi}^B_+(\vec{q})\}=\{\Psi^-_A(\vec{p}),
\bar{\Psi}^B_-(\vec{q})\}=(2\pi)^6(2p^0)\delta(\vec{p}-\vec{q})\delta^B_{~A}.}
Applying the same procedure to the tensor field $H^{(\alpha\beta)}$,
one will get the terms of the expansion modes of
$H^{(\alpha\beta)}$, with \eqn\modeHH{\int_{\vec{p}}|\left(p\cdot
\iota)|(\bar{H}_+H_++\bar{H}_-H_-+\bar{H}_0H_0\right),} which means
that $H_+$, $H_-$ and $H_0$ are the annihilation operators and their
complex conjugates are the creation operators, with the quantization
relations \eqn\shhcamodes{\eqalign{&[H_+(p), \bar{H}_+(q)]=(2\pi
)^6(2p^0)\delta(\vec{p}-\vec{q})\cr &[H_-(p), \bar{H}_-(q)]=(2\pi
)^6(2p^0)\delta(\vec{p}-\vec{q}) \cr &[H_{0\hskip 2.40pt}(p),
\bar{H}_{0\hskip 2.40pt}(q)]=(2\pi )^6(2p^0)\delta(\vec{p}-\vec{q})
.}}

In the present conventions, all the modes are defined on the
positive light cone $L_C^+$. One can use various relations between
negative energy modes and the conjugate of positive modes with
opposite helicity to extend the quantization relations to the full
modes algebra. For example, by using
$\Psi^{+}_{~A}(-p)=\tilde{\Psi}^-_{~A}(p)$, one can have
\eqn\fma{\{\Psi^+_{A}(p),
\Psi^+_B(q)\}=(2\pi)^6(2p\cdot\iota)\delta(\vec{p}+\vec{q})\omega_{AB}.}

By acting the creation operators on the Fock vacuum $|0\rangle$, one
can construct the full Fock space $\CH_{\rm F}$ of the oscillating
modes of free tensor multiplet. For example, the one particle
Hilbert space can be constructed as follows. $|AB,
\vec{p}\rangle=\bar{\phi}^{AB}(\vec{p})|0\rangle$ are the one
particle states of the scalars, which transformed as $({\bf 5})$
representation under the $\BR$ symmetry group $Sp(2, \H)_{\BR}$, and
$|\half, A, \vec{p}\rangle=\bar{\Psi}^A_{+}(\vec{p})|0\rangle$ are
the one particle states of the helicity $+\half$ fermions, which
transform as $|\half, A, \vec{p}\rangle\rightarrow e^{i\half
\theta}|\half, A, \vec{p}\rangle$ under the helicity transformation
$e^{i\theta}\times e^{i\varphi}\subset H_p$, and transform as $({\bf
4})$ representation under $Sp(2, \H)_{\BR}$. Similarly, $|-\half, A,
\vec{p}\rangle=\bar{\Psi}^A_-(\vec{p})|0\rangle$ are the one
particle states of the helicity $-\half$ fermions. Acting the
creation operators $\bar{H}_+(\vec{p})$, $\bar{H}_-(\vec{p})$ and
$\bar{H}_0(\vec{p})$ on $|0\rangle$ once will create the one
particle states $|+, \vec{p}\rangle$, $|-, \vec{p}\rangle$ and $|0,
\vec{p}\rangle$, respectively. Under $e^{i\theta}\times
e^{i\varphi}\in H_p$, $|+, \vec{p}\rangle$ is helicity $+1$ with
$|+, \vec{p}\rangle\rightarrow e^{i\theta}|+,\vec{p}\rangle$, $|-,
\vec{p}\rangle$ is helicity $-1$ with $|-, \vec{p}\rangle\rightarrow
e^{-i\theta}|-, \vec{p}\rangle$, and $|0, \vec{p}\rangle$ is
helicity $0$ with $|0, \vec{p}\rangle\rightarrow |0,
\vec{p}\rangle$, all these states are $Sp(2, \H)_{\BR}$ singlets.
These one particle states, $|AB, \vec{p}\rangle$, $|\pm\half, A,
\vec{p}\rangle$, $|\pm, \vec{p}\rangle$ and $|0, \vec{p}\rangle$
consist of a short representation of the $\EUN=(2, 0)$
supersymmetries.

We now can calculate out the modes expression of $Q^{\alpha}_{~A}$.
This can be achieved by substituting the mode expansions of free
tensor multiplet into the supercharges\foot{This can be got by
integrating the supercurrent $J^{\alpha,
\gamma\delta}_{~A}=J^{\alpha \mu}_{~A}\gamma^{\gamma\delta}_{~\mu}$
over $X$, with $Q^{\alpha}_{~A}=\int_X J^{\alpha,
\gamma\delta}_{~A}\iota_{\gamma\delta}$. For free tensor multiplet,
the supercurrent $J^{\alpha, \delta\gamma}_{~A}$ can be
schematically written as $J^{\alpha, \delta\gamma}_{~A}\sim
H^{(\alpha\delta)}\Psi^{\gamma}_{~A}-H^{(\alpha\gamma)}\Psi^{\delta}_{~A}+..$.
We note that this expression satisfies the trace condition
$J^{\alpha}_{A \alpha\beta}=0$ enforced by the conformal invariance,
since
$H^{(\alpha\delta)}\Psi^{\gamma}_{~A}\epsilon_{\delta\gamma\alpha\beta}=
H^{(\alpha\gamma)}\Psi^{\delta}_{~A}\epsilon_{\delta\gamma\alpha\beta}=0$,
automatically.} \eqn\ssupercharge{Q^{\alpha}_{~A}={1\over
2\pi}\int_X
\left(H^{(\alpha\beta)}\iota_{\beta\gamma}\Psi^{\gamma}_{~A}
+\p^{\alpha\beta}\Phi_{AB}\iota_{\beta\gamma}\Psi^{\gamma B}\right).
} The result is (by using the relation
$p^{\alpha\beta}=\lambda^{\alpha}\tilde{\lambda}^{\beta}-
\lambda^{\beta}\tilde{\lambda}^{\alpha}$)
\eqn\sschargem{\eqalign{Q^{\alpha}_{~A}=&\int_{\vec{p}}(p\cdot\iota)
\left[\lambda^{\alpha}(\bar{H}_-\Psi^-_{~A}+H_+\tilde{\Psi}^+_{~A})
+\tilde{\lambda}^{\alpha}(H_-\tilde{\Psi}^-_{~A}-\bar{H}_+\Psi^+_{~A})\right]\cr
+i&\int_{\vec{p}}(p\cdot\iota)\left[\lambda^{\alpha}(\Psi^+_{~A}\bar{H}_0
+\tilde{\Psi}^-_{~A}H_0)/2+
\tilde{\lambda}^{\alpha}(\Psi^-_{~A}\bar{H}_0-\tilde{\Psi}^+_{~A}H_0)/2\right]\cr
+i&\int_{\vec{p}}(p\cdot\iota)
\left[\lambda^{\alpha}(\bar{\phi}^{AB}\Psi^{+}_{~B}+\phi_{AB}\bar{\Psi}^B_{~-})
+\tilde{\lambda}^{\alpha}(\bar{\phi}^{AB}\Psi^{-}_{~B}-\phi_{AB}\bar{\Psi}^B_{~+})\right].}}
One can check that this expression satisfies real conditions
\superchargesmw. As a further check, one can also calculate the
$Q-Q$ anti-commutators by using the modes algebra \scc, \sfcc\ and
\shhcamodes, the results just give out the terms \modeHs, \modeHf\
and \shhcamodes\ of the Hamiltonian.

\nref\wittenolive{E. Witten and D. Olive, Phys. Lett. 78B(1978)97.}

However, the calculations in previous paragraph are restricted to
oscillating modes. In more general case, the $Q-Q$ anti-commutators
will contain additional terms ${\cal{Z}}^{\alpha\beta}_{~AB}$, which
come from the anti-commutator between the first and the second term
of the right hand side of \ssupercharge\ and serve as central
extensions of $\EUN=(2, 0)$ supersymmetry algebra (This mechanism is
familiar in four-dimensional supersymmetrical gauge theory
\wittenolive). By noticing the equation of motion
$\p_{\alpha\beta}H^{(\beta\gamma)}=0$ of $H^{(\alpha\beta)}$,
${\cal{Z}}^{\alpha\beta}_{~AB}$ can be written as
\eqn\zcharge{\eqalign{{\cal{Z}}^{\alpha\beta}_{~AB}&={1\over
2\pi}\int_X\iota^{\alpha\gamma}\p_{\gamma\delta}(H^{(\delta\beta)}\Phi_{AB})\cr
&={1\over 2\pi}\int_{\p
X}n_{\gamma\delta}\iota^{\alpha\gamma}H^{(\delta\beta)}\Phi_{AB},}}
where $\p X$ stands for the spatial boundary at the infinity, and
$n_{\alpha\beta}$ is the unit normal vector of $\p X$, with $n^2=1$
and $n\cdot \iota\sim n_{\alpha\beta}\iota^{\alpha\beta}=0$.
Therefore, if the scalars $\Phi_{AB}$ have nonzero vacuum
expectation $\Phi_0\psi_{AB}$ at the infinity, where $\psi_{AB}$ is
a unit vector of $Sp(2, \H)_{\BR}\simeq \Spin(5)_{\BR}$ with
$(1/4)\psi_{AB}\psi^{AB}=1$, ${\cal{Z}}^{\alpha\beta}_{~AB}$ will
serve as a measurement of the $H$-flux through $\p X$, as one can
see from \zcharge.  In fact, the integrand of \zcharge\ is just the
operator $J^{(\alpha\beta)}_{~AB}$ in $n=2$ chiral primary multiplet
of $OSp(2, 6|2)$, with $J^{(\alpha\beta)}_{~AB}\sim
H^{(\alpha\beta)}\Phi_{AB}$. Thus, ${\cal{Z}}^{\alpha\beta}_{~AB}$
can be viewed as the conservation charges of the current
$J^{(\alpha\beta)}_{~AB}$ of self-dual strings. In section 5, we'll
discuss the self-dual string excitations extending along some
$l^{\alpha\beta}$ direction. There, $\Phi_0$ will be interpreted as
the tension of the string, $H$-fluxes will measure the winding
number $w$ of the self-dual string, ${\cal{Z}}^{\alpha\beta}_{~AB}$
will then be given as $w|\Phi_0|\psi_{AB}l^{\alpha\beta}L$, where
$L$ is the string length along $l^{\alpha\beta}$ direction.

\bigskip\noindent{\it{Operator Product Expansions}}

Having quantized the free tensor multiplet. We can now calculate the
operator product expansion(OPE) of the theory. Since what we are
considering is a free field theory, we only need to consider the
$\Phi-\Phi$ OPE, the $\Psi-\Psi$ OPE and the $H-H$ OPE, the OPE of
other local operators will then follows.

Let's begin from the $\Phi-\Phi$ OPE. By using the mode expansions
of the scalar fields and the mode algebra \scc, one can get the most
singular term of the $\Phi-\Phi$ OPE, with
\eqn\ppope{\Phi_{AB}(\x)\Phi_{CD}(0)\sim
(2\pi)\epsilon_{ABCD}{i\over 4\pi^3|\x|^4}+...,} where the factor
$1/(4\pi^3|\x|^4)$ can be got by evaluating the propagator
$\Delta(\x)=\int_p{e^{-ip\cdot \x}\over p^2}$\foot{which may be
calculated by $\int_p{1\over p^2}e^{-ip\cdot \x}=\int_0^{+\infty}
d\tau\int_pe^{-\tau p^2}e^{-ip\cdot \x}$, integrating out $p$ gives
$\int_0^{+\infty} d\tau{1\over (4\pi\tau)^3}e^{-{~\x^2\over
4\tau}}={1\over 4\pi^3|\x|^4}$.}. Since we are considering a free
field theory, the ellipsis of \ppope\ is nothing but an operator
given by the normal ordering $:\Phi_{AB}(\x)\Phi_{CD}(0):$, which is
nonsingular and can be expanded as the Taylor series $\sum{1\over
n!}\x^{\mu_1}...\x^{\mu_n}\p_{\mu_1}...\p_{\mu_n}\Phi_{AB}\Phi_{CD}(0)$.

Similarly, by using the mode expansions \freepsi\ and the mode
algebra \sfcc, one can calculate the OPE of the fermions
$\Psi^{\alpha}_{~A}$, with the help of
$p^{\alpha\beta}=\lambda^{\alpha}\tilde{\lambda}^{\beta}
-{\lambda}^{\beta}\tilde{\lambda}^{\alpha}$, the result is
\eqn\ffope{\Psi^{\alpha}_{~A}(\x)\Psi^{\beta}_{~B}(0)=(2\pi)\omega_{AB}{\x^{\alpha\beta}\over
\pi^3|\x|^6}+:\Psi^{\alpha}_{~A}(\x)\Psi^{\beta}_{~B}(0):} Here, the
factor $\x^{\alpha\beta}/(\pi^3|\x|^6)$ is given by evaluating
$\p^{\alpha\beta}\Delta(\x)$. This result can also be checked by
noticing that $\langle \Psi^{\alpha}(\x)\Psi^{\beta}(0)\rangle\sim
\langle[Q^{\alpha}, \Phi(\x)][Q^{\beta}, \Phi(0)]\rangle$, the
supersymmetry invariance of the vacuum state then tells us that
$\langle\{Q^{\alpha},[Q^{\beta}, \Phi(\x)]\}\Phi(0)\rangle\sim
\langle[\{Q^{\alpha}, Q^{\beta}\}, \Phi(\x)]\Phi(0)\rangle\sim
\p^{\alpha\beta}\langle\Phi(\x)\Phi(0)\rangle$, where we have
omitted all the $\BR$-symmetry indices in this sketchy calculation.

The $H-H$ OPE can also be calculated by using the mode algebra
\shhcamodes\ and the transformation
$p^{\alpha\beta}=\lambda^{\alpha}\tilde{\lambda}^{\beta}-
\lambda^{\beta}\tilde{\lambda}^{\alpha}$, with
\eqn\hhope{H^{(\alpha\beta)}(\x)H^{(\gamma\delta)}(0)=(2\pi){3\over
8}{6\over
\pi^3|\x|^8}(\x^{\alpha\gamma}\x^{\beta\delta}+\x^{\beta\gamma}\x^{\alpha\delta})
+:H^{(\alpha\beta)}(\x)H^{(\gamma\delta)}(0):} where the factor
$6(\x^{\alpha\gamma}\x^{\beta\delta}+\x^{\beta\gamma}\x^{\alpha\delta})/
(\pi^3|\x|^8)$ is got by evaluating
$\left(\p^{\alpha\gamma}\p^{\beta\delta}+\p^{\beta\gamma}\p^{\alpha\delta}\right)\Delta(\x)$.

In fact, up to a normalization constant, all these OPE's can be
fixed by superconformal symmetries. We have related the OPE of free
fermions to the OPE of the scalars by using supersymmetries. The
relationship between the $H-H$ OPE and the OPE of the scalars is
similar, since the free field operators form a chiral multiplet of
$OSp(2,6|2)$ -- belonging to $n=1$ chiral primary representations,
as we have noted in section \superalgebra. The conformal weight of
$\Phi_{AB}$ is $2$, thus the most singular term (multiplying to
weight 0 unit operator) of the $\Phi-\Phi$ OPE must be $\sim
1/|x|^4$, just as \ppope\ shows us. Hence, our explicit calculations
of the OPEs of free tensor multiplet can be viewed as a consistency
check to our quantum theory of the Abelian gerbe.

\subsec{The Direct Calculations Of Anomalies}

By utilizing the OPEs of free tensor multiplet, we can now calculate
the $\BR$ symmetry anomaly and the gravitational anomaly of the
tensor multiplet. This is achieved by calculating the four-point
functions of the $\BR$ symmetry currents -- with schematic form
$\langle J^{\BR}J^{\BR}J^{\BR}J^{\BR}\rangle$ -- and four-point
functions of the energy-momentum tensor -- with schematic form
$\langle TTTT\rangle$. The mixed anomalies may also be calculated by
evaluating various four-point functions concerning both the
energy-momentum tensor and $\BR$ symmetry current. We then compare
our direct calculations to the results got by utilizing index
theorems and find agreements. We'll not consider the Weyl anomaly
separately, since it is related to the $\BR$ symmetry anomaly by
supersymmetry\foot{One can see this as follows, firstly
$\p^{\alpha\beta}J^{\BR}_{~\beta\gamma}\sim [\P^{\alpha\beta},
J^{\BR}_{~\beta\gamma}]$, using the superalgebra $\sim
[\{Q^{\alpha}, Q^{\beta}\}, J^{\BR}_{~\beta\gamma}]$, and by
noticing that $\{Q^{\alpha}[Q^{\beta}, J^{\BR}_{~\beta\gamma}]\}$
$\sim \epsilon_{\alpha'\beta'\gamma'\delta'}T^{(\alpha'\beta',
\gamma'\delta')}\delta^{\alpha}_{~\gamma}$, one finally has
$\p^{\alpha\beta}J^{\BR}_{\beta\gamma}\sim
T\delta^{\alpha}_{~\gamma}$.}.

By noticing that only the chiral fermions $\Psi^{\alpha}_{~A}$ of
the tensor multiplet contribute to the $\BR$ symmetry anomaly, we'll
focus on the terms of $J^{\alpha\beta}_{~\BR (AB)}$ that concern
$\Psi^{\alpha}_{~A}$ only $J^{\alpha\beta}_{\BR(AB)}\sim
(\Psi^{\alpha}_{~A}\Psi^{\beta}_{~B}+\Psi^{\alpha}_{~B}\Psi^{\beta}_{~A})+...$,
where the ellipsis stands for the terms contributed by the scalars
(roughly
$\omega^{CD}(\Phi_{AC}\p^{\alpha\beta}\Phi_{DB}+\Phi_{BC}\p^{\alpha\beta}\Phi_{DA})
$).

By using the $\Psi-\Psi$ OPE \ffope, the four-point function of the
$\Spin(5)_{\BR}$ currents can be calculated out, with schematic
terms \eqn\fourpoint{\eqalign{&\langle J^{\alpha_1\beta_1}_{\BR
(A_1B_1) }(\x_1)J^{\alpha_2\beta_2}_{\BR
(A_2B_2)}(\x_2)J^{\alpha_3\beta_3}_{\BR
(A_3B_3)}(\x_3)J^{\alpha_4\beta_4}_{\BR
(A_4B_4)}(\x_4)\rangle\cr\sim
&D_{AB}{~(\x_1-\x_2)^{\beta_1\alpha_2}(\x_2-\x_3)^{\beta_2\alpha_3}(\x_3-\x_4)^{\beta_3\alpha_4}
(\x_4-\x_1)^{\beta_4\alpha_1}\over
|\x_1-\x_2|^6|\x_2-\x_3|^6|\x_3-\x_4|^6|\x_4-\x_1|^6}, }} where
$D_{AB}$ is the abbreviation of
$D_{A_1B_1A_2B_2A_3B_3A_4B_4}\sim(\omega_{B_1A_2}\omega_{B_2A_3}\omega_{B_3A_4}\omega_{B_4A_1})
+{permutations}$. Other terms can be got by noticing the symmetries
of the spinor indices and $\BR$ symmetry indices.

\nref\fbeff{E. Witten, ¡°Five-brane effective action in M-theory,¡±
J. Geom. Phys. 22 (1997) 103 [arXiv:hep-th/9610234].}

We now couple $J^{\alpha\beta}_{\BR (AB)}$ to a nontrivial
background $Sp(2, \H)_{\BR}$ gauge field with potential
$A^{(AB)}_{~\alpha\beta}$. One can view $A^{(AB)}_{~\alpha\beta}$ as
the connection of a nontrivial $Sp(2, \H)_{\BR}$ vector bundle
$S_{\BR}\rightarrow \W$ and the configurations of
$\Psi^{\alpha}_{~A}$ as the sections of $S_{\BR}$. Now, we have
$\epsilon^{AB}\langle \p_{\alpha\beta}J^{\alpha\beta}_{\BR
(AB)}(\x)\rangle_A\sim \epsilon^{AB}(\x)\int_2\int_3\int_4\langle
\p_{\alpha\beta}J^{\alpha\beta}_{\BR (AB)}(\x)J^{\BR}\cdot
A(\x_2)J^{\BR}\cdot A(\x_3) J^{\BR}\cdot A(\x_4)\rangle$, where
$J^{\BR}\cdot A(\x)\sim J^{\alpha\beta}_{~\BR
(AB)}A^{(AB)}_{~\alpha\beta}(\x)$ and $\epsilon^{AB}(\x)$ are the
parameters of a infinite small $Sp(2, \H)_{\BR}$ gauge
transformation.  By using $\p_{\alpha\beta}\left({1/
|\x|^4}\right)\sim {\x_{\alpha\beta}/ |\x|^6}$,
$\p_{\alpha\beta}\left({\x^{\beta\gamma}/ |\x|^6}\right)\sim
\delta_{\alpha}^{~\gamma}\delta(\x)$, and after some tedious
calculations, one finally has \eqn\RsA{\eqalign{\epsilon^{AB}\langle
\p_{\alpha\beta}J^{\alpha\beta}_{\BR (AB)}\rangle\sim -&{1\over
3}\Tr_{\hskip -2.2pt R}\left[(dA)^{\alpha}_{~\beta}
(dA)^{\beta}_{~\gamma}(dA)^{\gamma}_{~\alpha}\epsilon\right]\cr +
&{1\over 4}\Tr_{\hskip -2.2pt
R}\left[(dA)^{\alpha}_{~\beta}\epsilon\right]\Tr_{\hskip -2.2pt R}
\left[(dA)^{\beta}_{~\gamma}(dA)^{\gamma}_{~\alpha}\right]}} where
$(dA)^{\alpha}_{~\beta}=\half
(\p_{\mu}A_{\nu}-\p_{\nu}A_{\mu})(\gamma^{\mu}\wedge
\gamma^{\nu})^{\alpha}_{~\beta}$, and we have omitted the $\BR$
symmetry indices since we are tracing over them. The Wess-Zumino
consistency conditions, combing equation \RsA, tell us that the full
$\BR$-symmetry anomalies can be got from the descendants of the
terms $I^{\BR}_{~A}\sim -{1\over 3}\Tr_{\hskip -2.2pt
R}(F'F'F'F')+{1\over 4}[\Tr_{\hskip -2.2pt R}(F'F')]^2$ (by
expanding it to the first order in $\epsilon$), where
$F'=F+\epsilon$. By using the Pontrjagin classes of the $Sp(2,
\H)_{\BR}\simeq\Spin(5)_{\BR}$ vector bundle $S_{\BR}\rightarrow
\W$, $p_1(S_{\BR})\sim -{1\over 2}\Tr_{\hskip -2.2pt R}(F^{'2})$ and
$p_2(S_{\BR})\sim -{1\over 4}\Tr_{\hskip -2.2pt R}(F^{'4})+{1\over
8}[\Tr_{\hskip -2.2pt R}(F^{'2})]^2$, one can then reorganize the
terms of $I^{\BR}_{~A}$ as \eqn\RsAr{I^{\BR}_{~A}\sim
p_2(S_{\BR})+{1\over 4}p_1(S_\BR)^2.} Up to a overall numerical
factor that we did not try to fix in our heuristic derivation,
\RsAr\ agrees with the results of \fbeff. A careful calculation may
give out the correct numerical factor ${1\over 48}$.

\nref\gaumewitten{L. Alvarez-Gaum¡äe and E. Witten, ¡°Gravitational
Anomalies,¡± Nucl. Phys. B234 (1983) 269.}

For completeness, we now include the index theoretical derivation of
the $\BR$ symmetry anomaly (following \fbeff). The beautiful
arguments of \gaumewitten\ tell us that this anomaly is given by
descending the fourth order terms of the Chern character
$\Ch(S_{\BR})$ of $S_{\BR}\rightarrow W$. To get $\Ch(S_{\BR})$, one
picks out a maximal torus $\BFT_{\BR}\subset Sp(2,
\H)_{\BR}\simeq\Spin(5)_{\BR}$, $\BFT_{\BR}=\Spin(2)_{\BR}\times
\Spin(2)_{\BR}$. Letting $(t_1, t_2)$ denote the pull back of the
two generators of cohomology ring $H^*(B\T_{\BR}, \BQ)$, where
$B\T_{\BR}$ is the classifying space of $\BFT_{\BR}$ vector bundles.
Since $\Psi^{\alpha}_{~A}$ transform as the spinor representation of
$\Spin(5)_{\BR}$, the four weights of $\Psi^{\alpha}_{~A}$ are
$(+\half, +\half), (-\half, -\half), (+\half, -\half), (-\half,
+\half)$. $\Ch(S_{\BR})$ can then be calculated as
$\Ch(S_{\BR})=\half (e^{\half t_1}+e^{-\half t_1})(e^{\half
t_2}+e^{-\half t_2})$, where the factor $\half$ is included since
$\Psi^{\alpha}_{~A}$ satisfy the real condition \fermionsmw.
Finally, one expands $\Ch(S_{\BR})$ to fourth order and gets
${1\over 192}(t_1^4+t_2^4)+{1\over 32}t_1^2t_2^2$. Hence, the $\BR$
symmetry anomaly is given by the descendant of
\eqn\ranomaly{I^{\BR}_{~A}={1\over 48}(p_2(S_{\BR})+{1\over
4}p_1(S_{\BR})^2).} Here, the Pontrjagin classes are given as
$p_1(S_{\BR})=t_1^2+t_2^2$ and $p_2(S_{\BR})=t_1^2t_2^2$.

\bigskip\noindent{\it{Gravitational Anomaly And Mixed Anomaly}}

The gravitational anomaly can also be calculated by computing the
four point functions of the energy-momentum tensor $T^{(\alpha\beta,
\gamma\delta)}$. But the result can be more directly got by
evaluating the Feynman diagrams. Figure 1 is one of the typical
anomalous diagram\foot{For a detailed discussion to other anomalous
diagrams, especially the diagrams with seagull vertices where two
gravitons are simultaneously attached, see \gaumewitten .}, where
the interior loop is for the chiral fermions $\Psi^{\alpha}_{~A}$ or
for the chiral tensor $H^{(\alpha\beta)}$. The interaction between
these fields and the background gravity is through $
(1/4)^2h_{(\alpha\beta, \gamma\delta)}T^{(\alpha\beta,
\gamma\delta)}$, where the energy-momentum tensor of the chiral
tensor $H^{(\alpha\beta)}$ is given as $T^{(\alpha\gamma,
\beta\delta)}\sim
H^{(\alpha\beta)}H^{(\gamma\delta)}-H^{(\gamma\beta)}H^{(\alpha\delta)}+...$,
$h^{(\alpha\beta,
\gamma\delta)}=h^{\mu\nu}\gamma^{\alpha\beta}_{~\mu}\gamma^{\gamma\delta}_{~\nu}$,
and the background metric is given as
$g_{\mu\nu}=\eta_{\mu\nu}+h_{\mu\nu}$. At every vertex of Figure 1
but the dotted one one inserts $(1/2)h^{\mu\nu}T_{\mu\nu}$, but at
the dotted vertex, one should use
$\varepsilon^{\mu}\p^{\nu}T_{\mu\nu}$.

\ifig\diagram{An anomalous diagram. The internal lines are the
chiral field $H^{(\alpha\beta)}$ (or the chiral fermions) and the
external lines are gravitons. The coupling of the dotted vertex is
$\varepsilon^{\mu}\p^{\nu}T_{\mu\nu}$ and the coupling of other
vertexes are $(1/2)h^{\mu\nu}T_{\mu\nu}$.}
{\epsfbox{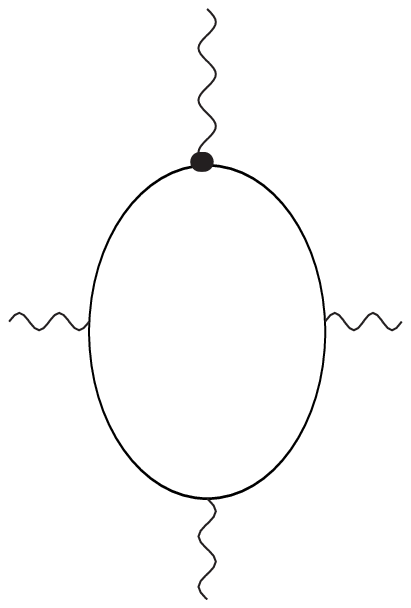}}

We now focus on the contributions of the chiral tensor
$H^{(\alpha\beta)}$ since the contributions of $\Psi^{\alpha}_{~A}$
are quite usual. The propagator of $H^{(\alpha\beta)}$ can be easily
calculated by using the quantization relations or from \hhope\
directly, \eqn\ppinmo{\langle
H^{(\alpha\beta)}(\x)H^{(\gamma\delta)}(0)\rangle\sim \int_p
e^{-ip\cdot
\x}{{p^{\alpha\gamma}p^{\beta\delta}+p^{\beta\gamma}p^{\alpha\delta}}\over
p^2+i\epsilon}.}

One can then evaluate the Feynman diagrams by using \ppinmo.
Fortunately, the calculations that we need to perform have been
essentially carried out in the original calculations of the
gravitational anomaly \gaumewitten. The underlying reason is because
that the spinor techniques have been employed there -- to simplify
the algebra -- in calculating gravitational anomaly of the self-dual
field. In fact, one can identify our chiral field
$H^{(\alpha\beta)}$ as the symmetrical part of the chiral projection
of the field $\phi_{\alpha\beta}$ introduced in \gaumewitten\
(restricted on six-dimensions). Moreover, upon truncating
$\phi_{\alpha\beta}$ to $H^{(\alpha\beta)}$, the momentum
independent term -- which is irrelevant for the gravitational
anomaly -- of \gaumewitten's formula $(48)$ automatically
disappears. The $\phi_{\alpha\beta}$ propagator there then becomes
our $H^{(\alpha\beta)}$ propagator \ppinmo. And the energy-momentum
tensor of projected $\phi_{\alpha\beta}$ (formula (51) of
\gaumewitten ) becomes the energy-momentum tensor of
$H^{(\alpha\beta)}$. Thus, one can borrow the calculations of
\gaumewitten . It turns out that the anomaly terms of the chiral
tensor are the descendants of $I^{H}_{~A}$, which is given by
$-{1\over 8}$ times the power 4 terms of the expansion of $L$-genus,
\eqn\gravitanor{I^{H}_{~A}={1\over
360}\left(p_1(T\W)^2-7p_2(T\W)\right).}

We now include the contributions of the chiral fermions
$\Psi^{\alpha}_{~A}$. According to the calculations of \fbeff , the
corresponding anomaly terms are given by the power 4 terms of the
expansion of $\half \hat{A}(\W)\times 4$, where $\hat{A}(\W)$ is the
$\hat{A}$-genus, $\half$ comes from the real condition \fermionsmw\
and the factor $4$ comes for four chiral fermions. By adding these
contributions and the terms $I^{H}_{~A}$ of the chiral tensor
$H^{(\alpha\beta)}$, one has the total gravitational anomaly
$I^{G}_{~A}$ \eqn\tgn{I^G_{~A}={1\over 48}\left({1\over
4}p_1(T\W)^2-p_2(T\W)\right).}

The mixed terms $I^M_{~A}$ of the $\BR$ symmetry anomaly and the
gravitational anomaly are contributed by the chiral fermions and can
also be calculated by evaluating the relevant Feynman diagrams.
According to \fbeff's results, $I^M_{~A}$ can be extracted from the
mixed terms of the expansion of $\half \hat{A}(\W)\Ch(S_{\BR})$,
where the factor $1/2$ comes from the real condition imposed on
$\Psi^{\alpha}_{~A}$, \eqn\Ma{I^M_{~A}=-{1\over
4}{p_1(S_{\BR})p_1(T\W)\over 24}.}

The total terms $I_A$ of the anomalies are then given by the
summation of the $\BR$-symmetry anomaly $I^{\BR}_{~A}$ \ranomaly,
the gravitational anomaly $I^G_{~A}$ \tgn\ and the mixed anomaly
$I^{M}_{~A}$ \Ma, $I_A=I^{G}_{~A}+I^{\BR}_{~A}+I^{M}_{~A}$.

\bigskip\noindent{\it{Digress To The Anomaly Cancelation Of $M_5$ Brane}}

\nref\hmooreg{D. Freed, J. A. Harvey, R. Minasian and G. W. Moore,
¡°Gravitational Anomaly Cancellation For M-theory Five-Branes,¡±
Adv. Theor. Math. Phys. 2 (1998) 601 [arXiv:hep-th/9803205].}
\nref\hmooren{J. A. Harvey, R. Minasian and G. W. Moore,
¡°Non-Abelian Tensor-Multiplet Anomalies,¡± JHEP 9809 (1998) 004
[arXiv:hep-th/9808060].}

We now digress to discuss the implications of the anomalies of
quantum Abelian gerbe theory when it is applied to the $M_5$ brane
dynamics. In what follows we will connect anomaly $I_A$ to the
gravitational anomaly of $M_5$ brane. And, for completeness, we'll
also include a treatment to the subtle cancelation mechanism (due to
\fbeff\hmooreg\hmooren\ ) of this anomaly.

It is well known that, the six-dimensional quantum Abelian gerbe
theory can be realized as the low energy world-volume theory of
$M_5$ brane. The five scalars $\Phi^{AB}$ extend to the five
coordinates $x^{AB}_{\perp}$ (in Plank unit $2\pi l_p=1$) of the
transverse position of five-brane world volume $\W$. The
$\Spin(5)_{\BR}$ vector bundle $N_{\BR}\rightarrow \W$ serves as the
normal bundle of the five-brane $\W$ in eleven-dimensional spacetime
$\M$, $\W\hookrightarrow \M$. And the $\BR$ symmetries
$\Spin(5)_{\BR}\subset GL(5, \RR)$ act as local diffeomorphisms
along the transverse directions. Hence, both the anomalies of the
diffeomorphisms along $\W$ and the $\BR$ symmetry anomaly are
gravitational anomalies of $M$-theory and must be canceled to
preserve the general covariances.

It turns out that this cancelation is achieved by including the
anomalies that inflow from the bulk. The inflowing anomalies come
from the Chern-Simons terms $I_{CS}$ of the bulk theory. These terms
concern the three-form potential $C$ and its four-form strength $G$
of the eleven-dimensional supergravity multiplet. After normalizing
the kinetic term of $C$ as $I_{\rm k}=-{1\over 4\pi}\int_{\M}
G\wedge
*G$, $I_{CS}$ can
be written as \eqn\tfa{I_{CS}=-{1\over 6}\int_{\M} {C}\wedge {G\over
2\pi}\wedge {G\over 2\pi}+\int_{\M} C\wedge *J,} where $J$ is given
by $*J={1\over 48}\left[p_2(T\M)-{1\over 4}p_1(T\M)^2\right]$ and
can be viewed as the current of dissolved $M_2$ branes.

One then views the $M_5$ brane as the magnetic source of the
potential $C$, its presence will modify the Bianchi identity of the
four-form strength $G$ to ${dG}=2\pi[\delta_W]$, where $[\delta_W]$
is the Poincare dual of $\W$.

To represent $[\delta_W]$ in terms of the Thom class $\Phi_{\W}$ of
the normal bundle $N_{\BR}\rightarrow \W$, one picks out a tubular
neighborhood $T_{\epsilon}$ of $\W$ in $\M$ by attaching to each
point of $\W$ a five-dimensional open sphere of sufficiently small
radius $\epsilon$ perpendicular to $\W$ at the center.
$T_{\epsilon}$ is diffeomorphic to the normal bundle $N_{\BR}$ of
$\W$ in $\M$. And the boundary of $T_{\epsilon}$'s closure forms a
sphere bundle $S_{\epsilon}\rightarrow \W$ with four-dimensional
sphere surface $S^4_{\epsilon}$(of radius $\epsilon$) as its fiber.
$[\delta_{\W}]$ can then be taken as the extension of $\Phi_{\W}$ in
$\M$. By using the Euler class $e(S_{\epsilon})$ of $S_{\epsilon}$,
this extension can be explicitly written as $[\delta_{\W}]=\half
e(S_r)\wedge dr\delta(r-\epsilon)=\half
d(e(S_r)\theta(r-\epsilon))$, where the factor $\half$ is inserted
by noticing that the integration of $e(S_r)$ along fiber $S^4_r$ is
$\int_{S^4_{r}}e(S_{r})=2$. Thus, the modified Bianchi identity can
be satisfied by modifying the four-form strength $G$ as
\eqn\gfield{G=dC+2\pi\theta(r-\epsilon)e(S_r)/2.}

The term -- due to the current of the dissolved $M_2$ branes (the
second term of \tfa ) -- of the inflowing anomaly is given by the
descendant of $(G\wedge
*J)|_{S_{\epsilon+0}}$. By substituting \gfield, integrating over $S^4_{\epsilon+0}$ and taking the
$\epsilon\rightarrow 0$ limit, one will have
$I_C=(2\pi)*J|_{\W}$\foot{By noticing that $T\M|_{\W}=T\W\oplus
N_{\BR}$, one has $p_1(T\M)=p_1(T\W)+p_1(N_{\BR}),
p_2(T\M)=p_2(T\W)+p_2(N_{\BR})+p_1(T\W)p_1(N_{\BR})$. Hence,
$*J|_{\W}$ can be given as $*J|_{\W}={1\over 48}[p_2(T\W)-{1\over
4}p_1^2(T\W)+p_2(N_{\BR})-{1\over 4}p_1^2(N_{\BR})+\half
p_1(T\W)p_1(N_{\BR})]$.}.

And the anomaly inflows from the first terms of \tfa\ is given by
the descendant of $-{1\over 24\pi^2}(G\wedge G\wedge
G)|_{S_{\epsilon+0}}$. By substituting the expression \gfield\ of
$G$, one has $-(2\pi){1\over 48}e(S_\epsilon)\wedge
e(S_\epsilon)\wedge e(S_\epsilon)$. After integrating over the fiber
$S^4_\epsilon$, one will have $I_S=-(2\pi){1\over 24}p_2(N_{\BR})$
\foot{To see this result, one notices
$S^4_{\epsilon}\simeq\Spin(5)_{\BR}/\Spin(4)_{\BR}$, and views
$N_{\BR}$ as the vector bundle of this $\Spin(4)_{\BR}$, hence,
$p_2(N_{\BR})=e^2(N_{\BR})$, which can be identified with the pull
back of $e^2(S_\epsilon)$. By further using $\int_{S^4_\epsilon}
e(S_\epsilon)=2$, the result follows. For a rigorous treatment of
the related mathematical facts see \nref\bottc{R. Bott and A. S.
Cattaneo, ¡°Integral Invariants of 3-Manifolds,¡± J. Diff. Geom. 48
(1998) 91, [arXiv: dg-ga/9710001].}\bottc\ Lemma. 2.1.}.

A little algebra shows us that all the anomalies cancel out after
summing up the anomalous terms $I_A$, $I_C$ and $I_S$, that is
\eqn\acel{I_A+I_C+I_S=0.}

\subsec{Quantization Of Fluxes}

The purpose of the present subsection is to discuss the quantum
mechanics of the zero modes of the free tensor multiplet defined on
a more general spacetime manifold. As one will see that these modes
should be classified by the cohomology group (or compact cohomology
group) of the (1+5)-dimensional spacetime $\W$. Thus, if $\W$ is
topologically trivial, for example $\W=\RR^{1, 5}$ as the most parts
of the present paper are focusing on, these zero modes are absent,
and one needs not to discuss their quantization. But, in applying
QNG theory to describe $M$ theory compactified on $\BFT^4$, in the
matrix theory context, $\W$ may be $\BFT^5\t\RR^1$, so that its
cohomology is nontrivial. Hence, it is deserved to quantize the zero
modes of $H^{(\alpha\beta)}$ canonically. These zero modes are the
static solutions of the following equations with nontrivial fluxes
\eqn\zmodeeq{\eqalign{\vec{\p}_{\alpha\beta}\vec{H}^{(\beta\gamma)}&=
0\cr
\iota^{\beta\gamma}\vec{\p}_{\alpha\beta}\vec{G}_{(\gamma\delta)}&=0,}}
the accurate definition of the magnetic field
$\vec{H}^{(\alpha\beta)}$ and electric field
$\vec{G}_{(\alpha\beta)}$ will be given later. On a curved $X$, the
partial differential operator $\vec{\p}_{\alpha\beta}$ in equation
\zmodeeq\ should be understood as a covariant derivative with some
spin connection.

It would be convenient to pick out an axial gauge. We denote the
connection of the $U(1)$ gerbe in this gauge as
$\vec{B}^{\alpha}_{~\beta}$ which satisfies the conditions
$\vec{B}^{\alpha}_{~\gamma}\iota^{\gamma\beta}=\vec{B}^{\beta}_{~\gamma}\iota^{\gamma\alpha}$
and
$\iota_{\alpha\gamma}\vec{B}^{\gamma}_{~\beta}=\iota_{\beta\gamma}\vec{B}^{\gamma}_{~\alpha}$.
We then decompose the tensor $H^{(\alpha\beta)}$ as time derivative
part $\vec{\Pi}^{(\alpha\beta)}$ plus spatial derivative part
$\vec{H}^{(\alpha\beta)}$. Here, $\vec{\Pi}^{(\alpha\beta)}=2
(\iota\cdot \p)\iota^{\alpha\gamma}\vec{B}^{\beta}_{~\gamma}$ and
$\vec{H}^{(\alpha\beta)}=
\vec{\p}^{\alpha\gamma}\vec{B}^{\beta}_{~\gamma}+
\vec{\p}^{\beta\gamma}\vec{B}^{\alpha}_{~\gamma}$. And the self dual
condition
$\p_{\alpha\gamma}B^{\gamma}_{~\beta}+\p_{\beta\gamma}B^{\gamma}_{~\alpha}=0$,
acting on the configuration space of $B^{\alpha}_{~\beta}$, now
becomes $\vec{\p}_{\alpha\gamma}\vec{B}^{\gamma}_{~\beta}
+\vec{\p}_{\beta\gamma}\vec{B}^{\gamma}_{~\alpha}=-2 (\iota\cdot
\p)\iota_{\alpha\gamma}\vec{B}^{\gamma}_{~\beta}=-\iota_{\alpha\gamma}\iota_{\beta\delta}\vec{\Pi}^{(\gamma\delta)}$.
This can be rewritten as
\eqn\recc{\vec{G}_{(\alpha\beta)}=\vec{\Pi}^{(\gamma\delta)}\iota_{\gamma\alpha}\iota_{\delta\beta},}
where we have denoted
$\vec{\p}_{\alpha\gamma}\vec{B}^{\gamma}_{~\beta}
+\vec{\p}_{\beta\gamma}\vec{B}^{\gamma}_{~\alpha}$ as $
-\vec{G}_{(\alpha\beta)}$. Now the equation of motion
$\p_{\alpha\beta}H^{(\beta\gamma)}=0$ can be rewritten as the form
of \zmodeeq.

To perform the canonical quantization procedure, we will view \recc\
as a constraint equation that acts on the Hilbert space after
finishing quantization. In doing so, we effectively enlarge the
chiral theory to a theory of ordinary tensor field and then throw
out the anti-self-dual part after quantizing it.

Now, we naturally identify $\vec{\Pi}^{(\alpha\beta)}$ with the
conjugate momentum operator of $\vec{B}^{\alpha}_{~\beta}$. By using
the canonical commutator between $\vec{B}^{\alpha}_{~\beta}$ and
$\vec{\Pi}^{(\alpha\beta)}$, one can calculate the commutator
between $\vec{H}^{(\alpha\beta)}$ and $\vec{\Pi}^{(\alpha\beta)}$.
For convenience, we view $\vec{H}^{(\alpha\beta)}$ as a three form
$\vec{H}$ and $\vec{\Pi}^{(\alpha\beta)}$ as a two form $\vec{\Pi}$
on $X$, then the $\vec{H}-\vec{\Pi}$ commutator can be elegantly
written as \eqn\canonicalquantization{\left[\int_X {\vec{H}\over
2\pi}\wedge\alpalp, \int_X \star{\vec{\Pi}\over
2\pi}\wedge\betbet\right]={1\over 2\pi i}\int_X\betbet\wedge
d\alpalp} where $\alpalp$ and $\betbet$ are two arbitrary $2$-forms
on $X$.

Since we are focusing on the quantization of \zmodeeq\ 's static
solutions with non-trivial fluxes, we can eliminate
$\vec{\Pi}^{(\alpha\beta)}$ by making use of equation \recc\ and
viewing $\vec{G}$ as the three form $\star\vec{\Pi}$. The fluxes
will consist of a system with Hamiltonian \eqn\rehh{{1\over
4\pi}\int_X\left(
\iota^{\alpha\beta}\iota^{\gamma\delta}\vec{G}_{(\alpha\gamma)}\vec{G}_{(\beta\delta)}
+\vec{H}^{(\alpha\gamma)}\vec{H}^{(\beta\delta)}\iota_{\alpha\beta}
\iota_{\gamma\delta}\right)+{1\over
2\pi}\int_X\vec{H}^{(\alpha\beta)}\vec{G}_{(\alpha\beta)},} and with
nontrivial $\vec{H}-\vec{G}$ commutators, \eqn\cqt{\left[\int_X
{\vec{H}\over 2\pi}\wedge\alpalp, \int_X {\vec{G}\over
2\pi}\wedge\betbet\right]={L(\alpalp, \betbet)\over 2\pi i},} where
the link number $L(\alpalp, \betbet)$ between $\alpalp$ and
$\betbet$ is given by $\int_X\betbet\wedge d\alpalp$, which is
antisymmetric about $\alpalp$ and $\betbet$, $L(\alpalp,
\betbet)=-L(\betbet, \alpalp)$.

For the compactness of the Abelian group $U(1)$, the magnetic fluxes
$\int{\vec{H}\over 2\pi}$ on some three cycles will be topologically
quantized, and the electric fluxes $\int {\vec{G}\over 2\pi}$ on
three cycles will be canonically quantized. Thus, roughly speaking,
both these fluxes will be classified by $H^3(X, \BZ)$, or by $H^2(X,
\RR/2\pi\BZ)=\Hom(H^3(X, \BZ), U(1))$ through Pontrjagin-Poincare
duality. Then, the phase space of the fluxes may be identified as
$H^2(X, \RR/2\pi\BZ)\oplus H^2(X, \RR/2\pi\BZ)$. And the commutator
\cqt\ gives us a sympletic form $L(\alpalp, \betbet)$ on this phase
space, which will enable us to properly quantize these fluxes. The
final results may be identical to the analysis to the quantum
self-duality of the fluxes of $H$ \wittenmaction .

\bigskip\noindent{\it Generalization}

It is interesting to generalize the above discussions of $U(1)$
Abelian gerbe theory to more general case with Abelian gerbe group
$\BFT_G$. The generalization of the quantization of oscillating
modes is trivial. We now focus on the generalization of the
quantization of $H$-fluxes. To specify the periods of the $r$
tensors of $\BFT_G$ theory, one should pick out an imbedding of
$U(1)$ into $\BFT_G$ firstly. All possible imbedding will define a
magnetic charge space $\BM=\Hom(U(1), \BFT_G)$. The periods of the
$r$ tensor fields should take value in
\eqn\periods{H^3(X,\BZ)\otimes \BM.}

To specify $\BM$, one puts a coordinates system $(z_1, z_2, ..z_r)$,
$|z_i|^2=1, (i=1,2...r)$ on the maximal torus $\BFT_G$. The
imbedding $U(1)\rightarrow \BFT_G$ can then be characterized by
\eqn\utot{z\rightarrow (z^{k_1}, z^{k_2},...,z^{k_r}),} where $z$ is
an element of $U(1)$ group. Thus, each imbedding is specified by a
sequence of $r$ integers $(k_1, k_2,......, k_r)$. The collection of
all possible imbedding form a $r$-dimensional lattice
$\Lambda_{\rm{cochar}}$, which is the cocharacter space of $G$.
Hence, $\BM=\Lambda_{\rm{cochar}}$.

The Poincare and Pontrjagin duality of $H^3(X, \BZ)\otimes \BM$ is
$H^2(X, \BFT_{\check{G}})$, where $\BFT_{\check{G}}$ is the maximal
torus of the Langlands dual group $\check{G}$ of $G$. Further more,
the flat connections are parameterized by $H^2(X,\BFT_G)$, which is
the Poincare and Pontrjagin duality of the electric fluxes
$H^3(X,\BZ)\otimes {\BE}$. Thus, the phase space of the $H$-fluxes
may be identified as \eqn\phases{H^2(X, \BFT_{\check{G}})\simeq
H^2(X, \BFT_{{G}}).} To study the quantum mechanical theory of these
fluxes, one should impose an appropriate sympletic form on this
phase space. This sympletic form should be some kind of natural
generalization of $L(\alpalp, \betbet)$. It seems that this can be
achieved only when $\BFT_{G}$ is identical to $\BFT_{\check{G}}$.
Clearly, the $\rm{A}-\rm{D}-\rm{E}$ series of Lie groups satisfy
this requirement.

\nref\thooftflux{G. 't Hooft, A property of electric and magnetic
flux in nonabelian gauge theories, Nucl. Phys. B 153 (1979) 141.}

\nref\henningsonnew{M. Henningson, ¡°The low-energy spectrum of
(2,0) theory on $\T^5\times \R$,¡± JHEP 0811 (2008) 028
[arXiv:hep-th/0809.4156]; M. Henningson, ¡°BPS states in (2,0)
theory on $\R\times\T^5$,¡± JHEP 0903 (2009) 021
[arXiv:hep-th/0901.0785].}

As an analog of 't Hooft's discussions for the Abelian fluxes of a
non-Abelian gauge theory \thooftflux, one can naturally guess that
the Hilbert space of Abelian flux states of a QNG theory with gerbe
group $G_{\ad}$ is given by $H^3(X, {\cal{Z}}(\hat{G}))$, where
$\hat{G}$ is the universal covering of $G$, ${\cal{Z}}(\hat{G})$ is
$\hat{G}$'s center, and $G_{\ad}=\hat{G}/{\cal{Z}}(\hat{G})$. This
result may be checked by firstly reducing the QNG theory on a $S^1$,
and then comparing $H^3(X, {\cal{Z}}(\hat{G}))$ to the Hilbert space
of the Abelian fluxes of the resulting $(4+1)$-dimensional gauge
theory \henningsonnew .

\newsec{$\EUN=4$ Gauge Theory From QNG Theory}
\seclab\emduality

\nref\kapustinwitten{A. Kapustin and E. Witten, ¡°Electric-Magnetic
Duality And The Geometric Langlands Program,¡±
[arXiv:hep-th/0604151].}

In last section, to study the QNG theory, we perturbed it into
quantum Abelian gerbe theory. In this section we will try to get a
glance at the non-Abelian nature of QNG theory by perturbing it into
a four-dimensional $\EUN=4$ super-Yang-Mills theory with gauge group
$G$. To do this, we compactify the six dimensional theory on a torus
$\BFT^2$, which is characterized by two parameters $(\tau,
A_{\BFT})$, where $\tau$ is its complex structure and $A_{\BFT}$ is
its world volume. We'll see that $\tau$ will become the complex
coupling constant of the gauge theory, and the Mantonen Olive
duality of $\EUN=4$ super-Yang-Mills theory can be explained as the
mapping class symmetry of $\BFT^2$. We'll understand how to get the
four-dimensional superconformal group $PSU(2, 2|4)$ (and its central
extensions) from $OSp(2, 6|2)$ (and its central extensions), under
the compactification.

All these results have been known (conjectured) in literatures. But,
by using the formalism in the present paper, we can get them quite
directly. Especially, for the Abelian theories, we can explicitly
write down the dimensional reduction. We further argue that the six
dimensional origin of $\EUN=4$ super-Yang-Mills theories and their
$SL(2, \BZ)$ dualities can be extended to the non-Abelian cases.

\subsec{The Dimension Reduction Of Superconformal Symmetries And
Dualities}

In this subsection we'll try to match the $\BFT^2$ reduction of the
algebra of superconformal group $OSp(2, 6|2)$ with the algebra of
$PSU(2,2|4)$ which is the superconformal group of four-dimensional
$\EUN=4$ super-Yang-Mills theory. The benefits of this matching are
two folds. Firstly, it is required by the matching between dimension
reduced QNG theory and the $\EUN=4$ super-Yang-Mills theory.
Moreover, with the six-dimensional interpretation of Mantonen Olive
duality in mind, one can conveniently derive the $SL(2, \BZ)$
transformations of the supersymmetries with the results agreeing
with \kapustinwitten .

\bigskip\noindent{\it The Dimension Reduction Of Superconformal Symmetries}

To see the dimension reduction of the (1+5)-dimensional
superconformal symmetry, we let the compactified $\BFT^2$ be along
the $\x^5, \x^6$ directions and decompose the Lorentz group
$\Spin(1,5)$ into $\Spin(1,3)\times \Spin(2)$, where $\Spin(1, 3)$
is the Lorentz group of four-dimensional theory and $\Spin(2)$ is
the rotation of $(5,6)$ directions. The chiral spinor of $\Spin(1,
5)\simeq SL(2, \H)\subset SL(4, \ct)$ can then be decomposed into
irreducable representations of $\Spin(1, 3)\times \Spin(2)$, with
the decomposing of indexes \eqn\deindex{(\alpha)\rightarrow (a,
+\half)\oplus (\dd{a}, -\half),} where $a=1, 2$ are the chiral
spinor indexes and $\dd{a}=1, 2$ are the anti-chiral spinor indexes
of the four-dimensional Lorentz group $\Spin(1,3)\subset \Spin(4,
\ct)\simeq SL(2, \ct)\otimes SL(2, \ct)$, $+1/2$, $-1/2$ are the
weights (the eigenvalues of the generator
$S_{56}=\half\gamma^5\gamma^6$) of $\Spin(2)$.

We then decompose the $Q$-charges and $S$-changes into
\eqn\superchargeinsix{\eqalign{&Q^{\alpha}_{~A}\rightarrow
(Q^{a}_{~A},~ \omega_{AB}\bar{Q}^{B}_{~\dd{a}})\cr
&S^A_{~~\alpha}\rightarrow (\omega^{AB}\bar{S}_{B\dd{a}}, S^{a
A}),}} according to their $\Spin(2)$ weights, where $Q^a_{~A}$
weights $+1/2$ and $S^{a A}$ weights $-1/2$. The real structure
$\hat{\tau}$ now acts as the complex conjugation that exchanges
$Q^{a}_{~A}$, $S^{a A}$ with $\bar{Q}^{B}_{~\dd{a}}$,
$\bar{S}_{B\dd{a}}$, respectively.

The six-dimensional momentum $\P^{\alpha\beta}$ can be decomposed
as, \eqn\pdecomp{\P^{\alpha\beta}\rightarrow
\P^{a\dd{b}}\oplus\bar{\P}_{\dd{a}b}\oplus{\cal Z}\oplus\bar{{\cal
Z}},} where $\bar{\P}_{\dd{a}b}$ is the conjugate of $\P^{a\dd{b}}$
with
$\bar{\P}_{\dd{a}b}=\epsilon_{bd}\epsilon_{\dd{a}\dd{c}}\P^{d\dd{c}}$,
and ${\cal Z}$, $\bar{{\cal Z}}$ are the momentums around the
internal space $\BFT^2$ with $\Spin(2)$ weight $+1$ and $-1$,
respectively. $\K_{\alpha\beta}$ can also be likewise decomposed.

One can then derive some of the superalgebras of $PSU(2,2|4)$ by
dimensional reducing the superalgebras of $OSp(2, 6|2)$ which have
been explicitly written out in section \superalgebra. For examples,
the $Q-Q$ anti-commutators are \eqn\fourqqss{\{Q^a_{~A},
\bar{Q}^{B\dd{a}}\}=\P^{a\dd{a}}\delta^{B}_{~A}.} Likewise, one can
have the $S-S$ anti-commutators $\{S^{a A},
\bar{S}^{\dd{a}}_{~B}\}=\K^{a\dd{a}}\delta^A_{~B}$. The $\K-S$
commutators and $\P-Q$ commutators are $i[\K^{a\dd{a}},
S^A_{~b}]=\delta^a_{~b}\bar{Q}^{A\dd{a}}$, $i[\P^{a\dd{a}},
Q_{bA}]=\delta^{a}_{~b}\bar{S}^{\dd{a}}_{~A}$ and their complex
conjugates.

Further more, one can decompose the angular momentum
$J^{\alpha}_{~\beta}$ as, \eqn\dej{J^{\alpha}_{~\beta}\rightarrow
J_{ab}\oplus J_{\dd{a}\dd{b}}\oplus {\cal
Z}^{a\dd{a}}\oplus\bar{{\cal Z}}_{\dd{a}a},} where $J_{ab}$ and
$J_{\dd{a}\dd{b}}$, which are complex conjugate to each other, are
the selfdual and anti-selfdual parts of the four-dimensional angular
momentums, and ${\cal Z}^{a\dd{a}}$ is a vector of $\Spin(4, \ct)$
(the complexification of the (1+3)-dimensional Lorentz group
$\Spin(1, 3)$) $\bar{{\cal Z}}_{\dd{a}a}$ is the complex conjugate
of ${\cal Z}^{a\dd{a}}$. One can have the $Q-S$ anti-commutators (by
using the results in section \superalgebra), \eqn\fourqs{\{Q_{a A},
S^{B}_{~b}\}=J_{ab}-i\epsilon_{ab}(D\delta^{B}_{~A}-\BR^{B}_{~A}).}
The $\bar{Q}-\bar{S}$ anti-commutators are the complex conjugation
of \fourqs .

Concerning the terms of ${\cal Z}$, $\bar{\cal Z}$, one will have
the additional $Q-Q$ anti-commutators, $\{Q^{a}_{~A},
Q^b_{~B}\}=\epsilon^{ab}\omega_{AB}{\cal Z}$ and its complex
conjugate $\{\bar{Q}^{A}_{~\dd{a}},
\bar{Q}^{B}_{~\dd{b}}\}=\epsilon_{\dd{a}\dd{b}}\omega^{AB}\bar{\cal
Z}$. States carrying nonzero ${\cal Z}$ charge will break the
$SU(4)_{\BR}$ symmetry down to the $\Spin(5)\simeq Sp(2, \H)_{\BR}$
subgroup that preserves $\omega_{AB}$. The corresponding BPS states
will satisfy the condition $|\P|^2-|{\cal Z}|^2=0$. In terms of
six-dimensional theory, this condition is just the on-sell condition
of massless particles, $\P^{\alpha\beta}\P_{\alpha\beta}=0$.
(Concerning ${\cal Z}^{a\dd{a}}$, $\bar{{\cal Z}}_{\dd{a}a}$, they
will contribute additional $Q-\bar{S}$ anti-commutators $\{Q^a_{~A},
\bar{S}^{\dd{a}}_{~B}\}={\cal Z}^{a\dd{a}}\omega_{AB}$ and its
complex conjugate $\{\bar{Q}^A_{~\dd{a}}, {S}^{B}_{~a}\}=\bar{{\cal
Z}}_{\dd{a}a}\omega^{AB}$.)

To see the meaning of ${\cal Z}$, one takes $\BFT^2$ as $S\times S'$
with radius $U$ and $V$ respectively. Let $u$, $v$ denote the
coordinates of $S$ and $S'$ with $(u\sim u+2\pi U, v\sim v+2\pi V)$.
${\cal Z}$ can then be written as \eqn\calzg{{\cal Z}={n\over
V}+i{m\over U},} where $n\in \BZ, m\in \BZ$, $n/V$ are the momentum
modes around the $v$ cycle and $m/U$ are the momentum modes around
the $u$ cycle. To generalize to the case of a general $\BFT^2$ with
complex structure $\tau$, one introduce complex coordinates
$dz=dv+i(V/U)du$, $d\bar{z}=dv-i(V/U)du$ to the special cases that
we just considered, and identify $V/U$ as $\Im\tau$ (in these
special cases $\Re\tau=0$). The general cases correspond to set
$dz=dv+\tau du$, $d\bar{z}=dv+\bar{\tau}du$, and make the
replacement of $iV/U\rightarrow \tau$ and $VU\rightarrow A_{\BFT}$,
where $A_{\BFT}=(i/2)\int_{\BFT^2}(dz\wedge d\bar{z})/\Im\tau$ is
the volume\foot{Since one can easily carry out the integration $(i/
2)\int_{\BFT^2}dz\wedge d\bar{z}$ to get $(i/
2)(\oint_vdz\oint_ud\bar{z}-\oint_udz\oint_vd\bar{z})=(\Im\tau)VU$.}
of $\BFT^2$. ${\cal Z}$ can then be generalized as \eqn\calz{{\cal
Z}={(n+\tau m)\over \sqrt{\Im\tau A_{\BFT}}}\ .}

To rewrite the parameter $1/\sqrt{A_{\BFT}}$ of \calz\ in terms of
the variables of the resulting four-dimensional gauge theory. We
first recall that the quantum states of the fluxes $H^2(X,
\BFT_G)\simeq H^2(\BFT^2, \BFT_G)\simeq \BFT_G$, thus, for a
$\BFT_G$ quantum Abelian gerbe theory compactified on $\BFT^2$, we
will have $r$ additional scalars coming from the Wilson-t'Hooft
surfaces of $r$ two forms $B^i, i=1,...,r$. Now, we set $r=1$ for
simplicity (this case will be further discussed in more details in
next section) with the additional scalar $\Phi\sim
B\sqrt{A_{\BFT}}$. Noticing that the flux quantization means $B
A_{\BFT}\sim 1$, we have $\Phi\sim 1/\sqrt{A_{\BFT}}$. In another
words, the central charge ${\cal Z}$ is proportional to the vacuum
expectation of the additional scalar.

Further more, the $SU(4)_{\BR}\subset SL(4, \ct)_{\BR}$ invariance
of the underlying four-dimensional $\EUN=4$ superconformal field
theory will tells us that general central terms ${\cal Z}_{AB}$ of
four-dimensional $\EUN=4$ superalgebra must take the form of
\eqn\zab{{\cal Z}_{AB}=\langle\Phi_{AB}\rangle(n+\tau
m)/\sqrt{\Im\tau},} where $\langle\Phi_{AB}\rangle$ stand for the
vacuum expectations of the scalars $\Phi_{AB}$ of the $\EUN=4$ gauge
theory. These scalars are the dimensional reduction of the original
five scalars in $\EUN=(2, 0)$ tensor multiplet plus an additional
scalar that comes from the tensor field. We shall discuss these in
details in next subsection.

Now, the problem is what are the six dimensional explanations about
the terms of ${\cal Z}_{AB}$ other than ${\cal Z}$? It turns out
that they come from the dimensional reduction of the central
extension of six-dimensional $\EUN=(2, 0)$ superalgebra.

As we have calculated (for free tensor multiplet, by using the
expressions \ssupercharge ), the $Q-Q$ anti-commutators of the six
dimensional $\EUN=(2,0)$ theory have a central extension ${\cal
Z}^{\alpha\beta}_{~AB}$ contributed by the self-dual strings,
\eqn\sixalgebra{\{Q^{\alpha}_{~A},
Q^{\beta}_B\}=\omega_{AB}\P^{\alpha\beta}+
{\cal{Z}}^{\alpha\beta}_{~AB},} where
${\cal{Z}}^{\alpha\beta}_{~AB}$ is carried by a infinite long string
extended along some direction of $X$ or by a winding string wound
around a super-symmetric 1-cycle of $X$.
${\cal{Z}}^{\alpha\beta}_{~AB}$ transform as the ${\bf{5}}$
representation of $Sp(2,\H)_{\BR}$, thus
\eqn\spcc{{\cal{Z}}^{\alpha\beta}_{~AB}\omega^{AB}=0.}

We now decompose ${\cal Z}^{\alpha\beta}_{~AB}\rightarrow
\epsilon^{ab}{\cal Z}_{AB}\oplus\epsilon^{\dot{a}\dot{b}}\bar{\cal
Z}_{AB}\oplus {\cal Z}^{a\dot{c}}_{~AB}\oplus{\bar{\cal
Z}}^{\dot{a}c}_{~AB}$, with the constraint \spcc\ acts on them
respectively. Combing $\epsilon^{ab}{\cal Z}_{AB}$ with
$\omega_{AB}\epsilon^{ab}{\cal Z}$ that comes from the decomposing
of $\P^{\alpha\beta}$, one has a $SU(4)_{\BR}\subset SL(4,
\ct)_{\BR}$ vector ${\cal Z}_{AB}$, with the constraint \spcc\ now
dropped (One can likewise form the $\bar{\bf 6}$ representation
$\bar{\cal Z}_{AB}$ of $SU(4)_{\BR}\subset SL(4, \ct)_{\BR}$).
Moreover, the combinations of the $Sp(2, \H)_{\BR}\subset Sp(4,
\ct)_{\BR}$ ${\bf 5}$ representation ${\cal Z}^{a\dot{c}}_{~AB}$ and
singlet representation ${\cal Z}^{a\dot{c}}\omega_{AB}$ will form a
${\bf 6}$ representation ${\cal Z}^{a\dot{c}}_{~AB}$ of
$SU(4)_{\BR}\subset SL(4, \ct)_{\BR}$, and one can likewise form the
complex conjugation $\bar{\cal Z}^{AB}_{~a\dot{c}}$ of ${\cal
Z}^{a\dot{c}}_{~AB}$. Thus, we have
\eqn\baqq{\eqalign{\{{Q}^{a}_{~A},Q^{b}_{~B}\}&=\epsilon^{ab}{{\cal
Z}}_{AB},\cr \{Q^a_{~A}, \bar{S}^{\dd{a}}_{~B}\}&=\ {\cal
Z}^{a\dd{a}}_{~~AB},}} and their complex conjugations. Charges
${\cal Z}^{a\dd{a}}_{~~AB}$ are carried by the self-dual strings
along the four dimensions $\RR^3\times\R^1$.

\bigskip\noindent{\it{The Duality Transformations Of The Superalgebra}}

Having reduced the superconformal group $OSp(2, 6|2)$ of QNG theory
into the superconformal group $PSU(2, 2|4)$ of $\EUN=4$ gauge
theory, and matched their central extensions, we now turn to derive
the $SL(2, \BZ)$ transformations of the four-dimensional
supersymmetries, with the six-dimensional explanation of
Mantonen-Olive duality in mind.

Firstly, one notice that the transformation $S: u\rightarrow v$,
$v\rightarrow -u$ acts as $dz\rightarrow -{1\over \tau}dz$,
$d\bar{z}\rightarrow -{1\over \bar{\tau}}d\bar{z}$ and
$\tau\rightarrow -{1\over \tau}$, that is,
${dz/\sqrt{\tau}}\rightarrow i{dz/\sqrt{\tau}}$ and
${d\bar{z}/\sqrt{\bar{\tau}}}\rightarrow
-i{d\bar{z}/\sqrt{\bar{\tau}}}$. Clearly, this transformation
implies a $\Spin(2)$ rotation $\exp(i\phi_S)=-{|\tau|/\tau}$.
Further more the six-dimensional origin \superchargeinsix\ tells us
that $Q^{a}_{~A}$ is weight $1/2$ under $\Spin(2)$, thus one has the
$S$-duality transformation of $Q^{a}_{~A}$, ${Q^{a}_{~A}\rightarrow
e^{i{\phi_S/2}}Q^{a}_{~A}=\sqrt{-{|\tau|/\tau}}Q^{a}_{~A}}$. By
noticing that $T: \tau\rightarrow \tau+1$ is trivially acting on
$Q^{a}_{A}$, one can get the transformations of $Q^{a}_{~A}$ under
full $SL(2, \BZ)$ mapping class symmetries, $\tau\rightarrow
{(a\tau+b)}/{(c\tau +d)}, ad-bc=1$,  with
\eqn\sdualityq{Q^{a}_{~A}\rightarrow \left(|c\tau+d|\over
c\tau+d\right)^{1/2}Q^{a}_{~A}.} This result is exactly identical to
the corresponding results of \kapustinwitten\ (section 2).

Further more, one can consider the central extension of the $\EUN=4$
superalgebra by the anti-commutators $\{Q^{a}_{~A},
Q^{b}_{~B}\}=\epsilon^{ab}{\cal Z}_{AB}$, which breaks the
$SU(4)_{\BR}$ symmetry. The six-dimensional explanation of this
extension has been given in last subsection. We can see that
\sdualityq\ implies the $SL(2, \BZ)$ transformation of
${\cal{Z}}_{AB}$ \eqn\ztrans{{\cal Z}_{AB}\rightarrow
\left({|c\tau+d|\over c\tau+d}\right){\cal Z}_{AB},} and $|{\cal
Z}|$ is $SL(2, \BZ)$ invariant.

In next section, we'll see that $\tau$ should be identified as the
complex coupling constant of $\EUN=4$ super-Yang-Mills theory. And
the mapping class symmetry of $\BFT^2$ should be identified as the
$SL(2, \BZ)$ Montonen-Olive duality. Thus, the quantum numbers $(m,
n)$ of equation \zab\ should be interpreted as the quantum numbers
of magnetic-electric charges. By noticing the transformation law of
$\tau$, $\tau\rightarrow (a\tau+b)/(c\tau+d)$, and
$\Im\tau\rightarrow \Im\tau/|c\tau+d|^2$, equation \ztrans\ will
tell us that the $SL(2, \BZ)$ duality transformations act on the
magnetic and electric charges as \eqn\mntrans{(m, n)\rightarrow (m,
n) \left(\matrix{a &b \cr c &d}\right)^{-1}=(m, n)\left(\matrix{d
&-b\cr -c &~a}\right).}

\bigskip
\subsec{$\EUN=4$ Super-Yang-Mills Theory From QNG Theory And $SL(2,
\BZ)$ Dualities}

We now compactify tensor multiplet on $\BFT^2$, $X=\RR^3\times
\BFT^2$. We'll show that the reduced Hamiltonian of the $U(1)$
quantum Abelian gerbe theory is identical to the one of the
four-dimensional $\EUN=4$ $U(1)$ gauge theory. The self-duality of
$H$ plus the mapping class transformation of $\BFT^2$ lead to the
$S$ duality of four dimensional gauge theory which is found through
a five dimensional gauge theory argument \wittenoncft\witten .

For simplicity, we'll take $\BFT^2=S\times S'$. $S$ and $S'$ are
circles with radius $U$ and $V$, respectively. Let $u$ and $v$
denote the coordinates of $S$ and $S'$ with $(u\sim u+2\pi U, v\sim
v+2\pi V)$. We now can pull the tensor field $\vec{H}$ down to
$\RR^3$ by assuming \eqn\reduction{ \eqalign{\vec{H}=Fdu/2\pi
U+\check{F}dv/2\pi V+g du\wedge dv+*f,}} where $\check{F}={1\over
2}\check{F}_{ij}d\x^i\wedge d\x^j (i=1,2,3)$ stands for the electric
2-form on the three dimensional space $\RR^3$, and $F={1\over
2}F_{ij}d\x^i\wedge d\x^j$ stands for the magnetic 2-form.

We'll focus on the $\check{F}$ and $F$ fields. By substituting
\reduction\ into the Hamiltonian $\bf{H}$ and integrating over
$\BFT^2$, one can get \eqn\fourhamilton{ \BFH={1\over
4\pi}\int_{\RR^3}\left({V\over U}||F||^2+{U\over
V}||\check{F}||^2+....\right),} where $||F||^2$ and
$||\check{F}||^2$ stand for $F\wedge
*F$ and $\check{F}\wedge *\check{F}$, respectively,
and the ellipses refer to terms involving other fields of the theory.

One can compare \fourhamilton\ with the Hamiltonian of four
dimensional electro-magnetic field. Here, we normalize the action of
the electro-magnetic field as ${1\over e^2}\int
(\vec{E}^2-\vec{B}^2)$. In terms of the canonical variables, the
associated Hamiltonian is of the form $\BFH=\int_{\RR^3}
\left(e^2{1\over 4}\vec{\EUP}^2+{1\over e^2}\vec{B}^2\right)$.
Clearly, this Hamiltonian can be identified with \fourhamilton\ by
setting $F={1\over 2}B_{ij}d\x^i\wedge d\x^j, {\check{F}\over
2\pi}=*(\EUP_id\x^i)$ and identifying the electric coupling constant
as \eqn\esqure{{e^2\over 4\pi}={U\over V}.} From \reduction\ and
\esqure, one can see that the transformation $u\rightarrow v$
$v\rightarrow -u$ acts as the $S$ duality of the four dimensional
theory, which interchanges the magnetic field $F$ with the electric
field $\check{F}$ and maps the coupling constant $e^2$ to $g^2$,
\eqn\sduality{e^2\rightarrow g^2=(4\pi)^2/e^2.}

\bigskip\noindent{\it Generalized To The Full Supersymmetrical
Theory}

To study the dimensional reduction of the 4-fermions
$\Psi^{\alpha}_{A}$, one decomposes $\Psi^{\alpha}_{~A}$ as
$\Psi^{(a,+1/2)}_{~A}\oplus \Psi^{(\dd{a}, -1/2)}_{~A}$. One can
then make the following identifications \eqn\psichi{\eqalign{{i\over
2}\int_{\BFT^2} {dz\wedge d\bar{z}\over \Im\tau}\bar{\Psi}^{(\dd{a},
-\half)A}L_{(\dd{a}, -\half)(a,
+\half)}\Psi^{(a,+\half)}_{~A}\rightarrow
\bar{\chi}^{\dd{a}A}L_{\dd{a}a}\chi^{a}_{~A}}} where,
$\chi^{a}_{~A}$ and $\bar{\chi}^{A}_{~\dd{a}}$ are the chiral and
anti-chiral fermions of the four-dimensional theory, and $L$ stands
for some operator acting on the fermions. Since there is no real
problem for the Lagrangian description of the chiral fermions, we'll
not explicitly write down the reduction in terms of the Hamiltonian,
but write out the terms of action of $\chi^{a}_{~A}$ and
$\bar{\chi}^{A}_{~\dd{a}}$ \eqn\fermionicerms{I_{\rm{F}}={1\over
4\pi}\int_{\RR^3\times
\R^1}\left(\bar{\chi}^{\dd{a}A}\p_{\dd{a}a}\chi^{a}_{~A}\right).}
These terms can be got by Lergende transforming the Hamiltonian got
from the dimensional reduction.

Now, we can study the action of $S$-duality on $\chi^{a}_{~A}$ and
$\bar{\chi}^{A}_{~\dd{a}}$ by utilizing their six dimensional
origins that we have just explored. Firstly, by viewing \psichi, one
can see that $\chi^{a}_{~A}$ is roughly identified with $\Psi^{(a,
+1/2)}_{~A}$, and $\bar{\chi}^{A}_{~\dd{a}}$ is roughly identified
with $\bar{\Psi}^{(\dd{a}, -\half)A}$. One then notices that the
transformations $u\rightarrow v$, $v\rightarrow -u$, $dz\rightarrow
-{1\over\tau}dz$, $d\bar{z}\rightarrow -{1\over\bar{\tau}}d\bar{z}$
imply a $\Spin(2)$ rotation $\exp(i\phi_S)=-|\tau|/ \tau$. Under
this rotation, $dz\rightarrow \exp(i\phi_S)dz$, $d\bar{z}\rightarrow
\exp(-i\phi_S)d\bar{z}$, and $\Psi^{(a, +1/2)}_{~A}\rightarrow
e^{i\phi_S/2}\Psi^{(a, +1/2)}_{~A}$, $\Psi^{(a,
-1/2)}_{~A}\rightarrow e^{-i\phi_S/2}\Psi^{(a, -1/2)}_{~A}$, thus
\eqn\schi{\eqalign{\chi^{a}_{~A}\rightarrow \exp(+i{1\over
2}\phi_S)\chi^{a}_{~A}\cr \bar{\chi}^{A}_{~\dd{a}}\rightarrow
\exp(-i{1\over 2}\phi_S)\bar{\chi}^{A}_{~\dd{a}}.}}

The full mapping class symmetries $SL(2, \BZ)$ are generated by the
$S$ transformation, which we have just investigated, and the $T$
transformation $\tau\rightarrow \tau+1$ which is trivially acting on
these fermions.  Thus, a general transformation $\tau\rightarrow
(a\tau+b)/(c\tau+d)$ must act as
\eqn\sschi{\eqalign{&\chi^{a}_{~A}\rightarrow \left(|c\tau+d|\over
c\tau+d\right)^{1/2}\chi^{a}_{~A}\cr
&\bar{\chi}^{A}_{~\dd{a}}\rightarrow \left(|c\bar{\tau}+d|\over
c\bar{\tau}+d\right)^{1/2}\bar{\chi}^{A}_{~\dd{a}}\ .}} This may be
compared to the transformations of $\chi^{a}_{~A}$ and
$\bar{\chi}^{A}_{~\dd{a}}$ under Montonen-Olive dualities.

We now turn to the dimension reduction of the tensor field and the
scalar fields. One first decompose $H^{(\alpha\beta)}$ as
\eqn\FFFF{\eqalign{ +&(1/\sqrt{A}_{\BFT})\left(\check{F}_+^{ ab}{dz/
\sqrt{\tau}}+\check{F}_-^{\dd{a}\dd{b}}{d\bar{z}/
\sqrt{\bar{\tau}}}\right)\cr -i&(1/\sqrt{A}_{\BFT})\left( F^{ab}_+{\
\sqrt{\tau}\ dz}-F_-^{\dd{a}\dd{b}}\
{\sqrt{\bar{\tau}}d\bar{z}}\right)}} plus the terms concerning the
derivatives of an additional scalar field $\Phi$, which is connected
with the one form $g$ in \reduction\ through $g\sim d\Phi$, with
$\p^{a\dd{b}}\Phi+\p^{\dd{a}b}\Phi$. The real condition now tells us
that $F^{\dd{a}\dd{b}}_-$ and $\check{F}^{\dd{a}\dd{b}}_-$ are the
complex conjugations of $F^{ab}_+$ and $\check{F}^{ab}_+$,
respectively.

By substituting \FFFF\ into the Hamiltonian $\BFH$ \shamiltionian ,
one has the terms concerning the electro-magnetic fields of the
resulting four dimensional theory \eqn\FFH{\eqalign{&{i\over
4\pi}\int_{\RR^3}\left(\ \tau\parallel F_+\parallel^2-\
\bar{\tau}\parallel F_-\parallel^2\right)\cr -&{i\over
4\pi}\int_{\RR^3}\left({1\over\tau}\parallel
\check{F}_+\parallel^2-{1\over\bar{\tau}}\parallel
\check{F}_-\parallel^2\right),}} where $\parallel F_{+}\parallel^2$
and $\parallel F_{-}\parallel^2$ stand for $F^{ab}_{+}F_{ab+}$ and
$F^{\dd{a}\dd{b}}_{-}F_{\dd{a}\dd{b}-}$, the meaning of the
notations $\parallel\check{F}_{\pm}\parallel^2$ are likewise.
Comparing to the Hamiltonian of four dimensional $U(1)$ gauge theory
one can see that $F_{+}$, $F_{-}$ are the selfdual and anti-selfdual
part of the magnetic field $F$, and $\check{F}_{+}$, $\check{F}_{-}$
are the selfdual and anti-selfdual part of the electric field
$\check{F}$. Thus, $\tau$ will be identified as the complex coupling
constant of the gauge theory \eqn\coplexcoupling{\tau={4\pi i\over
e^2}+{\theta\over 2\pi},} where $\theta$ is the theta angle. And, by
noticing the decomposition \FFFF, one will see that the
$v\rightarrow -u, u\rightarrow v$ transformation will exchange the
electric field and the magnetic field
\eqn\fvee{\check{F}_+\leftrightarrow\tau F_+,\
\check{F}_-\leftrightarrow\bar{\tau}F_-.} Thus, this transformation
can indeed be identified with the $S$-duality.

Now, we turn to the dimensional reduction of the scalars. We have
five-scalars $\Phi_{AB}$ which transform as the $\bf{5}$
representation of $Sp(2,\H)_{\BR}$. The dimensional reduction of the
tensor field $H^{(\alpha\beta)}$ will give us an additional scalar
$\omega_{AB}\Phi$ as we have mentioned. The combination of
$\omega_{AB}\Phi$ and the previous five scalars forms the $\bf{6}$
representation of the ${\BR}$-symmetry $SU(4)_{\BR}\subset SL(4,
\ct)_{\BR}$ of the resulting four dimensional $\EUN=4$ gauge theory.
Without confusing, we'll denote these scalars as $\Phi_{AB}$ with
the constraint equation $\Phi_{AB}\omega^{AB}=0$ now dropped. These
six scalars should satisfy the following real conditions
\eqn\sixscalarreal{\Phi_{AB}=\half\epsilon_{ABCD}\bar{\Phi}^{CD},}
where $\epsilon_{ABCD}$ is the volume form of the complex space
$\ct^4_{\BR}$ which represents $SL(4, \ct)_{\BR}$.

The dimensional reduction of $\Phi_{AB}$ is achieved through
${i\over 2}(\Im\tau)^{-1}\int{dz\wedge d\bar{z}}\
\bar{\Phi}^{AB}\Phi_{AB}\rightarrow \bar{\Phi}^{AB}\Phi_{AB}$. With
an appropriate normalization, the corresponding terms of the
Hamiltonian concerning these scalars are \eqn\scalarhamiltonian{
{1\over
2\pi}\int_{\RR^3}\left(\bar{\Pi}^{AB}\Pi_{AB}+\p\bar{\Phi}^{AB}\p\Phi_{AB}\right).}
The mapping class symmetry $u\rightarrow v$, $v\rightarrow -u$
trivially acts on these scalars, $\Phi_{AB}\rightarrow \Phi_{AB}$.

Now, we would like to transform the Hamiltonian, got from the
dimensional reduction of the six dimensional theory, into the action
expressions, $I=I_{\rm{F}}+I_B$. Here, $I_{\rm{F}}$ of the fermions
has been written out \fermionicerms, and $I_{B}$ of the bosonic
fields can be written as \eqn\fouraction{I=\int_{\RR^3\times
\R^1}\left({i\tau\over 4\pi}F^{ab}F_{ab}-{i\bar{\tau}\over
4\pi}F^{\dd{a}\dd{b}}F_{\dd{a}\dd{b}}+{1\over
4\pi}\p_{a\dd{b}}\bar{\Phi}^{AB}\p^{a\dd{b}}\Phi_{AB}\right),} where
the self dual part $F^{ab}$ of four-dimensional gauge field is
$F^{ab}=F^{ab}_{+}+{i\over \tau}\check{F}^{ab}_{+}$, and the
anti-self dual part $F^{\dd{a}\dd{b}}$ is
$F^{\dd{a}\dd{b}}_--{i\over \bar{\tau}}\check{F}^{\dd{a}\dd{b}}_-$.
As one can see that $I$ is indeed the action of four-dimensional
$\EUN=4$ $U(1)$ gauge theory.

We have shown that the dimensional reduction on $\BFT^2$ of the
quantum $U(1)$ gerbe theory exactly reproduce the Hamiltonian of
four dimensional $\EUN=4$ super-symmetrical $U(1)$ gauge theory. And
the mapping class symmetry of $\BFT^2$ is transmuted into the $SL(2,
\BZ)$ duality of the four-dimensional $\EUN=4$ gauge theory. These
results can be trivially generalized to the case of $\BFT_G$ Abelian
gerbe theory, which is the low energy effective theory of
six-dimensional $\EUN=(2, 0)$ theory at a generic point of the
moduli space $\RR^{5r}/W_G$. Thus, by including the additional $r$
scalars coming from the dimensional reduction of $r$ $H$ fields, the
reduced theory can be reasonably identified with the low energy
effective theory of four-dimensional $\EUN=4$ super-Yang-Mills
theory at a generic point of the moduli space $\RR^{6r}/W_G$.
Further more, as it is well known that at the singularities of
$\RR^{6r}/W_G$ some BPS states with nontrivial magnetic-electric
charges become massless, and the corresponding theory becomes an
interacting $\EUN=4$ superconformal field theory with some
non-Abelian gauge group $G'\subset G$. By noticing the matching
between the superconformal symmetries of the four-dimensional theory
and of the six-dimensional theory, one naturally expects that this
four-dimensional interacting superconformal field theory may be the
dimensional reduction of a six-dimensional QNG theory with the
non-Abelian gerbe group $G'\subset G$. The symmetries are enhanced
due to some BPS self-dual strings becoming tensionless at the
singularities. In next section, we will turn to investigate these
self-dual string excitations, which consist another ingredient of
the QNG theory.

\bigskip
\newsec{The Self-Dual Strings}
\seclab\selfdualstring

\nref\stringsoliton{P. S. Howe, N. D. Lambert, P. C. West, The
self-dual string soliton, Nucl. Phys. B515 (1998) 203-216.}
\nref\MC{Meng-Chwan Tan, Five-Branes in M-Theory and a
Two-Dimensional Geometric Langlands Duality, [arXiv:
hep-th/0807.1107]. }

In the present section, we'll study the self-dual string excitations
of the $\EUN=(2, 0)$ theory and their tensionless limit. We'll begin
from the classical self-dual string solution \stringsoliton\ of the
Bogomolnyi equation, and then we'll try to propose a world sheet
$\EUN=(4, 4)$ superconformal field theory to describe their long
wavelength oscillations and their coupling with the free tensor
multiplets. We'll argue that under the tensionless limit these
self-dual strings will become some objects moving in supertwisor
space $\hat{\BT}$, which should be discussed in more details in the
next section.

\bigskip\noindent{\it{Bogomolnyi Equations And Self-Dual Strings}}

We begin from the Bogomolnyi equation that should be fulfilled by
the configurations of the BPS states, \eqn\sstsecond{
\delta\Psi^{\alpha}_{~A}=0\Leftrightarrow\left(\half
H^{(\alpha\beta)}\epsilon^{A}_{~\beta}+\p^{\alpha\beta}\Phi^{A}_{~B}\epsilon^{B}_{~\beta}\right)=0.}
Here, we have used the second equation of \sst.

A straight self-dual string configurations will break the
(1+5)-dimensional Lorentz symmetry to $SO(1, 1)\times
SO(4)_{\perp}$, where $SO(1, 1)$ is generated by the boost along the
string and $SO(4)_{\perp}$ is the rotation around it. Moreover, this
self-dual string will also break the $\BR$ symmetry down to
$SO(4)_{\BR}$. Thus, the static configurations of such straight
strings are parameterized by the imbedding of
$SO(4)_\perp\hookrightarrow SO(5)$ (the spatial rotation group) and
$SO(4)_{\BR}\hookrightarrow SO(5)_{\BR}$.

To specify a static straight BPS string, one can pick out a
$SO(5)_{\BR}\simeq Sp(2, \H)_{\BR}$ vector $\psi_{AB}$ -- with
$\psi_{AB}\omega^{AB}=0$, $\psi^A_{~B}\psi^{B}_{~C}=\delta^{A}_{~C}$
-- and a spatial direction $l_{\alpha\beta}$ with
$l_{\alpha\beta}t^{\alpha\beta}=0$. Now, let the string lie along
$l_{\alpha\beta}$ direction, and set the scalars to take the form of
$\Phi_{AB}=\psi_{AB}\Phi$ which will break the $SO(5)_{\BR}$
symmetry down to $SO(4)_{\BR}$, here $\Phi$ is a scalar function of
$l_{\alpha\beta}$, $t_{\alpha\beta}$ and the transverse coordinates
$\x_{\perp}$. It is convenient to take the $\sigma_+, \sigma_-$
coordinates along the world sheet of the string, with $\sigma_+=l+t$
and $\sigma_-=l-t$ ($l$ and $t$ are the vectors whose components are
$l_{\alpha\beta}$ and $t_{\alpha\beta}$, respectively). On these
coordinates, the $SO(1, 1)$ boost acts as $\sigma_+\rightarrow
e^{\theta}\sigma_+, \sigma_{-}\rightarrow e^{-\theta}\sigma_-$.

Clearly, $SO(4)_{\perp}$ preserves $l_{\alpha\beta}$, and
$SO(4)_{\BR}$ preserves $\psi_{AB}$. One then decomposes the $SO(1,
5)$ anti-chiral spinor $\epsilon^A_{~\alpha}$ into
$\epsilon^{A}_{-,\a}\oplus\epsilon^{A}_{+,\dot{\a}}$ according to
$SO(1, 5)\rightarrow SO(1, 1)\times SO(4)_{\perp}$, where $\a,
\dot{\a}=1,2$ are used to label the chiral and anti-chiral spinors
of $SO(4)_{\perp}$, and $+$ ($-$) labels the $+1/2$ ($-1/2$) weight
of $SO(1, 1)$. One can further decompose the tensor fields
$H^{(\alpha\beta)}$ into $H_{+1}^{(\a\c)}\oplus
H_{-1}^{(\dd{\a}\dd{\c})}\oplus H^{\dd{\a}\c}$, where the
subscriptions $+1$ and $-1$ are the weights of $SO(1, 1)$
($H^{\dd{\a}\c}$ is invariant under $SO(1, 1)$). Thus \sstsecond\
can be rewritten as \eqn\bps{\eqalign{\half
H_{+1}^{(\a\b)}\epsilon^A_{-\b}+\epsilon^{\a\b}\p^+\Phi\psi^{A}_{~B}\epsilon^{B}_{-\b}&=0\cr
\half
H_{-1}^{(\dd{\a}\dd{\b})}\epsilon^{A}_{+\dd{\b}}+\epsilon^{\dd{\a}\dd{\b}}
\p^{-}\Phi\psi^{A}_{~B}\epsilon^{B}_{+\dd{\b}}&=0,}} which are a
pair of equations concerning the derivatives along the string world
sheet. Here, $\epsilon^{\a\b}$ and $\epsilon^{\dd{\a}\dd{\b}}$ are
the symplectic forms of the $SO(4)_{\perp}\subset SO(4, \ct)\simeq
SL(2, \ct)\otimes SL(2, \ct)$ chiral and anti-chiral spinors,
respectively, and we'll use them to raise and lower the
$SO(4)_{\perp}$ spinor indexes. The decomposition of \sstsecond\
also gives us a pair of equations concerning the derivatives of the
transverse coordinates, \eqn\bpsb{\eqalign{\half
H^{\dd{\a}\b}\epsilon^{A}_{-\b}+
(\p^{\dd{\a}\b}\Phi)\psi^{A}_{~B}\epsilon^B_{-\b}&=0\cr \half
H^{\dd{\a}\b}\epsilon^A_{+\dd{\a}}-(\p^{\dd{\a}\b}\Phi)\psi^{A}_{~B}\epsilon^{B}_{+\dd{\a}}&=0\
.}} In getting these equations, we have used the decompositions of
$\p^{\alpha\beta}\rightarrow\p^{\dd{\a}\b}\oplus\epsilon^{\a\b}\p^+
\oplus\epsilon^{\dd{\a}\dd{\b}}\p^{-}$, according to their
representations under $SO(1, 5)\rightarrow SO(1, 1)\times
SO(4)_{\perp}$.

For the static configurations, $\p_+\Phi=\p_-\Phi=0$. If the
considered configuration preserve some supersymmetries, the pair of
equations \bps\ will enforce $H_{+1}^{(\a\b)}$ vanish or
$H_{-1}^{(\dd{\a}\dd{\b})}$ vanish or both. To satisfy the pair of
equations \bpsb, one set
\eqn\bpshphi{H^{\dd{\a}{\b}}/2=\p^{\dd{\a}{\b}}\Phi\ .} By
substituting \bpshphi\ into \bpsb, one can get
\eqn\bpseps{\eqalign{\epsilon^A_{-\a}+\psi^{A}_{~B}\epsilon^B_{-\a}&=0\cr
\epsilon^{A}_{+\dd{\a}}-\psi^{A}_{~B}\epsilon^{B}_{+\dd{\a}}&=0\ .}}
Combining the equations of motion
$\p_{\alpha\beta}H^{(\beta\gamma)}=0$, \bpshphi\ implies that $\Phi$
should be harmonic function of $\x_{\perp}$, and can be solved as
$\Phi(\x_{\perp})=\Phi_0+ Q/\pi\x^2_{\perp}$, where $Q$ is the
charge of the self-dual string, with
$\lim_{\x_{\perp}\rightarrow\infty}\int_{S^3(\x_{\perp})}{H\over
2\pi}=\int_{S^3(\x_{\perp})}\star{H\over 2\pi}=Q$, $S^3(\x_{\perp})$
is the 3-dimensional sphere surrounding the string at transversal
radius $|\x_{\perp}|$.

\nref\dvv{R. Dijkgraaf, E. Verlinde and H. Verlinde, 5-D Black Holes
And Matrix Strings, Nucl.Phys. B506 (1997) 121-142, [arXiv:
hep-th/9704018]; E. Witten, ¡°On the Conformal Field Theory of the
Higgs Branch,¡± JHEP. 9707 (1997) 003, [arXiv:hep-th/9707093].}

\bigskip\noindent{\it{World Sheet Theory Of The Self-Dual String}}

The $\half$-BPS states correspond to the case of
$H_{+1}^{(\a\b)}=0$, $H_{-1}^{(\dd{\a}\dd{\b})}=0$ of the static
configurations. In this case, half of the super-symmetries that
satisfy \bpseps\ can be preserved. By viewing $\psi^{A}_{~B}$ as
$\gamma^5_{\BR}$ of the remaind $\BR$ symmetry $SO(4)_{\BR}$, one
can take that the preserved super-symmetries as anti-chiral and
chiral spinors of $SO(4)_{\BR}$, with
$\gamma^5_{\BR}\epsilon^{\dd{\A}}_{-\a}=-\epsilon^{\dd{\A}}_{-\a}$,
$\gamma^5_{\BR}\epsilon^{{\A}}_{+\dd{\a}}=\epsilon^{{\A}}_{+\dd{\a}}$,
where $\A, \dd{\A}=1,2$ are the indexes of $SO(4)_{\BR}$ chiral and
anti-chiral spinors (should not confuse with the indexes of
4-components of $Sp(2, \H)_{\BR}$ which are labeled as $A$). And the
broken super-symmetries are $\epsilon^{\A}_{-\a}$ and
$\epsilon^{\dd{\A}}_{+\dd{\a}}$.

We'll denote the generators of the preserved supersymmetries as
$G^{{\a}}_{-\half\dd{\A}}$, and $G^{\dd{\a}}_{-\half{\A}}$, which
are left-moving charges and right-moving charges along the string.
Here the subscription $-1/2$ is used to denote the $1/2$ dimension
of these supercharges.

Further more, we suppose that the global superconformal group of the
IR superconformal field theory, describing the long wavelength
oscillating of the string world sheet, is given by the dimensional
reduction of $OSp(2, 6|2)\subset OSp(8|4, \ct)$. Thus, the world
sheet IR theory is a $\EUN=(4, 4)$ superconformal field theory. Its
global superconformal group is $PSU(1, 1|2)\subset PSL(2|2,
\ct)\hookrightarrow  OSp(8|4, \ct)\supset OSp(2, 6|2)$. Here we are
focusing on the left-moving part of this superconformal field
theory, the investigation to the right-moving part is quite similar,
and we will leave it to the reader. In this reduction, the $SU(2)$
$\BR$-symmetry group of $PSU(1, 1|2)$ is identified as
$SU(2)^L_{\BR}\subset SO(4)_{\BR}\hookrightarrow SO(5)_{\BR}$ acting
on the left-moving modes, and the $SU(2)$ exterior automorphism of
the algebra of $PSU(1, 1|2)$ is identified as
$SU(2)^L_{\perp}\subset SO(4)_{\perp}$. We'll denote the generators
of $SU(2)^{L}_{\BR}$ as $R_{\dot{\A}\dot{\B}}$, which is the
reduction of $\BR_{(AB)}$. The reduction of the $S$-charges give us
the dimension $-1/2$ fermionic generators $G^{{\a}}_{+\half
\dd{\A}}$, with the $G-G$ anti-commutators
\eqn\ggalge{\{G^{{\a}}_{-\half\dd{\A}},
G^{{\c}}_{+\half\dd{\B}}\}=\epsilon^{{\a}{\c}}\epsilon_{\dd{\A}\dd{\B}}L_0
+\epsilon^{{\a}{\c}}R_{\dot{\A}\dot{\B}},} where $L_0$ is the
dimension $0$ generator of the left-moving conformal group
$SL(2,\R)$.

The localization, on the string, of the broken left moving fermionic
transformations $\epsilon^{\dd{\A}}_{+\dd{\a}}$ gives out
4-fermionic left moving fields $\psi^{\dd{\A}}_{~\dd{\a}}$, and $\p
x_{\dd{\a}\a}$ are their super-partners. Also, the broken right
moving fermionic transformations $\epsilon^{{\A}}_{-{\a}}$ give us
$\bar{\p} x_{\dd{\a}\a}$ and $\tilde{\psi}^{\A}_{~\a}$. These fields
localize on the self-dual string and form the collective coordinates
describing its long scale oscillations. Now, we can write out the
most simple possible action of the world sheet theory,
\eqn\sws{I={1\over 4\pi}\int_{\Sigma}|\Phi_0|\left(\p
x^{\dd{\a}\a}\bar{\p}x_{\dd{\a}\a}+\psi^{\dd{\a}}_{~\dd{\A}}\bar{\p}\psi^{\dd{\A}}_{~\dd{\a}}
+\tilde{\psi}^{{\a}}_{~{\A}}\p\tilde{\psi}^{{\A}}_{~{\a}}\right),}
where the string tension $\Phi_0$ is the expectation value of $\Phi$
at the infinity of the string, and we have rotated the world sheet
coordinates into complex coordinates $z, \bar{z}$ with
$\sigma_+\rightarrow z$, $\sigma_-\rightarrow \bar{z}$. This is a
$c=6$ $\EUN=(4, 4)$ superconformal field theory with level $k=1$
current algebra $R_{\dot{\A}\dot{\B}}(z)$. One can work out the
explicit expressions of virous currents, for example, the
left-moving supercurrents are given by
\eqn\ttpsiphi{G^{{\a}}_{~\dd{\A}}(z)=|\Phi_0|\p
x^{\dd{\a}\a}\psi_{\dd{\a}\dd{\A}}(z).} And one can calculate the
OPE of these currents, for example, the $G-G$ OPE is
\eqn\ggcope{\eqalign{G^{\a}_{~\dd{\A}}(z)G^{\b}_{~\dd{\B}}(0)&\sim
{2\over z^3}\epsilon^{\a\b}\epsilon_{\dd{\A}\dd{\B}}+
\epsilon^{\a\b}\epsilon_{\dd{\A}\dd{\B}}{1\over z}T(0)\cr
&+\epsilon^{\a\b}{2\over
z^2}R_{\dd{\A}\dd{\B}}(0)+\epsilon^{\a\b}{1\over z}\p
R_{\dd{\A}\dd{\B}}(0).}}

The interactions between the self-dual string and the free tensor
multiplets can be given by calculating the correlations of various
appropriate operators inserted in the path integration of the world
sheet theory. The operators that correspond to the incident waves of
the tensor field are $\exp(i
|\Phi_0|^{-1}\int_{\Sigma}H^{\dd{\a}\c}x_{\dd{\a}\c})$,
$\exp(i|\Phi_0|^{-1/2}\int_{\Sigma}H^{(\a\c)}\tilde{\psi}_{\a\A}\tilde{\psi}^{\A}_{~\c})$,
or
$\exp(i|\Phi_0|^{-1/2}\int_{\Sigma}H^{(\dd{\a}\dd{\c})}{\psi}_{\dd{\a}\A}{\psi}^{\A}_{~\dd{\c}})$
according to their polarizations. The operators that correspond to
the incident waves of $\Psi^{\alpha}_{~A}$ are
$\exp(i|\Phi_0|^{-3/4}\int_{\Sigma}\Psi^{{\a}}_{~{\A}}\tilde{\psi}^{{\A}}_{~{\a}})$,
$\exp(i|\Phi_0|^{-3/4}\int_{\Sigma}\Psi^{\dd{\a}}_{~\dd{\A}}\psi^{\dd{\A}}_{~\dd{\a}})$,
and
$\exp(i|\Phi_0|^{-1/4}\int_{\Sigma}\Psi^{\a}_{~\dd{\A}}\psi^{\dd{\A}\dd{\a}}\p
x_{\dd{\a}\a})$,
$\exp(i|\Phi_0|^{-1/4}\int_{\Sigma}\Psi^{\dd{\a}}_{~{\A}}\psi^{\a\A}\bar{\p}
x_{\dd{\a}\a})$, where we have decomposed the incident waves of
$\Psi^{\alpha}_{~A}$ as
$\Psi^{{\a}}_{~{\A}}\oplus\Psi^{\dd{\a}}_{~\dd{\A}}\oplus\Psi^{\a}_{~\dd{\A}}\oplus\Psi^{\dd{\a}}_{~{\A}}$.
And the operators that correspond to the incident waves of the
scalars can be discussed likewise.

Now, we want to understand the tensionless limit
$|\Phi_0|\rightarrow 0$ of the self-dual string. The author did not
have a clear idea about how to take this limit appropriately. But
there are some clues that may be notable. Firstly, by noticing the
imbedding $PSL(2|2, \ct)\hookrightarrow OSp(8|4, \ct)$ that we have
defined, especially by noticing the actions of the R-symmetries, one
can see that the world sheet $\EUN=(4, 4)$ superconformal field
theory may be identified as the IR conformal field theory of the
Coulomb branch of some vector and hypermultiplets with $\EUN=(4, 4)$
supersymmetry. Thus, one may try to identify the tensionless limit
of the world sheet theory with the IR conformal field theory of the
Higgs branch, with the transverse rotation $SO(4)_{\perp}\simeq
SU(2)^L_{\perp}\otimes SU(2)^R_{\perp}$ acting as the R-symmetry.
The situation is quite similar with the Matrix-String proposal \dvv\
describing the dynamics of IIA $NS_5$ branes (little string theory).
But, in the present situation, the tensionless strings cannot move
along the space-time since the transverse rotations act as the
R-symmetries. Furthermore, as we have argued in last section that
under taking the tensionless string limit, we will have the full QNG
theory which realizes the superconformal symmetry $OSp(2,
6|2)\subset OSp(8|4, \ct)$. Thus, we naturally expect that the world
sheet theory should be explicitly $OSp(2, 6|2)\subset OSp(8|4, \ct)$
invariant. A conjecture is that {\it{these tensionless strings are
in fact moving in the supertwistor space $\hat{\BT}$ of $OSp(8|4,
\ct)$.}}

\newsec{Toward A Formulation In Supertwistor Space}
\seclab\twistor

\nref\penroserindle{R. Penrose and W. Rindler, ¡°Spinors And
Space-Time. Vol. 2: Spinor And Twistor Methods In Space-Time
Geometry,¡± Cambridge University Press, 1986 .} \nref\berk{N.
Berkovits, S. A. Cherkis, Higher-dimensional twistor transforms
using pure spinors, JHEP 0412 (2004) 049, [arXiv: hep-th/0409243].}

We now try to unify the various elements that we have investigated
in the previous sections into a unique framework to formulate the
QNG theory in terms of the variables of supertwistor space. We
haven't finished this program, but some results may be notable. In
this section, we'll construct the supertwistor space $\hat{\BP\BT}$
corresponding to the QNG theory, and implement the superconformal
symmetry $U^*Sp(4|2, \H)\subset OSp(8|4, \ct)$ in it. We'll then
encode the information of the full free tensor multiplet into a
superfield $\Psi$ in $\hat{\BP\BT}$, by utilizing the Penrose
transformation. And we'll propose a super-twistor space effective
action of the $\Psi$ field.

Some speculations concerning the possible non-Abelian generalization
and the super-twistor formulation of the QNG theory are also
presented in this section.

\subsec{The Twistor and Supertwistor Space}

\bigskip\noindent{\it{Twistor Space And Conformal Symmetry}}

We can construct the twistor space $\BT$ of the $(1+5)$-dimensional
spacetime $\W$ as follows. Firstly, we complexify $\W$ as $\BFW$. We
then conformally compactify $\BFW$ as the standard quadric $\BFQ$ of
the 7-dimensional complex projective space $\BFP$ of 8-dimensional
vector space $\BFV$. In terms of the homogeneous coordinates
${\v}^{I} \{I=1,2...8\}$ of $\BFP$, $\BFQ$ can be given as
\eqn\defineq{(\v^1)^2+(\v^2)^2+...+(\v^8)^2=0.} The subgroup of the
automorphism of $\BFP$ that preserves $\BFQ$ is $\Spin(8, \ct)$,
which is the complexification of the conformal symmetry
$\Spin(2,6)$.

On the local patch $\BFW$ of $\BFQ$, one can take the local
coordinates $\{\x^{\mu}, \mu=1,2...,6\}$. To see the action of
$\Spin(8, \ct)$ on these local coordinates explicitly, we set
$u=\v^7+i\v^8$, $v=\v^7-i\v^8$, and rewrite the equation
\defineq\ as \eqn\newq{(\v^1)^2+(\v^2)^2+...+(\v^6)^2+uv=0.} Let
$\BFW$ be the $u\neq 0$ local patch, and let the local coordinates
$\x^{\mu}$ be $\x^{\mu}=\v^{\mu}/u$. In terms of these coordinates,
$\BFW$ can be described as $\x^2+u/v=0$.

On $\BFW$, the generator $i(\v^7{\p \over \p \v^8}-\v^8{\p \over \p
\v^7})=-(u{\p \over\p u}-v{\p \over\p v})$ of the rotation in
$\{7,8\}$ plane acts as $D=-\x\cdot{\p \over\p \x}$. And the
symmetry subgroup $\Spin(6, \ct)$ that commutes with $D$ acts as the
complexified Lorentz group of $\BFW$. Finally, the translation
$\P^{\mu}$ and the special conformal translation $\K^{\mu}$ can be
determined as $\P_{\mu}=(u{\p\over\p \v^{\mu}}-2\v_{\mu}{\p\over\p
v})$ and $\K_{\mu}=(v{\p\over\p \v^{\mu}}-2\v_{\mu}{\p \over \p
u})$. Obviously, $[D, \P^{\mu}]=\P^{\mu},\ [D, \K^{\mu}]=-\K^{\mu}$.

The twistor space $\BT$ is the 8-dimensional complex space $\CS^+$
of the chiral spinors of $\Spin(8, \ct)$. The associated
7-dimensional projective space $\BPT$ is the associated projective
twistor space. One can equip $\BT$ with a symmetrical quadratic form
$(\phi, \phi')_{\BT}$ that preserves $\Spin(8, \ct)$, where $\phi
\in \BT$ and $\phi' \in \BT$ are two arbitrary chiral spinors of
$\Spin(8, \ct)$. This determines a 6-dimensional quadric $\BQ$ in
projective twistor space $\BP\BT$, with $\phi^{\sigma}\in \BQ$
satisfying $(\phi,\phi)_{\BT}=0$.

The standard correspondence between $\BFQ$ and $\BQ$ (due to the
triality of $\Spin(8,\ct)$) tells us that the points of complex
spacetime $\BFQ$ are parameterizing the 3-plane in twistor quadric
$\BQ$ (notice that any 6-dimensional complex quadric contains
3-plane) and the 3-planes in $\BFQ$ are parameterized by the points
of $\BQ$.

We now decompose the chiral spinor $\phi^{\sigma}$ into
$\mu_{\alpha}$ and $\lambda^{\alpha}$ according to the different
eigenvalues under $D$, $\phi^{\sigma}=(\lambda^{\alpha},
\mu_{\beta})$, where \eqn\dd{[D,
\lambda^{\alpha}]=\half\lambda^{\alpha},\ [D,
\mu_{\alpha}]=-\half\mu_{\alpha}.} And under this decomposition the
quadratics $\BQ$ will be described as
\eqn\qqnewdefine{\lambda^{\alpha}\mu_{\alpha}=0.} Under the action
of the $\Spin(6, \ct)$, $\lambda^{\alpha}$ transforms as a chiral
spinor, while $\mu_{\alpha}$ transforms as an anti-chiral spinor.
\dd\ indicates that the dilation generator $D$ should act as
\eqn\dspin{D=\half\left(\lambda^{\alpha}{\p\over \p
\lambda^{\alpha}}-\mu_{\alpha}{\p \over \p\mu_{\alpha}}\right).}

The relationship between $\BFW$ and the twistor quadric $\BQ'$ can
be explicitly formulated as a generalized Penrose equation
\eqn\penrose{\mu_{\alpha}+\x_{\alpha\beta}\lambda^{\beta}=0.} By
noticing that the $\alpha\beta$ indexes of $\x_{\alpha\beta}$ are
antisymmetric, one can easily see that for a given $\x^{\mu}$,
\penrose\ satisfies \qqnewdefine\ automatically. This means that the
3-plane $\BD_{\x}$ of $\BPT$ described by \penrose\ are in fact the
3-plane of twistor quadric $\BQ'$. On the other hand, given
$(\lambda^{\alpha}, \mu_{\alpha})$ fixed, the correspond
$\x_{\alpha\beta}$ is determined by \penrose\ up to a shift
$V_{\alpha\beta}$ that satisfies
\eqn\alphaplane{V_{\alpha\beta}\lambda^{\beta}=0.} This equation
always have three linear independent solutions since any chiral
spinor $\lambda^{\alpha}$ of $\Spin(6,\ct)$ is a pure spinor. Thus,
any point $(\lambda^{\alpha}, \mu_{\beta})$ of $\BQ$ corresponds to
a 3-plane of $\BFQ$, such a 3-plane can be named as a
$\alpha$-plane, in terms of the terminology of twistor theory.

By utilizing \penrose, one can write out the expressions of the
other generators of the conformal symmetry group in terms of the
twistor coordinates. For examples,
\eqn\pkspin{\eqalign{&\P^{\alpha\beta}=\lambda^{\alpha}{\p\over\p
\mu_{\beta}}-\lambda^{\beta}{\p\over\p \mu_{\alpha}},\cr
&\K_{\alpha\beta}=\mu_{\alpha}{\p \over
\p\lambda^{\beta}}-\mu_{\beta}{\p \over \p\lambda^{\alpha}}.}} The
$\Spin(6, \ct)$ rotations act as
\eqn\jspin{J^{\alpha}_{~\beta}=\left(\lambda^{\alpha}{\p \over
\p\lambda^{\beta}}-\mu_{\beta}{\p\over\p
\mu_{\alpha}}\right)-{1\over
4}\delta^{\alpha}_{\beta}\left(\lambda^{\delta}{\p\over
\p\lambda^{\delta}}-\mu_{\delta}{\p\over\p \mu_{\delta}}\right).}

To return to the real $(1+5)$-dimensional spacetime $\W$, one should
impose an appropriate real structure $\tau$ on the twistor quadric
$\BQ$. The action of $\tau$ can be explicitly written as
\eqn\realsix{\eqalign{\tau: \lambda^{\alpha}\rightarrow
\tilde{\lambda}^{\alpha}=\tau^{\alpha}_{~\beta}\bar{\lambda}^{\beta},\
\mu_{\alpha}\rightarrow
\tilde{\mu}_{\alpha}=\tau_{\alpha}^{~\beta}\bar{\mu}_{\beta}.}} By
using Penrose equation \penrose, one can get the action of $\tau$ on
$\x^{\alpha\beta}$. All the $\x^{\alpha\beta}$ that commute with
$\tau$ form the real slice $\W$ of the complex spacetime $\BFQ$. In
terms of the homogeneous coordinates, $\W$ is described by
\eqn\realw{-(\v^1)^2+(\v^2)^2+...+(\v^7)^2-(\v^8)^2=0.}

\bigskip\noindent{\it{Supertwistor}}

To generalize above construction of the twistor space $\BQ$ to the
super-symmetrical case, with the super-conformal symmetry group
$OSp(2,6|2)$, we will introduce four fermionic homogenous
coordinates $\psi^A$ to the $\ct^4_{\BR}$, which appear in section
\qagt. The construction in section \qagt\ indicates that $\psi^A$
transforms as the $\bf 4$ representation under $Sp(2,\H)_{\BR}$.

The full super-twistor space $\hat{\BT}$ now can be constructed as
the super-linear space $\ct^{8|4}$ whose bosonic part is coordinated
as $(\lambda^{\alpha}, \mu_{\beta})$ and whose fermionic part is
coordinated as $\psi^A$. The projective space of $\hat{\BT}$ is
denoted as $\hat{\BP\BT}$. The correspond super-twistor quadric
$\hat{\BQ}$ can be given as
\eqn\hatq{\lambda^{\alpha}\mu_{\alpha}+\half\psi^{A}\psi_{A}=0.}

Besides the previous generators \dspin,\pkspin,\jspin, there are
some additional generators that preserve \hatq. The additional
bosonic generators are
$\BR^{A}_{~B}=\psi^A{\p\over\p\psi^{B}}-\psi_B{\p\over\p\psi_A}$,
the additional fermionic $Q$-generators are
\eqn\qscharges{\eqalign{Q^{\alpha}_{~A}=\psi_A{\p\over\p
\mu_{\alpha}}+\lambda^{\alpha}{\p\over\p\psi^A},}} and the
additional fermionic $S$-generators are $S_{\alpha
A}=\mu_{\alpha}{\p\over \p\psi^A}-\psi_A{\p\over\p
\lambda^{\alpha}}$. Here, we have used the same symbols for the
superconformal generators and the generators of the super-twistor
symmetries since we'll identified them in what follows.

Such an identification can be achieved by generalizing the Penrose
equation \penrose\ to the fermionic coordinates
\eqn\penrosefermion{\psi_A+\theta_{A\alpha}\lambda^{\alpha}=0,} as
one can see by comparing \qscharges\ to \qderivative.

Given $\theta_{A\alpha}$ satisfying the symplectic-Majorana
condition, \penrosefermion\ is preserved by the real structure
$\hat{\tau}$ which acts as
$\psi_A\rightarrow\tilde{\psi}_{A}=\omega_{AB}\bar{\psi}^{B}$,
$\lambda^{\alpha}\rightarrow\tilde{\lambda}^{\alpha},
\mu_{\alpha}\rightarrow\tilde{\mu}_{\alpha}$. Clearly, $\hat{\tau}$
preserves the twistor quadric $\hat{\BQ}$. This gives a natural real
structure to $\hat{\BP\BT}$ that gives out the real superspace of
the $5+1$ dimensional spacetime. And the $\BR^{A}_{~B}$ generators
that commute with this real structure naturally give out
$\BR^{(AB)}$. Hence, we get the full $OSp(2,6|2)$ superconformal
symmetries from the supertwistor space $\hat{\BQ}$.

\subsec{Towards A Formulation Of The $\rm{QNG}$ Theory}

\bigskip\noindent{\it{Encoding The Tensor Multiplet}}

Having constructed the supertwistor space of the $\EUN=(2,0)$
super-conformal field theory, we now try to transform the free
tensor multiplet into twistor space by using the six-dimensional
Penrose transformation, which has been developed in
\penroserindle\berk .

According to the Penrose transformation, solutions of the wave
equations for helicity $h$ are equivalent to the element of the
sheaf cohomology group $H^3(\BQ', \CO(-2h-4))$. To employ it, one
picks out an element $g(\lambda, \mu)$ of $H^3(\BQ', \CO(-2h-4))$,
restricts $g(\lambda, \mu)$ to be an element of the restricted
cohomology group $H^3(\BD_{\x}, \CO(-2h-4))$ for a fixed spacetime
point $\x$, and notices the natural measurement $\mu(\BD)$ on the
3-plane $\BD_{\x}$, with
$\mu(\BD)=\epsilon_{\alpha\beta\gamma\delta}\lambda^{\alpha}
d\lambda^{\beta}\wedge d\lambda^{\gamma}\wedge
d\lambda^{\delta}=\bigvev{\lambda, d\lambda, d\lambda, d\lambda}$.
The Penrose transformations can then be given out by appropriate
contour integrations over $\BD_{\x}$.

Now, we apply the Penrose transformation to the linear equations of
motion of the tensor multiplet that we have studied in section
\qagt. For the five scalars, $g(\lambda, \mu)$ can be written as
$\Phi_{AB}(\lambda, \mu)\in H^3(\BQ', \CO(-4))$, with
\eqn\phitwistor{\Phi_{AB}(\x)={1\over (2\pi
i)^3}\int_{\BD_{\x}}\mu(\BD)\cdot\Phi_{AB}(\lambda, \mu)\ .}

As an example, we set $\Phi_{AB}(\lambda,\mu)\sim
\omega_{AB}{\bigvev{\ld_1,\ld_2,\ld_3,\ld_4}/[(\mu,\ld_1)(\mu,\ld_2)(\mu,\ld_3)(\mu,\ld_4)]}\in
H^3(\BQ', \CO(-4))$, where $\lambda_1, \lambda_2, \lambda_3,
\lambda_4$ are four different chiral spinors of $\BPT$,
$\bigvev{\ld_1, \ld_2, \ld_3,
\ld_4}=\epsilon_{\alpha\beta\gamma\delta}\ld^{\alpha}_1
\ld^{\beta}_{2}\ld^{\gamma}_3\ld^{\delta}_{4}$. By performing the
contour integration we can get \berk\
\eqn\eg{\Phi_{AB}(\x)=\int_{\BD_{\x}}{\bigvev{\lambda, d\lambda,
d\lambda, d\lambda}\over (2\pi
i)^3}{\bigvev{\ld_1,\ld_2,\ld_3,\ld_4}\over{(\mu,\ld_1)(\mu,\ld_2)(\mu,\ld_3)(\mu,\ld_4)}}\omega_{AB}\sim
{1\over |\x|^4}\omega_{AB}\ ,} which is a wave function
$\Phi_{AB}(\x)$ on spacetime $\BFW$, with a delta function source
supported on the origin.

For the chiral fermion $\Psi^{\alpha}_{~A}$, we need to decompose it
into the $+1/2$ helicity part $\Psi^{\alpha}_{+A}$ and the $-1/2$
helicity part $\Psi^{\alpha}_{-A}$ as we have done in section \tm.
For the positive helicity part, the Penrose transformation can be
simply given as \eqn\psitwistorp{\Psi^{\alpha}_{+A}(\x)={1\over
(2\pi
i)^3}\int_{\BD_{\x}}\mu(\BD)\cdot\lambda^{\alpha}\Psi_{+A}(\lambda,
\mu),} where $\Psi_{+A}(\lambda, \mu)\in H^3(\BQ', \CO(-5))$ is the
correspond twistor wave function. To get the Penrose transformation
of the negative helicity part $\Psi^{\alpha}_{-A}$, one notices that
on $\BD_{\x}$ the plane wave $\tilde{\lambda}^{\alpha}\exp(i{1\over
2}\lambda^{\alpha}\wedge \tilde{\lambda}^{\beta}\x_{\alpha\beta})$
can be rewritten as
$\tilde{\lambda}^{\alpha}\exp(\mu_{\beta}\tilde{\lambda}^{\beta})
\sim(\p/\p\mu_{\alpha})\exp(\mu_{\beta}\tilde{\lambda}^{\beta})$ by
using the Penrose equation \penrose\ . Thus, the Penrose
transformation can be written as
\eqn\psitwistorn{\Psi^{\alpha}_{-A}(\x)={1\over (2\pi
i)^3}\int_{\BD_{\x}}\mu(\BD)\cdot{\p\over
\p\mu_{\alpha}}\Psi_{-A}(\lambda, \mu),} where the twistor wave
function $\Psi_{-A}(\lambda, \mu)$ is an element of $H^3(\BQ',
\CO(-3))$.

Quite similar arguments will convince us that the Penrose
transformations for the chiral tensor field $H^{(\alpha\beta)}$
should be given as follows. Firstly, the transformations for the
$+1$ helicity part $H^{(\alpha\beta)}_{+}$ and $-1$ helicity part
$H^{(\alpha\beta)}_-$ can be written as
\eqn\htwistor{\eqalign{H^{(\alpha\beta)}_{+}(\x)&={1\over (2\pi
i)^3}\int_{\BD_{\x}}\mu(\BD)\cdot\lambda^{\alpha}\lambda^{\beta}H_+(\lambda,
\mu)\cr H^{(\alpha\beta)}_{-}(\x)&={1\over (2\pi
i)^3}\int_{\BD_{\x}}\mu(\BD)\cdot{\p\over \p\mu_{\alpha}}{\p\over
\p\mu_{\beta}}H_-(\lambda, \mu),}} where the twistor wave functions
$H_{+}(\lambda, \mu)$ and $H_-(\lambda, \mu)$ are the elements of
$H^3(\BQ', \CO(-6))$ and $H^3(\BQ', \CO(-2))$, respectively.
Secondly, the Penrose transformation of the helicity $0$ part
$H^{(\alpha\beta)}_0$ can be written as
\eqn\htwistorz{H^{(\alpha\beta)}_{0}(\x)={1\over (2\pi
i)^3}\int_{\BD_{\x}}\mu(\BD)\cdot\left(\lambda^{\alpha}{\p\over
\p\mu_{\beta}}+\lambda^{\beta}{\p\over
\p\mu_{\alpha}}\right)H_0(\lambda, \mu),} where $H_0(\lambda,
\mu)\in H^3(\BQ', \CO(-4))$.

Now, we can encode the full free tensor multiplet as a superfield
$\Psi$ defined on the supertwistor space $\hat{\BP\BT}$,
\eqn\superftwistor{\eqalign{{\Psi}&=H_-+\psi^A\Psi^-_{A}+\half
\psi^A\psi_AH_0+\half\psi^A\psi^B\Phi_{AB}\cr &+{1\over
3!}\epsilon_{ABCD}\psi^A\psi^B\psi^C\Psi^D_{+}+{1\over
4!}\epsilon_{ABCD}\psi^A\psi^B\psi^C\psi^DH_+.}} Clearly,
$\Psi(\phi, \psi)$ is an element of the cohomology group
$H^3(\hat{\BP\BT}', \CO(-2))$. By viewing the sheaf cohomology group
$H^3(\hat{\BP\BT}')$ as the $\bar{\p}$ cohomology group
$H^3_{\bar{\p}}(\hat{\BP\BT}')$, one can identify the wave function
$\Psi$ as a $(0, 3)$ form on $\hat{\BP\BT}$.

There is a natural holomorphic measure
$\hat{\Omega}=\phi^1d\phi^2\wedge d\phi^2\wedge...\wedge
d\phi^8\wedge d\psi^1\wedge d\psi^2\wedge ...\wedge d\psi^4$ on
$\hat{\BP\BT}$. $\hat{\Omega}$ takes value in the line bundle
$\CO(4)$ since under $(\phi, \psi)\rightarrow t(\phi, \psi)$, $t\in
\ct^*$, $\hat{\Omega}\rightarrow t^4\hat{\Omega}$. Thus, at the
linear level, one can write out a well defined effective action for
the twistor wave function $\Psi$,
\eqn\twistoraction{I_{\Psi}={1\over
2}\int_{\hat{\BP\BT}}\hat{\Omega}\wedge
\left({\Psi}\bar{\p}{\Psi}\right).}

\bigskip\noindent{\it{Towards A Super-twistor Formulation Of The QNG Theory}}

We now suppose that for the non-Abelian cases with non-Abelian gerbe
group $G=U(N)\subset G_{\ct}=GL(N, \ct)$ the correspond supertwistor
wave function $\Psi$ (which is a $(0, 3)$ form) takes value in
$\a\d(E)\otimes\CO(-2)$, where $\a\d(E)$ denotes the adjoint bundle
of the $GL(N, \ct)$ holomorphic vector bundle $E\rightarrow
\hat{\BP\BT}$.

One may generalize \twistoraction\ to ${1\over
2}\int_{\hat{\BP\BT}}\hat{\Omega}\wedge
\Tr\left({\Psi}\bar{\p}{\Psi}\right)$, but this cannot be the
correct action since it is essentially a linear theory (and we have
known that the QNG can not be a linear theory). To cure this
problem, the simplest thing that we can do is to introduce a $(0,
1)$ connection ${\cal{A}}$ which takes value in $\a\d(E)\otimes\CO$,
and generalize the effective action $I_{\Psi}$ to some $I_{\Psi,
{\cal A}}$. One of the terms of $I_{\Psi, {\cal A}}$, that
corresponds to the term of $I_{\Psi}$, may be \eqn\psiaction{{1\over
2}\int_{\hat{\BP\BT}}\hat{\Omega}\wedge
\Tr\left({\Psi}\bar{\p}_{\cal A}{\Psi}\right),} where
$\bar{\p}_{\cal A}\Psi=\bar{\p}\Psi+[{\cal A}, \Psi]$ . The equation
of motion of $\Psi$ will give us $\bar{\p}_{\cal A}\Psi=0$, which
will define an element of $H^3(\hat{\BP\BT}',
\a\d(E)\otimes\CO(-2))$. And one may need to study holomorphic
3-planes $\hat{\BD}$ (and integrate over the moduli space of them)
of $\hat{\BP\BT}$, with additional terms
\eqn\Aaction{\half\int_{\hat{\BD}}\mu(\hat{\BD})\wedge \Tr({\cal
A}\bar{\p}{\cal A}+{2\over 3}{\cal A}\wedge {\cal A}\wedge {\cal
A})\ . }Here, $\mu(\hat{\BD})$ is the natural holomorphic measure on
$\hat{\BD}$, which takes value in the holomorphic line bundle $\CO$
of $\hat{\BP\BT}$. One can get the equation of motion
$\bar{\p}{\cal{A}}+{\cal A}\wedge{\cal A}=0$, which defines an
element of $H^1(\hat{\BD}, \a\d(E)\otimes\CO)$ on each holomorphic
3-plane.

The investigation in section \selfdualstring\ tells us that the
contributions of ${\cal A}$ may come from the objects that
correspond to the tensionless strings. Presently, the author is
totally ignorant to these potential objects.

\nref\hodge{A. Hodges, Eliminating spurious poles from
gauge-theoretic amplitudes, [arXiv: hep-th/0905.1473].
}\nref\work{Working in progress. }

To shed light on this obscure situation, trying to extract out the
twistor string theory -- proposed to capture the perturbative
structure of $\EUN=4$ super-Yang-Mills theory -- from some known
aspects of the super-twistor formulation for the QNG theory may be
desired, since the four-dimensional theory has a natural QNG theory
origin as we have seen in section \emduality\ . The similarity
between Hodges's momentum-twistor coordinates $(Z^{\alpha},
W_{\beta})$, introduced in \hodge\ to cure the problem of spurious
poles in gauge-theoretical scattering amplitudes, and our twistor
coordinates $(\lambda^{\alpha}, \mu_{\beta})$ may suggest some
interesting connections, we'll leave it to the future \work\ .

\bigskip

\cta The work was supported by the NSF of China under Grant NO.
10575106. I am grateful to Prof Z. Chang for his constant
encouragements and useful discussions. I would like to thank Prof J.
X. Lu for useful discussions, and my gratitude will go to H. Q.
Zhang for reading manuscript and making comments. The conversations
with X. Li and Y. Yue are also helpful to the author. I would like
to thank Prof E. Witten for comments on the non-Abelian version of
the $H^3$ cohomology and for suggesting the calculation to transform
the variables on twistor space to spacetime.

\listrefs
\end